\documentclass[11pt]{article}

\usepackage{amsmath,amssymb,amsthm}
\usepackage[dvipdfm]{graphicx}

\DeclareMathOperator*{\Tr}{Tr}
\newcommand{\ket}[1]{\left| #1 \right>}
\newcommand{\bra}[1]{\left< #1 \right|}
\newcommand{\rmd}{{\rm d}}
\newcommand{\rme}{{\rm e}}
\newcommand{\rmi}{{\rm i}}
\newcommand{\Vv}{\boldsymbol{v}}
\newcommand{\Vp}{\boldsymbol{p}}
\newcommand{\Vq}{\boldsymbol{q}}
\newcommand{\Vzero}{\boldsymbol{0}}
\DeclareMathOperator*{\osc}{osc}

\newtheorem{thm}{Theorem}[section]
\newtheorem{lem}[thm]{Lemma}
\newtheorem{cor}[thm]{Corollary}
\theoremstyle{remark}
\newtheorem*{rem}{Remark}

\title{Mathematical Foundation of Quantum Annealing}
\author{Satoshi Morita\thanks{International School for Advanced Studies (SISSA),
Via Beirut 2-4, I-34014 Trieste, Italy}
~and Hidetoshi Nishimori\thanks{Department of Physics, Tokyo 
Institute of Technology,
Oh-okayama, Meguro-ku, Tokyo 152-8551, Japan}}
\date{}
\begin{document}
\maketitle

\abstract{
Quantum annealing is a generic name of quantum algorithms to use
quantum-mechanical fluctuations to search for the solution of optimization problem.
It shares the basic idea with quantum adiabatic evolution
studied actively in quantum computation.
The present paper reviews the mathematical and theoretical
foundation of quantum annealing.
In particular, theorems are presented for convergence conditions
of quantum annealing to the target optimal state after an infinite-time evolution
following the Schr\"odinger or stochastic (Monte Carlo) dynamics.
It is proved that the same asymptotic behavior of the control parameter
guarantees convergence both for the Schr\"odinger
dynamics and the stochastic dynamics in spite of the essential difference
of these two types of dynamics.
Also described are the prescriptions to reduce errors in the final approximate
solution obtained after a long but finite dynamical evolution of quantum annealing.
It is shown there that we can reduce errors significantly by an ingenious
choice of annealing schedule (time dependence of the control parameter)
without compromising computational complexity qualitatively.
A review is given on the derivation of the convergence condition for classical
simulated annealing from the view point of quantum adiabaticity
using a classical-quantum mapping.
}

\section{Introduction}

An optimization problem is a problem to minimize or maximize a real
single-valued function of multivariables called the cost function
\cite{GareyJ, HartmannW}.  If the problem is to maximize the cost function $f$,
it suffices to minimize $-f$.
It thus does not lose generality to consider minimization only.
In the present paper we consider combinatorial optimization, in which
variables take discrete values.  Well-known examples are satisfiability
problems (SAT), Exact Cover, Max Cut, Hamilton graph, and Traveling
Salesman Problem.  In physics, the search of the ground state of spin
systems is a typical example, in particular, systems with quenched
randomness like spin glasses.

Optimization problems are classified roughly into two types, easy and hard ones.
Loosely speaking, easy problems are those for which we have algorithms to
solve in steps(=time) polynomial in the system size (polynomial complexity).
In contrast, for hard problems, all known algorithms take exponentially
many steps to reach the exact solution (exponential complexity).
For these latter problems it is virtually impossible to find the exact
solution if the problem size exceeds a moderate value.
Most of the interesting cases as exemplified above belong
to the latter hard class.

It is therefore important practically to devise algorithms which give
approximate but accurate solutions efficiently, {\em i.e.} with 
polynomial complexity.
Many instances of combinatorial optimization problems have such 
approximate algorithm.
For example, the Lin-Kernighan algorithm is often used to solve
the traveling salesman problem within a reasonable time \cite{Helsgaun}.

In the present paper we will instead discuss generic algorithms,
simulated annealing (SA) and quantum annealing (QA).
The former was developed from the analogy between
optimization problems and statistical physics
\cite{KirkpatrickGV,AartsK}.  In SA, the cost function to be
minimized is identified with the energy of a statistical-mechanical
system.
The system is then given a
temperature, an artificially-introduced control parameter, by reducing
which slowly from a high value to zero, we hope to drive the system to
the state with the lowest value of the energy (cost function), reaching
the solution of the optimization problem.  The idea is that the system
is expected to stay close to thermal equilibrium during time evolution
if the rate of decrease of temperature is sufficiently slow, and is thus
lead in the end to the zero-temperature equilibrium state, the
lowest-energy state.  In practical applications SA is immensely popular
due to its general applicability, reasonable performance, and
relatively easy implementation in most cases.  SA is usually used as a
method to obtain an approximate solution within a finite computation
time since it needs an infinitely long time to reach the exact solution
by keeping the system close to thermal equilibrium.

Let us now turn our attention to quantum annealing
\cite{Finnila,KadowakiN,Kadowaki,DasC,SantoroT,DasC08}
\footnote{
The term quantum annealing first appeared in 
\cite{ApolloniCdF1989,ApolloniCdF1990},
in which the authors used quantum transitions for state search and the
dynamical evolution of control parameters were set by hand as an algorithm.
Quantum annealing in the present sense using natural Schr\"odinger dynamics
was proposed later independently in \cite{Finnila} and \cite{KadowakiN}.
}
.
In SA, we make use of thermal (classical) fluctuations to let the system
hop from state to state over intermediate energy barriers to search
for the desired lowest-energy state.
Why then not try quantum-mechanical fluctuations (quantum tunneling) for
state transitions if such may lead to better performance?
In QA we introduce artificial degrees of freedom of quantum nature,
non-commutative operators, which induce quantum fluctuations.
We then ingeniously control the strength of these quantum fluctuations
so that the system finally reaches the ground state, just like SA
in which we slowly reduce the temperature.
More precisely, the strength of quantum fluctuations is first set to
a very large value for the system to search for the global structure
of the phase space, corresponding to the high-temperature situation in SA.
Then the strength is gradually decreased to finally vanish
to recover the original system hopefully in the lowest-energy state.
Quantum tunneling between different classical states replaces
thermal hopping in SA.
The physical idea behind such a procedure is to keep the system
close to the instantaneous ground state of the quantum system,
analogously to the quasi-equilibrium state to be kept during
the time evolution of SA.
Similarly to SA, QA is a generic algorithm applicable, in principle,
to any combinatorial optimization problem and is used as a method to
reach an approximate solution within a given finite amount of time.

The reader may wonder why one should invent yet another generic
algorithm when we already have powerful SA.
A short answer is that QA outperforms SA in most cases, at least theoretically.
Analytical and numerical results indicate that the computation time needed to
achieve a given precision of the answer is shorter in QA than in SA.
Also, the magnitude of error is smaller for QA than SA if we run the algorithm
for a fixed finite amount of time.
We shall show some theoretical bases for these conclusions in this paper.
Numerical evidence is found in
\cite{DasC, SantoroT, DasC08,SantoroMTC, MartonakST2002, SuzukiO, SarjalaPA, SuzukiNS,
MartonakST2004, StellaST2005, StellaST2006, DasCS}.

A drawback of QA is that a full practical implementation should rely on the
quantum computer because we need to solve the time-dependent Schr\"odinger
equation of very large scale.
Existing numerical studies have been carried out either for small prototype
examples or for large problems by Monte Carlo simulations using the
quantum-classical mapping by adding an extra (Trotter or imaginary-time)
dimension \cite{Trotter, Suzuki, LandauB}.
The latter mapping involves approximations, which inevitably introduces
additional errors as well as the overhead caused by the extra dimension.
Nevertheless, it is worthwhile to clarify the usefulness and limitations
of QA as a theoretical step towards a new paradigm of computation.
This aspect is shared by quantum computation in general whose practical
significance will be fully exploited on the quantum computer.

The idea of QA is essentially the same as quantum adiabatic evolution 
(QAE), which
is now actively investigated as an alternative paradigm of quantum
computation \cite{FarhiGGS}.
It has been proved that QAE is equivalent to the conventional circuit
model of quantum computation \cite{Mizel}, but QAE is sometimes
considered more useful than the circuit model for several reasons
including robustness against external disturbance.
In the literature of quantum computation, one is often interested in 
the computational
complexity of the QAE-based algorithm for a given specific problem
under a fixed value of acceptable error.
QAE can also be used to find the final {\em quantum} state when the 
problem is not
a classical optimization.

In some contrast to these situations on QAE, studies of QA are often focused
not on computational complexity but on the theoretical convergence conditions
for infinite-time evolution and on the amount of errors in the final state
within a fixed evolution time.
Such a difference may have lead some researchers to
think that QA and QAE are to be distinguished from each other.
We would emphasize that they are essentially the same and worth investigations
by various communities of researchers.

The structure of the present paper is as follows.
Section \ref{sec:QA} discusses the convergence condition of QA, in particular the
rate of decrease of the control parameter representing quantum fluctuations.
It will be shown there that a qualitatively faster decrease of the 
control parameter
is allowed in QA than in SA to reach the solution.
This is one of the explicit statements of the claim more vaguely stated above
that QA outperforms SA.
In Sec.~\ref{sec:SA} we review the performance analysis of SA using 
quantum-mechanical tools.
The well-known convergence condition for SA will be rederived from 
the perspective
of quantum adiabaticity.  The methods and results in this section
help us strengthen the interrelation between QA, SA and QAE.
The error rate of QA after a finite-time dynamical evolution
is analyzed in Sec.~\ref{sec:finiteQA}.
There we explain how to reduce the final residual error after evolution
of a given amount of time.  This point of view is unique in the sense that
most references of QAE study the time needed to reach a given amount of
tolerable error, {\em i.e.} computational complexity.
The results given in this section can be used to qualitatively reduce
residual errors for a given algorithm without compromising 
computational complexity.
Convergence conditions for stochastic implementation of QA are
discussed in Sec.~\ref{sec:QMC}.  The results are surprising in that
the rate of decrease of the control parameter for the system to reach the solution
coincides with that found in Sec.~\ref{sec:QA} for the pure quantum-mechanical
Schr\"odinger dynamics.  The stochastic (and therefore classical) dynamics
shares the same convergence conditions as fully quantum dynamics.
Summary and outlook are described in the final section.

The main parts of this paper (Secs. \ref{sec:QA}, \ref{sec:finiteQA}
and \ref{sec:QMC}) are based on the PhD Thesis of one of the authors
(S.M.) \cite{MoritaThesis} as well as several original papers of the present
and other authors as will be referred to appropriately.
The present paper is not a comprehensive review of QA since an emphasis is
given almost exclusively to the theoretical and mathematical aspects.
There exists an extensive
body of numerical studies and the reader is referred to \cite{DasC,
SantoroT,DasC08} for reviews.

\section{Convergence condition of QA -- Real-time Schr\"odinger evolution}
\label{sec:QA}

The convergence condition of QA with the real-time Schr\"{o}dinger dynamics
is investigated in this section, following \cite{MoritaN07}.
We first review the proof of the adiabatic theorem \cite{Messiah}
to be used to derive the convergence condition.
Then introduced is the Ising model with transverse field as a simple
but versatile implementation of QA. The convergence condition is
derived by solving the condition for adiabatic transition
with respect to the strength of the transverse field.

\subsection{Adiabatic theorem}

Let us consider the general Hamiltonian which depends on time $t$ only
through the dimensionless time $s=t/\tau$,
\begin{equation}
  H(t)=\tilde{H}\left(\frac{t}{\tau}\right)\equiv \tilde{H}(s).
   \label{eq:general Hamiltonian}
\end{equation}
The parameter $\tau$ is introduced to control the rate of change of
the Hamiltonian. In natural quantum systems, the state vector $\ket{\psi(t)}$
follows the real-time Schr\"{o}dinger equation,
\begin{equation}
  {\rm i}\frac{{\rm d}}{{\rm d}t}\ket{\psi(t)}=H(t)\ket{\psi(t)},
   \label{eq:Schrodinger equation}
\end{equation}
or, in terms of the dimensionless time,
\begin{equation}
  {\rm i}\frac{{\rm d}}{{\rm d}s}|\tilde{\psi}(s)\rangle
   =\tau \tilde{H}(s)|\tilde{\psi}(s)\rangle,\label{eq:TFIM_Schrodinger}
\end{equation}
where we set $\hbar =1$. We assume that the initial state is chosen to
be the ground state of the initial Hamiltonian $H(0)$ and that the
ground state of $\tilde{H}(s)$ is not degenerate for $s\geq 0$.
We show in the next section that the transverse-field Ising model, to
be used as $H(t)$ in most parts of this paper, has no
degeneracy in the ground state (except possibly in the limit of $t\to\infty$).
If $\tau$ is large, the Hamiltonian changes slowly and it is expected
that the state vector keeps track of the instantaneous ground state. The
adiabatic theorem provides the condition for adiabatic evolution. To
see this, we derive the asymptotic form of the state vector with respect
to the parameter $\tau$.

Since we wish to estimate how close the state vector is to the ground
state, it is natural to expand the state vector by the instantaneous
eigenstates of $\tilde{H}(s)$. Before doing so, we derive useful formulas for
the eigenstates. The $k$th instantaneous eigenstate of $\tilde{H}(s)$
with the eigenvalue $\varepsilon_k(s)$ is denoted as $\ket{k(s)}$,
\begin{equation}
  \tilde{H}(s)\ket{k(s)}=\varepsilon_k(s)\ket{k(s)}.
   \label{eq:eigenvalue equation}
\end{equation}
We assume that $\ket{0(s)}$ is the ground state of $\tilde{H}(s)$ and
that the eigenstates are orthonormal,
$\bra{j(s)}k(s)\rangle=\delta_{jk}$. From differentiation of
(\ref{eq:eigenvalue equation}) with respect to $s$, we obtain
\begin{equation}
  \bra{j(s)}\frac{\rmd}{\rmd s}\ket{k(s)}
   =\frac{-1}{\varepsilon_j(s)-\varepsilon_k(s)}
   \bra{j(s)}\frac{\rmd \tilde{H}(s)}{\rmd s}\ket{k(s)},
   \label{eq:derivative of eigenstate}
\end{equation}
where $j\neq k$. In the case of $j=k$, the same calculation does not
provide any meaningful result. We can, however, impose the following
condition,
\begin{equation}
  \bra{k(s)}\frac{\rmd}{\rmd s}\ket{k(s)}=0.
   \label{eq:eigenstate condition}
\end{equation}
This condition is achievable by the time-dependent phase shift:
If $|\tilde{k}(s)\rangle=\rme^{\rmi \theta(s)}\ket{k(s)}$, we find
\begin{equation}
  \langle \tilde{k}(s)|\frac{\rmd}{\rmd s}|\tilde{k}(s)\rangle
  =\rmi \frac{\rmd \theta}{\rmd s}
  +\bra{k(s)}\frac{\rmd}{\rmd s}\ket{k(s)}.
\end{equation}
The second term on the right-hand side is purely imaginary because
\begin{equation}
  \left[\bra{k(s)}\frac{\rmd}{\rmd s}\ket{k(s)}\right]^*+
   \bra{k(s)}\frac{\rmd}{\rmd s}\ket{k(s)}
  =\frac{\rmd}{\rmd s}\langle k(s)|k(s)\rangle=0.
\end{equation}
Thus, the condition (\ref{eq:eigenstate condition}) can be satisfied by
tuning the phase factor $\theta(s)$ even if the original eigenstate does
not satisfy it.

\begin{thm}
  \label{theorem:adiabatic}
  If the instantaneous ground state of the Hamiltonian $\tilde{H}(s)$ is
  not degenerate for $s\geq 0$ and the initial state is the ground state
  at $s=0$, {\it i.e.}  $|\tilde{\psi}(0)\rangle =\ket{0(0)}$, the state
  vector $|\tilde{\psi}(s)\rangle$ has the asymptotic form in the limit of
  large $\tau$ as
  \begin{equation}
   |\tilde{\psi}(s)\rangle=\sum_{j}c_j(s)
   \rme^{-\rmi \tau\phi_j(s)}\ket{j(s)},
   \label{eq:expansion of state vector}
  \end{equation}
  \begin{equation}
   c_0(s)\approx 1+\mathcal{O}(\tau^{-2}),
  \end{equation}
  \begin{equation}
   c_{j\neq0}(s)\approx \frac{\rmi}{\tau}
    \left[A_j(0)-\rme^{\rmi\tau[\phi_j(s)-\phi_0(s)]}A_j(s)
                    \right]+\mathcal{O}(\tau^{-2}),
    \label{eq:excitation amplitude}
  \end{equation}
  where $\phi_j(s)\equiv\int_{0}^{s}\rmd s' \varepsilon_j(s')$ ,
  $\Delta_j(s)\equiv\varepsilon_j(s)-\varepsilon_0(s)$ and
  \begin{equation}
   A_j(s)\equiv \frac{1}{\Delta_j(s)^2}\bra{j(s)}
    \frac{\rmd \tilde{H}(s)}{\rmd s}\ket{0(s)}.
    \label{eq:definition of A_j(s)}
  \end{equation}
\end{thm}

\begin{proof}
  Substitution of (\ref{eq:expansion of state vector}) into the
  Schr\"{o}dinger equation (\ref{eq:TFIM_Schrodinger}) yields
  the equation for the coefficient $c_j(s)$ as
  \begin{equation}
   \frac{\rmd c_j}{\rmd s}=\sum_{k\neq j}c_k(s)
    \frac{\rme^{\rmi\tau [\phi_j(s)-\phi_k(s)]}}
    {\varepsilon_j(s)-\varepsilon_k(s)}\bra{j(s)}
    \frac{\rmd \tilde{H}(s)}{\rmd s}\ket{k(s)},
  \end{equation}
  where we used (\ref{eq:derivative of eigenstate}) and (\ref{eq:eigenstate
  condition}). Integration of this equation yields
  \begin{equation}
   c_j(s)=c_j(0)+\sum_{k\neq j}\int_{0}^{s}\!\rmd \tilde{s}\, c_k(\tilde{s})
    \frac{\rme^{\rmi\tau[\phi_j(\tilde{s})-\phi_k(\tilde{s})]}}
    {\varepsilon_j(\tilde{s})-\varepsilon_k(\tilde{s})}\bra{j(\tilde{s})}
    \frac{\rmd \tilde{H}(\tilde{s})}{\rmd \tilde{s}}\ket{k(\tilde{s})}.
    \label{eq:integral equation for amplitudes}
  \end{equation}
  Since the initial state
  is chosen to be the ground state of $H(0)$, $c_0(0)=1$ and $c_{j\neq
  0}(0)=0$. The second term on the right-hand side is of the order of
  $\tau^{-1}$ because its integrand rapidly oscillates for large $\tau$.
  In fact, the integration by parts yields the $\tau^{-1}$-factor.  Thus,
  $c_{j\neq 0}(0)$ is of order $\tau^{-1}$ at most. Hence only the $k=0$
  term in the summation remains up to the order of $\tau^{-1}$,
  \begin{equation}
   c_{j\neq0}(s)\approx \int_{0}^{s}\!\rmd \tilde{s}
    \frac{\rme^{\rmi\tau[\phi_j(\tilde{s})-\phi_0(\tilde{s})]}}
    {\Delta_j(\tilde{s})}\bra{j(\tilde{s})}
    \frac{\rmd \tilde{H}(\tilde{s})}{\rmd 
\tilde{s}}\ket{0(\tilde{s})}+\mathcal{O}(\tau^{-2}),
  \end{equation}
  and the integration by parts yields (\ref{eq:excitation amplitude}).
\end{proof}

\begin{rem}
  The condition for the adiabatic evolution is given by the smallness of
  the excitation probability. That is, the right-hand side of
  (\ref{eq:excitation amplitude}) should be much smaller than unity. This
  condition is consistent with the criterion of the validity of the above
  asymptotic expansion. It is represented by
  \begin{equation}
  \tau \gg \left|A_j(s)\right|.
   \label{eq:adiabatic condition for tau}
  \end{equation}
  Using the original time variable $t$, this adiabaticity condition is
  written as
  \begin{equation}
  \frac{1}{\Delta_j(t)^2}
   \left|\bra{j(t)}\frac{\rmd H(t)}{\rmd t}\ket{0(t)}\right| =\delta
   \ll 1.
  \label{eq:adiabatic condition}
  \end{equation}
  This is the usual expression of adiabaticity condition.
\end{rem}

\subsection{Convergence conditions of quantum annealing}
In this section, we derive the condition which guarantees the
convergence of QA. The problem is what annealing schedule (time dependence
of the control parameter) would satisfy the adiabaticity condition
(\ref{eq:adiabatic condition}).
We solve this problem on the basis of the idea of Somma {\it et al} \cite{Somma}
developed for the analysis of SA in terms of quantum adiabaticity
as reviewed in Sec. \ref{sec:SA}.

\subsubsection{Transverse field Ising model}
\label{subsub:TFIgap}
Let us suppose that the optimization problem we wish to solve can be
represented as the ground-state search of an Ising model of general form
\begin{equation}
  H_{\rm Ising}\equiv -\sum_{i=1}^{N} J_i \sigma_i^z
  -\sum_{ij} J_{ij}\sigma_i^z \sigma_j^z
  -\sum_{ijk} J_{ijk}\sigma_i^z \sigma_j^z \sigma_k^z -\cdots,
  \label{eq:Ising potential}
\end{equation}
where the $\sigma_i^{\alpha}$ $(\alpha=x,y,z)$ are the Pauli matrices,
components of the spin $\frac{1}{2}$ operator at site $i$.
The eigenvalue of $\sigma_i^z$ is $+1$ or $-1$, which corresponds the
classical Ising spin.
Most combinatorial optimization problems can be written in this form
by, for example, mapping binary variables (0 and 1) to spin
variables ($\pm 1$).
Another important assumption is that the Hamiltonian (\ref{eq:Ising potential})
is extensive, {\em i.e.} proportional to the number of spins $N$ for large $N$.

To realize QA, a fictitious kinetic energy is introduced typically
by the time-dependent transverse field
\begin{equation}
  H_{\rm TF}(t) \equiv -\Gamma(t)\sum_{i=1}^{N}\sigma_i^x,
   \label{eq:transverse field}
\end{equation}
which induces spin flips, quantum fluctuations or quantum tunneling, between
the two states $\sigma_i^z=1$ and $\sigma_i^z=-1$, thus allowing a quantum
search of the phase space.
Initially the strength of the transverse field $\Gamma(t)$
is chosen to be very large, and the total Hamiltonian
\begin{equation}
  H(t)=H_{\rm Ising}+H_{\rm TF}(t)
   \label{eq:TFIM Hamiltonian}
\end{equation}
is dominated by the second kinetic term.
This corresponds to the high-temperature limit of SA.
The coefficient $\Gamma(t)$ is then gradually and monotonically 
decreased toward 0,
leaving eventually only the potential term $H_{\rm Ising}$.
Accordingly the state vector $\ket{\psi(t)}$, which follows the real-time
Schr\"{o}dinger equation, is expected to evolve from the trivial initial
ground state of the transverse-field term (\ref{eq:transverse field})
to the non-trivial ground state of (\ref{eq:Ising potential}),
which is the solution of the optimization problem.
An important issue is how slowly we should decrease
$\Gamma(t)$ to keep the state vector arbitrarily close to the
instantaneous ground state of the total Hamiltonian (\ref{eq:TFIM Hamiltonian}).
The following Theorem provides a solution to this problem as a 
sufficient condition.

\begin{thm} \label{theorem:TFIM_convergence}
  The adiabaticity (\ref{eq:adiabatic condition}) for the transverse-field Ising model
  (\ref{eq:TFIM Hamiltonian}) yields the time dependence of $\Gamma(t)$ as
  \begin{equation}
   \Gamma(t)= a (\delta t+c)^{-1/(2N-1)}
    \label{eq:adiabatic AS}
  \end{equation}
  for $t>t_0$ (for a given positive $t_0$) as a sufficient condition
  of convergence of QA.
  Here $a$ and $c$ are constants of $\mathcal{O}(N^0)$ and $\delta$ is
  a small parameter to control adiabaticity appearing in 
(\ref{eq:adiabatic condition}).
\end{thm}

The following Theorem proved by Hopf \cite{Hopf} will be useful to prove this
Theorem. See Appendix \ref{sec:Hopf} for the proof.
\begin{thm}
  \label{theorem:TFIM_Hopf}

  If all the elements of a square matrix $M$ are strictly
  positive, $M_{ij}>0$, its maximum eigenvalue $\lambda_0$ and any
  other eigenvalues $\lambda$ satisfy
  \begin{equation}
   |\lambda|\leq \frac{\kappa-1}{\kappa+1}\lambda_0,
    \label{eq:Hopf inequality}
  \end{equation}
  where $\kappa$ is defined by
  \begin{equation}
   \kappa\equiv\max_{i,j,k}\frac{M_{ik}}{M_{jk}}.
  \end{equation}
\end{thm}

\begin{proof} [{\bf Proof of Theorem \ref{theorem:TFIM_convergence}}]
  We show that the power decay (\ref{eq:adiabatic AS}) satisfies the
  adiabaticity condition (\ref{eq:adiabatic condition})
  which guarantees convergence to the ground state of $H_{\rm Ising}$ as $t\to\infty$.
  . For this purpose we
  estimate the energy gap and the time derivative of the Hamiltonian. As
  for the latter, it is straightforward to see
  \begin{equation}
   \left|\bra{j(t)} \frac{\rmd H(t)}{\rmd t}\ket{0(t)}\right|
    \leq -N\frac{\rmd \Gamma(t)}{\rmd t},
    \label{eq:derivative of Hamiltonian}
  \end{equation}
  since the time dependence of $H(t)$ lies only in the kinetic term 
$H_{\rm TF}(t)$,
  which has $N$ terms. Note that $\rmd\Gamma/\rmd t$ is negative.

  To estimate a lower bound for the energy gap, we apply Theorem
  \ref{theorem:TFIM_Hopf} to the operator $M\equiv(E_{+}-H(t))^N$. We assume
  that the constant $E_{+}$ satisfies $E_{+}>E_{\rm max}+\Gamma_0$, where
  $\Gamma_0\equiv\Gamma(t_0)$ and $E_{\rm max}$ is the maximum eigenvalue
  of the potential term $H_{\rm Ising}$. All the elements of the matrix
  $M$ are strictly positive in the representation that diagonalizes
  $\{\sigma_i^z\}$ because $E_{+}-H(t)$ is non-negative and
  irreducible, that is, any state can be reached from any other state
  within at most $N$ steps.

  For $t>t_0$, where $\Gamma(t) < \Gamma_0$, all the diagonal elements of
  $E_{+}-H(t)$ are larger than any non-zero off-diagonal element
  $\Gamma(t)$. Thus, the minimum element of $M$, which is between two
  states having all the spins in mutually opposite directions, is equal
  to $N!\Gamma(t)^N$, where $N!$ comes from the ways of permutation to
  flip spins. Replacement of $H_{\rm TF}(t)$ by $-N\Gamma_0$ shows that the
  maximum matrix element of $M$ has the upper bound $(E_{+}-E_{\rm
  min}+N\Gamma_0)^N$, where $E_{\rm min}$ is the lowest eigenvalue of
  $H_{\rm Ising}$. Thus, we have
  \begin{equation}
   \kappa\leq\frac{(E_{+}-E_{\rm min}+N\Gamma_0)^N}{N!\Gamma(t)^N}.
    \label{eq:kappa}
  \end{equation}
  If we denote the eigenvalue of $H(t)$ by $\varepsilon_j(t)$,
  (\ref{eq:Hopf inequality}) is rewritten as
  \begin{equation}
   \left[E_{+}-\varepsilon_j(t)\right]^N\leq\frac{\kappa-1}{\kappa+1}
    \left[E_{+}-\varepsilon_0(t)\right]^N.
  \end{equation}
  Substitution of (\ref{eq:kappa}) into the above inequality yields
  \begin{equation}
   \Delta_j(t)\geq\frac{2[E_{+}-\varepsilon_0(t)]N!}
    {N(E_{+}-E_{\rm min}+N\Gamma_0)^N}\Gamma(t)^N\equiv A\Gamma(t)^N,
    \label{eq:energy gap}
  \end{equation}
  where we used $1-((\kappa-1)/(\kappa+1))^{1/N}\geq
  2/N(\kappa+1)$ for $\kappa\geq 1$ and $N\geq 1$. The
  coefficient $A$ is estimated using the Stirling formula as
  \begin{equation}
   A\approx\frac{2\sqrt{2\pi N}[E_{+}-\varepsilon_0^{\rm max}]}{N \rme^N}
    \left(\frac{N}{E_{+}-E_{\rm min}+N\Gamma_0}\right)^N,
    \label{eq:coefficient_A}
  \end{equation}
  where $\varepsilon_0^{\rm max}$ is $\max_{t>t_0}\{\varepsilon_0 (t)\}$.
  This expression implies that $A$ is exponentially small for large $N$.

  Now, by combination of the above estimates (\ref{eq:derivative of
  Hamiltonian}) and (\ref{eq:energy gap}), we find that the sufficient
  condition for convergence for $t>t_0$ is
  \begin{equation}
   -\frac{N}{A^2 \Gamma(t)^{2N}}\frac{\rmd \Gamma(t)}{\rmd t}=\delta \ll 1,
  \end{equation}
  where $\delta$ is an arbitrarily small constant. By integrating this
  differential equation, we obtain (\ref{eq:adiabatic AS}).
\end{proof}

\begin{rem}
 The asymptotic power decay of the transverse field guarantees that the
 excitation probability is bounded by the arbitrarily small constant
 $\delta^2$ at each instant.  This annealing schedule is not valid when
 $\Gamma(t)$ is not sufficiently small because we evaluated the energy
 gap for $\Gamma(t)<\Gamma_0~(t>t_0)$. 
 If we take the limit $t_0\to 0$, $\Gamma_0$ increases indefinitely and
 the coefficient $a$ in  (\ref{eq:adiabatic AS}) diverges.
 Then the result (\ref{eq:adiabatic AS}) does not make sense.
 This is the reason why a finite positive time $t_0$ should be introduced
 in the statement of Theorem \ref{theorem:TFIM_convergence}.
\end{rem}

\subsubsection{Transverse ferromagnetic interactions}

The same discussions as above apply to QA using the transverse ferromagnetic
interactions in addition to a transverse field,
\begin{equation}
  H_{\rm TI}(t)\equiv -\Gamma_{\rm TI}(t)\left(
  \sum_{i=1}^{N}\sigma_i^x+\sum_{ij}\sigma_i^x \sigma_j^x\right).
\end{equation}
The second summation runs over appropriate pairs of sites that satisfy
extensiveness of the Hamiltonian.
A recent numerical study shows the effectiveness of this type of quantum kinetic
energy \cite{SuzukiNS}. The additional transverse interaction widens the
instantaneous energy gap between the ground state and the first excited
state. Thus, it is expected that an annealing schedule faster than
(\ref{eq:adiabatic AS}) satisfies the adiabaticity condition.
The following Theorem supports this expectation.

\begin{thm} \label{theorem:TFIM_TI}
  The adiabaticity for the quantum system $H_{\rm Ising}+H_{\rm TI}(t)$
  yields the time dependence of $\Gamma(t)$ for $t>t_0$ as
  \begin{equation}
   \Gamma_{\rm TI}(t) \propto t^{-1/(N-1)}.
  \end{equation}
\end{thm}

\begin{proof}
  The transverse interaction introduces non-zero off-diagonal elements to
  the Hamiltonian in the representation that diagonalizes
  $\sigma_i^z$. Consequently, any state can be reached from any other
  state within $N/2$ steps at most. Thus, the strictly positive operator
  is modified to $(E_{+}-H_{\rm Ising}-H_{\rm TI}(t))^{N/2}$, which leads to
  the lower bound for the energy gap as a quantity proportional to
  $\Gamma_{\rm TI}(t)^{N/2}$. The rest of the proof is the same as
  Theorem \ref{theorem:TFIM_convergence}.
\end{proof}

The above result implies that additional non-zero off-diagonal elements
of the Hamiltonian accelerates the convergence of QA. It is thus interesting
to consider the many-body transverse interaction of the form
\begin{equation}
  H_{\rm MTI}(t) = -\Gamma_{\rm MTI}(t)
   \prod_{i=1}^N \left(1+\sigma_i^x\right).
\end{equation}
All the elements of $H_{\rm MTI}$ are equal to $-\Gamma_{\rm MTI}(t)$ in
the representation that diagonalizes $\sigma_i^z$. In this system, the
following Theorem holds.

\begin{thm} \label{theorem:TFIM_TMI}
  The adiabaticity for the quantum system $H_{\rm Ising}+H_{\rm TMI}(t)$
  yields the time dependence of $\Gamma(t)$ for $t>t_0$ as
  \begin{equation}
   \Gamma_{\rm MTI}(t) \propto \frac{2^{N-2}}{\delta\, t}.
    \label{eq:adiabatic AS MTI}
  \end{equation}
\end{thm}

\begin{proof}
  We define the strictly positive operator as $M=E_{+}-H_{\rm
  Ising} -H_{\rm MTI}(t)$. The maximum and minimum matrix elements of $M$
  are $E_{+}-E_{\rm min}+\Gamma_{\rm MTI}(t)$ and $\Gamma_{\rm
  MTI}(t)$, respectively. Thus we have
  \begin{gather}
   \kappa = \frac{E_{+}-E_{\rm min}+\Gamma_{\rm MTI}(t)}
   {\Gamma_{\rm MTI}(t)},\\
   \frac{\kappa-1}{\kappa+1}=
   \frac{E_{+}-E_{\rm min}}{E_{+}-E_{\rm min}+2\Gamma_{\rm
   MTI}(t)}
   \geq 1-\frac{2\Gamma_{\rm MTI}(t)}{E_{+}-E_{\rm min}},
  \end{gather}
  The inequality for the strictly positive operator (\ref{eq:Hopf
  inequality}) yields
  \begin{equation}
   \Delta_j(t)\geq
    \frac{2\Gamma_{\rm MTI}(t) (E_{+}-\varepsilon_0^{\rm max})}
    {E_{+}-E_{\rm min}}\equiv \tilde{A}\, \Gamma_{\rm MTI}(t),
  \end{equation}
  where $\tilde{A}$ is $\mathcal{O}(N^0)$.  Since the matrix element of
  the derivative of the Hamiltonian is bounded as
  \begin{equation}
   \left|\bra{j(t)}\frac{\rmd H(t)}{\rmd t}\ket{0(t)}\right|\leq -2^N
   \frac{\rmd \Gamma_{\rm MTI}}{\rmd t},
  \end{equation}
  we find that the sufficient condition for convergence with the many-body
  transverse interaction is
  \begin{equation}
   -\frac{2^N}{\tilde{A}^2 \Gamma_{\rm MTI}(t)^2}
    \frac{\rmd \Gamma_{\rm MTI}}{\rmd t}=\delta \ll 1.
  \end{equation}
  Integrating this differential equation yields the annealing schedule
   (\ref{eq:adiabatic AS MTI}).
\end{proof}

\subsubsection{Computational complexity}

The asymptotic power-low annealing schedules guarantee the adiabatic
evolution during the annealing process.
The power-law dependence on $t$ is much faster than the log-inverse law for the
control parameter in SA, $T(t)= pN/\log (\alpha t+1)$, to be discussed in
the next section, first proved by Geman and Geman \cite{GemanG}.
However, it does not mean
that QA provides an algorithm to solve NP problems in polynomial
time. In the case with the transverse field only, the time for
$\Gamma(t)$ to reach a sufficiently small value $\epsilon$
(which implies that the system is sufficiently close to the final ground state
of $H_{\rm Ising}$ whence $H_{\rm TF}$ is a small perturbation)
is estimated from (\ref{eq:adiabatic AS}) as
\begin{equation}
  t_{\rm TF}\approx \frac{1}{\delta}\left(\frac{1}{\epsilon}\right)^{2N-1}.
\end{equation}
This relation clearly shows that the QA needs a time exponential in $N$ to
converge.

For QA with many-body transverse interactions, the exponent of $t$ in
the annealing schedule (\ref{eq:adiabatic AS MTI}) does not depend
on the system size $N$. Nevertheless, it also does not mean that QA
provides a polynomial-time algorithm because of the factor $2^N$. The
characteristic time for $\Gamma_{\rm MTI}$ to reach a sufficiently small
value $\epsilon$ is estimated as
\begin{equation}
  t_{\rm MTI}\approx \frac{2^{N-2}}{\delta \epsilon},
\end{equation}
which again shows exponential dependence on $N$.

These exponential computational complexities do not come as a surprise
because Theorems \ref{theorem:TFIM_convergence}, \ref{theorem:TFIM_TI}
and \ref{theorem:TFIM_TMI} all apply to any optimization problems
written in the generic form (\ref{eq:Ising potential}), which includes
the worst cases of most difficult problems.
Similar arguments apply to SA \cite{NishimoriI}.

Another remark is on the comparison of $\Gamma (t)(\propto t^{-1/(2N-1)})$
in QA with $T(t)(\propto N/\log (\alpha t+1))$ in SA to conclude
that the former schedule is faster than the latter.
The transverse-field coefficient $\Gamma$ in a quantum system plays the same role
qualitatively and quantitatively as the temperature $T$ does in a corresponding
classical system at least in the Hopfield model in a transverse field \cite{NishimoriNono}.
When the phase diagram is written in terms of $\Gamma$ and $\alpha$ (the
number of embedded patterns divided by the number of neurons) for the ground
state of the model, the result has precisely the same structure as the $T$-$\alpha$
phase diagram of the finite-temperature version of the Hopfield model
without transverse field.
This example serves as a justification of the direct comparison of $\Gamma$
and $T$ at least as long as the theoretical analyses of QA and SA are concerned.

\section{Convergence condition of SA and quantum adiabaticity}
\label{sec:SA}

We next study the convergence condition of SA to be compared with QA.
This problem was originally solved by Geman and Geman \cite{GemanG} using
the theory of inhomogeneous Markov chain as described in the Quantum
Monte Carlo context in Sec.~\ref{sec:QMC}.
It is quite surprising that their result is reproduced using the 
quantum adiabaticity
condition applied after a classical-quantum mapping \cite{Somma}.
This approach is reviewed in this section, following \cite{Somma},
to clarify the correspondence between the quasi-equilibrium condition for SA
in a classical system and the adiabaticity condition in the 
corresponding quantum system.
The analysis will also reveal an aspect related to the equivalence of QA and QAE.

\subsection{Classical-quantum mapping}

The starting point is an expression of a classical thermal 
expectation value in terms
of a quantum ground-state expectation value.
A well-known mapping between quantum and classical systems is to 
rewrite the former
in terms of the latter with an extra imaginary-time (or Trotter) 
dimension \cite{Suzuki}.
The mapping discussed in the present section is a different one, which allows us
to express the thermal expectation value of a classical system in terms of the
ground-state expectation value of a corresponding quantum system
{\em without} an extra dimension.

Suppose that the classical Hamiltonian, whose value we want to minimize, is
written as an Ising spin system as in (\ref{eq:Ising potential}):
\begin{equation}
  H= -\sum_{i=1}^{N} J_i \sigma_i^z
  -\sum_{ij} J_{ij}\sigma_i^z \sigma_j^z
  -\sum_{ijk} J_{ijk}\sigma_i^z \sigma_j^z \sigma_k^z -\cdots.
  \label{eq:Ising potential2}
\end{equation}
The thermal expectation value of a classical physical
quantity $Q(\{\sigma_i^z\})$ is
\begin{equation}
   \langle Q \rangle_T =\frac{1}{Z(T)}\sum_{\{ \sigma\}} e^{-\beta 
H}Q(\{\sigma_i\}),
   \label{eq:A-thermal-expectation}
\end{equation}
where the sum runs over all configurations of Ising spins, {\em i.e.} over
the values taken by the $z$-components of the Pauli matrices,
$\sigma_i^z=\sigma_i(\pm 1)~(\forall i)$.
The symbol $\{\sigma_i\}$ stands for the set 
$\{\sigma_1,\sigma_2,\cdots ,\sigma_N\}$.

An important element is the following Theorem.
\begin{thm}
\label{thm:classical-quantum-mapping}
The thermal expectation value (\ref{eq:A-thermal-expectation}) is 
equal to the expectation
value of $Q$ by the quantum wave function
\begin{equation}
   |\psi (T)\rangle=e^{-\beta H/2}\sum_{\{\sigma\}} |\{\sigma_i\}\rangle,
   \label{eq:psiT}
\end{equation}
where $|\{\sigma_i\}\rangle$is the basis state diagonalizing each $\sigma_i^z$ as
$\sigma_i$.  The sum runs over all such possible assignments.

Assume $T>0$.
The wave function (\ref{eq:psiT}) is the ground state of the quantum Hamiltonian
\begin{equation}
   H_q(T)=-\chi \sum_j H_q^j(T)\equiv -\chi\sum_j(\sigma_x^j-e^{\beta H_j}),
\end{equation}
where $H_j$ is the sum of the terms of the Hamiltonian
(\ref{eq:Ising potential2}) involving site $j$,
\begin{equation}
  H_j=-J_j \sigma_j^z- \sum_k J_{jk}\sigma_j^z \sigma_k^z
  -\sum_{kl}J_{jkl}\sigma_j^z\sigma_k^z\sigma_l^z-\cdots.
\end{equation}
The coefficient $\chi$ is defined by $\chi=e^{-\beta p}$ with $p=\max_j |H_j|$.
\end{thm}

\begin{proof}
The first half is trivial:
\begin{equation}
  \frac{\langle \psi (T)|Q|\psi (T)\rangle}{\langle \psi (T)|\psi (T)\rangle}
  =\frac{1}{Z(T)}\sum_{\{\sigma\}} e^{-\beta H} \langle 
\{\sigma_i\}|Q| \{\sigma_i\}\rangle
  =\langle Q\rangle_T.
\end{equation}
To show the second half, we first note that
\begin{equation}
   \sigma_x^j \sum_{\{\sigma\}}| 
\{\sigma_i\}\rangle=\sum_{\{\sigma\}}| \{\sigma_i\}\rangle
   \label{eq:sigma_xj}
\end{equation}
since the operator $\sigma_x^j$ just changes the order of the above summation.
It is also easy to see that
\begin{equation}
   \sigma_x^j e^{-\beta H/2}=e^{\beta H_j} e^{-\beta H/2} \sigma_x^j
   \label{eq:sigma_xjH}
\end{equation}
because
\begin{equation}
   \sigma_x^j e^{-\beta H/2}\sigma_x^j = e^{-\beta 
(H-H_j)/2}\,\sigma_x^j e^{-\beta H_j/2}
    \sigma_x^j
    = e^{-\beta (H-H_j)/2}e^{\beta H_j/2}=e^{\beta H_j}e^{-\beta H/2}.
\end{equation}
as both $H$ and $H_j$ are diagonal in the present representation
and $H-H_j$ does not include $\sigma_j^z$, so $[H-H_j,\sigma_j^x]=0$.
We therefore have
\begin{equation}
   H_q^j(T) |\psi (T)\rangle =(\sigma_x^j -e^{\beta H_j})\,e^{-\beta H/2}
    \sum_{\{\sigma\}} |\{\sigma_i\}\rangle =0.
\end{equation}
Thus $|\psi (T)\rangle$ is an eigenstate of $H_q(T)$ with eigenvalue 0.
In the present representation, the non-vanishing off-diagonal 
elements of $-H_q(T)$
are all positive and the coefficients of $|\psi (T)\rangle$ are also all positive
as one sees in (\ref{eq:psiT}).
Then $|\psi (T)\rangle$ is the unique ground state of $H_q(T)$ according
to the Perron-Frobenius Theorem \cite{Seneta}.
\end{proof}

A few remarks are in order.  In the high-temperature limit, the quantum
Hamiltonian is composed just of the transverse-field term,
\begin{equation}
   H_q(T\to\infty) =-\sum_j (\sigma_x^j-1).
\end{equation}
Correspondingly the ground-state wave function $|\psi (T\to\infty)\rangle$ is the
simple summation over all possible states
with equal weight. In this way the thermal fluctuations in the original
classical system are mapped to the quantum fluctuations.
The low-temperature limit has, in contrast, the purely classical Hamiltonian
\begin{equation}
   H_q(T\approx 0)\to \chi \sum_j e^{\beta H_j}
\end{equation}
and the ground state of $H_q(T\approx 0)$ is also the ground state of $H$
as is apparent from the definition (\ref{eq:psiT}).
Hence the decrease of thermal fluctuations in SA is mapped to the decrease
of quantum fluctuations.
As explained below, this correspondence allows us to analyze the
condition for quasi-equilibrium in the classical SA
using the adiabaticity condition for the quantum system.

\subsection{Adiabaticity and convergence condition of SA}

The adiabaticity condition applied to the quantum system introduced
above leads to the condition of convergence of SA.
Suppose that we monotonically decrease the temperature as a function of time,
$T(t)$, to realize SA.

\begin{thm}
\label{thm:SA_convergence}
The adiabaticity condition for the quantum system of $H_q(T)$ yields 
the time dependence of $T(t)$ as
\begin{equation}
  T(t)= \frac{pN}{\log (\alpha t+1)}
   \label{eq:SA_convergence}
\end{equation}
in the limit of large  $N$.  The coefficient $\alpha$
is exponentially small in $N$.
\end{thm}

A few Lemmas will be useful to prove this Theorem.
\begin{lem}
\label{lemma:gap2}
The energy gap $\Delta(T)$ of $H_q(T)$ between the ground state and the first
excited state is bounded below as
\begin{equation}
  \Delta (T)\ge a \sqrt{N} e^{-(\beta p+c)N},
\end{equation}
where $a$ and $c$ are $N$-independent positive constants, in the asymptotic
limit of  large $N$.
\end{lem}
\begin{proof}
The analysis of Sec.~\ref{subsub:TFIgap} applies with the replacement
of $\Gamma (t)$ by $\chi =e^{-\beta p}$ and $\varepsilon_0(t)=0$.
This latter condition comes from $H_q(T)|\psi (T)\rangle =0$.
The condition $\Gamma (t)<\Gamma_0~(t>t_0)$ is unnecessary here because
the off-diagonal element $\chi$ can always be chosen smaller than
the diagonal elements by adding a positive constant to the diagonal.
Equation (\ref{eq:energy gap}) gives
\begin{equation}
  \Delta_j(t)\geq  A e^{-\beta pN}
\end{equation}
and $A$ satisfies, according to (\ref{eq:coefficient_A}),
\begin{equation}
   A\approx b\sqrt{2\pi N}  e^{-cN}
\end{equation}
with $b$ and $c$ positive constants of $\mathcal{O}(N^0)$.
\end{proof}

\begin{lem}
\label{lem:matrix_element}
The matrix element of the derivative of $H_q(T)$, relevant to the adiabaticity
condition, satisfies
\begin{equation}
   \langle \psi_1(T)|\partial_{ T} H_q(T) | \psi (T)\rangle
    =-\frac{\Delta (T) \langle \psi_1(T)|H|\psi (T)\rangle }{2k_B T^2},
\end{equation}
where $\psi_1(T)$ is the normalized first excited state of $H_q(T)$.
\end{lem}
\begin{proof}
By differentiating the identity
\begin{equation}
   H_q(T)|\psi (T)\rangle =0
\end{equation}
we find
\begin{equation}
  \left( \frac{\partial}{\partial T} H_q(T)\right) |\psi (T)\rangle
  =-H_q(T) \frac{\partial}{\partial T}|\psi (T)\rangle
  =H_q(T) \left(-\frac{1}{2k_BT^2}H\right) |\psi (T)\rangle.
\end{equation}
This relation immediately proves the Lemma if we notice that the
ground state energy of $H_q(T)$ is zero and therefore
$H_q(T)|\psi_1 (T)\rangle =\Delta (T) |\psi_1 (T)\rangle$.
\end{proof}

\begin{lem}
\label{lem:matrix_element_bound}
The matrix element of $H$ satisfies
\begin{equation}
  |\langle \psi_1(T)|H|\psi (T)\rangle | \le pN\sqrt{Z(T)}.
\end{equation}
\end{lem}
\begin{proof}
There are $N$ terms in $H=\sum_j H_j$, each of which is of norm of at most $p$.
The factor $\sqrt{Z(T)}$ appears from normalization of $|\psi (T)\rangle$.
\end{proof}

\begin{proof}
[{\bf Proof of Theorem \ref{thm:SA_convergence}}]
The condition of adiabaticity for the quantum system $H_q(T)$ reads
\begin{equation}
  \frac{1}{\Delta(T)^2 \sqrt{Z(T)}} \left|\langle \psi_1(T) |
  \partial_TH_q(T) |\psi (T)\rangle \, \frac{dT}{dt}\right|=\delta
\end{equation}
with sufficiently small $\delta$.  If we rewrite the matrix element
 by Lemma \ref{lem:matrix_element} , the left-hand side is
\begin{equation}
   \frac{|\langle \psi_1(T)|H|\psi (T)\rangle |}{2k_BT^2 \Delta (T)\sqrt{Z(T)}}
   \left|\frac{dT}{dt}\right|.
\end{equation}
By replacing the numerator by its bound in Lemma 
\ref{lem:matrix_element_bound} we have
\begin{equation}
  \frac{pN}{2k_BT^2\Delta (T)}\left|\frac{dT}{dt}\right|=\tilde{\delta}\ll 1
\end{equation}
as a sufficient condition for adiabaticity.
Using the bound of Lemma \ref{lemma:gap2} and integrating the above differential
equation for $T(t)$ noticing $dT/dt<0$, we reach the statement of
Theorem \ref{thm:SA_convergence}.
\end{proof}

\subsection{Remarks}

Equation (\ref{eq:SA_convergence}) reproduces the Geman-Geman condition for
convergence of SA \cite{GemanG}.  Their method of proof is to use the
theory of classical inhomogeneous ({\it i.e.} time-dependent) Markov chain
representing non-equilibrium processes.
It may thus be naively expected that the classical system under consideration may
not stay close to equilibrium during the process of SA
since the temperature always changes.
It therefore comes as a surprise that the adiabaticity condition, which is
equivalent to the quasi-equilibrium condition according to Theorem
\ref{thm:classical-quantum-mapping}, leads to Theorem \ref{thm:SA_convergence}.
The rate of temperature change in this latter Theorem is slow enough to guarantee
the quasi-equilibrium condition even when the temperature keeps changing.

Also, Theorem \ref{thm:SA_convergence} is quite general, covering the 
worst cases,
as it applies to any system written as the Ising model of 
(\ref{eq:Ising potential2}).
This fact means that one may apply a faster rate of temperature decrease
to solve a given specific problem with small errors.
The same comment applies to the QA situation in Sec.~\ref{sec:QA}.

Another remark is on the relation of QA and QAE.
Mathematical analyses of QA often focus their attention to the 
generic convergence
conditions in the infinite-time limit as seen in Secs. \ref{sec:QA} 
and \ref{sec:QMC}
as well as in the early paper \cite{KadowakiN}, although the residual energy
after finite-time evolution
has also been extensively investigated mainly in numerical studies.
This aspect may have lead some researchers to think that QA is different from
QAE, since the studies using the latter mostly concern the 
computational complexity
of finite-time evolution for a given specific optimization problem using
adiabaticity to construct an algorithm of QAE.
As has been shown in the present and the previous sections, the
adiabaticity condition also leads to the convergence condition
in the infinite-time limit for QA and SA. In this sense
QA, QAE and even SA share essentially the  same mathematical background.

\section{Reduction of errors for finite-time evolution}
\label{sec:finiteQA}

In Sec.~\ref{sec:QA}, we discussed the convergence condition of QA implemented
for the transverse-field Ising model. The power decrease of the
transverse field guarantees the adiabatic evolution.
This annealing schedule, however, does not provide practically useful
algorithms because infinitely long time is necessary to reach the exact solution.
An approximate algorithm for finite annealing time $\tau$
should be used in practice.
Since such a finite-time algorithm does not satisfy the generic convergence
condition, the answer includes a certain amount of errors.
An important question is how the error depends on the annealing time $\tau$.

Suzuki and Okada showed that the error after adiabatic evolution for time
$\tau$ is generally proportional to $\tau^{-2}$ in the limit of large $\tau$
with the system size $N$ kept finite \cite{SuzukiO}.
In this section, we analyze their results in detail and propose new
annealing schedules which show smaller errors proportional to $\tau^{-2m}~(m>1)$
\cite{Morita}.
This method allows us to reduce errors by orders of magnitude without 
compromising
the computational complexity apart from a possibly moderate numerical factor.

\subsection{Upper bound for excitation probability}
Let us consider the general time-dependent Hamiltonian
(\ref{eq:general Hamiltonian}).
The goal of this section is to evaluate the excitation probability (closely
related with the error probability) at the final time $s=1$ under the 
adiabaticity
condition (\ref{eq:adiabatic condition for tau}).

This task is easy because we have already obtained the asymptotic form
of the excitation amplitude (\ref{eq:excitation amplitude}). The upper
bound for the excitation probability is derived as
\begin{equation}
  \bigl|\langle j(1)|\tilde{\psi}(1)\rangle\bigr|^2=
  \left|c_{j\neq 0}(1)\right|^2\lesssim
   \frac{1}{\tau^2}
   \Bigl[\left|A_j(0)\right|+\left|A_j(1)\right|\Bigr]^2
   +\mathcal{O}(\tau^{-3}).
   \label{eq:upper bound for excitation probability}
\end{equation}
This formula indicates that the coefficient of the $\tau^{-2}$ term is
determined only by the state of the system at $s=0$ and 1 and vanishes if
$A_j(s)$ is zero at $s=0$ and 1.

When the $\tau^{-2}$-term vanishes, a similar calculation yields the
next order term of the excitation probability. If
$\tilde{H}'(0)=\tilde{H}'(1)=0$, the excitation amplitude $c_{j\neq
0}(1)$ is at most of order $\tau^{-2}$ and then $c_0(1)\approx
1+\mathcal{O}(\tau^{-3})$. Therefore we have
\begin{align}
  c_{j\neq 0}(1)&\approx \int_{0}^{1}\rmd s
  \frac{\rme^{\rmi \tau [\phi_j(s)-\phi_0(s)]}}
  {\Delta_j(s)}
  \bra{j(s)}\frac{\rmd \tilde{H}(s)}{\rmd s}\ket{0(s)}
  +\mathcal{O}(\tau^{-3})\nonumber \\
  &\approx \frac{1}{\tau^2}\left[A_j^{(2)}(0)-
  \rme^{\rmi \tau[\phi_j(s)-\phi_0(s)]}A_j^{(2)}(1)\right]
  +\mathcal{O}(\tau^{-3}),
  \label{eq:estimate for excitation amplitude}
\end{align}
where we defined
\begin{equation}
  A_j^{(m)}(s)\equiv \frac{1}{\Delta_j(s)^{m+1}} \langle j(s)|
  \frac{{\rm d}^m\tilde{H}(s)}{{\rm d}s^m}|0(s)\rangle.
\end{equation}
To derive the second line of (\ref{eq:estimate for excitation
amplitude}), we used integration by parts twice, and
(\ref{eq:derivative of eigenstate}) and (\ref{eq:eigenstate condition}).
The other $\tau^{-2}$ terms vanish because of the assumption
$\tilde{H}'(0)=\tilde{H}'(1)=0$.  Thus the upper bound of the next order for
the excitation probability under this assumption is obtained as
\begin{equation}
  \left|\langle j(1)|\tilde{\psi}(1)\rangle\right|^2\lesssim
   \frac{1}{\tau^4}\left[\left|A_j^{(2)}(0)\right|
    +\left|A_j^{(2)}(1)\right|\right]^2
   +\mathcal{O}(\tau^{-5}).
\end{equation}
It is easy to see that the $\tau^{-4}$-term also vanishes when
$\tilde{H}''(0)=\tilde{H}''(1)=0$.
It is straightforward to generalize these results to prove
the following Theorem.

\begin{thm}
  If the $k$th derivative of $\tilde{H}(s)$ is equal to zero at
  $s=0$ and 1 for all $k=1,2,\cdots, m-1$, the excitation probability has
  the upper bound
  \begin{equation}
  \left|\langle j(1)|\tilde{\psi}(1)\rangle\right|^2\lesssim
   \frac{1}{\tau^{2m}}\left[\left|A_j^{(m)}(0)\right|
                       +\left|A_j^{(m)}(1)\right|\right]^2
   +\mathcal{O}(\tau^{-2m-1}).
   \label{eq:UB}
  \end{equation}
\end{thm}

\subsection{Annealing schedules with reduced errors}
Although we have so far considered the general time-dependent Hamiltonian,
the ordinary Hamiltonian for QA with finite annealing time is composed of the
potential term and the kinetic energy term,
\begin{equation}
  \tilde{H}(s)=f(s) H_{\rm pot}
   +\left[1-f(s)\right] H_{\rm kin}.\label{eq:Hamiltonian},
\end{equation}
where $H_{\rm pt}$ and $H_{\rm kin}$ generalize $H_{\rm Ising}$
and $H_{\rm TF}$ in Sec. \ref{sec:QA}, respectively.
The function $f(s)$, representing the annealing schedule, satisfies
$f(0)=0$ and $f(1)=1$.
Thus $\tilde{H}(0)=H_{\rm kin}$ and $\tilde{H}(1)=H_{\rm pot}$.
The ground state of $H_{\rm pot}$ corresponds to the solution
of the optimization problem.
The kinetic energy is chosen so that its ground state is trivial.
The above Hamiltonian connects the trivial initial state and the
non-trivial desired solution after evolution time $\tau$.

The condition for the $\tau^{-2m}$-term to exist in the error
is obtained
straightforwardly from the results of the previous section because the
Hamiltonian (\ref{eq:Hamiltonian}) depends on time only through the
annealing schedule $f(s)$. It is sufficient that the $k$th derivative of
$f(s)$ is zero at $s=0$ and 1 for $k=1,2,\cdots, m-1$. We note that
$f(s)$ should belong to $C^m$, that is, $f(s)$ is an $m$th differentiable
function whose $m$th derivative is continuous.

\begin{figure}
  \begin{center}
   \includegraphics[scale=0.7]{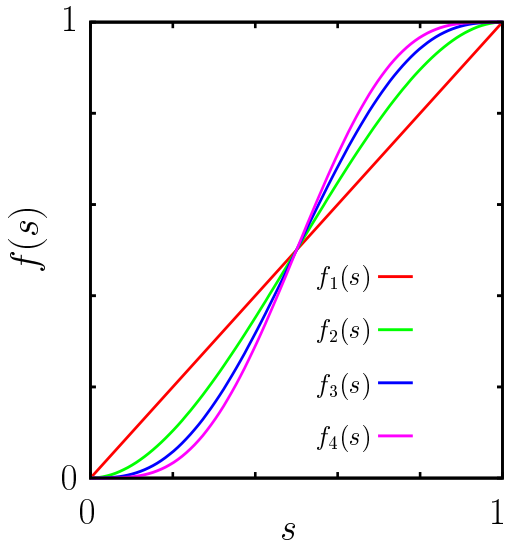}
  \caption{Examples of annealing schedules with reduced errors listed in
  (\ref{eq:f1})-(\ref{eq:f4}).} \label{fig:as}
  \end{center}
\end{figure}

Examples of the annealing schedules $f_m(s)$ with the $\tau^{-2m}$ error
rate are the following polynomials:
\begin{equation}
  f_1(s)=s,\label{eq:f1}
\end{equation}
\begin{equation}
  f_2(s)=s^2 (3-2s),
\end{equation}
\begin{equation}
  f_3(s)=s^3 (10-15s+6s^2),
\end{equation}
\begin{equation}
  f_4(s)=s^4 (35-84s+70s^2-20s^3).\label{eq:f4}
\end{equation}
The linear annealing schedule $f_1(s)$, which shows the $\tau^{-2}$
error, has been used in the past studies. Although we here
list only polynomials symmetrical with respect to the point $s=1/2$,
this is not essential. For example, $f(s)=(1-\cos (\pi s^2))/2$ also has
the $\tau^{-4}$ error rate because $f'(0)=f'(1)=f''(0)=0$ but $f''(1)=
-2\pi^2$.

\subsection{Numerical results}

\subsubsection{Two-level system}
To confirm the upper bound for the excitation probability discussed
above, it is instructive to study the two-level system, the Landau-Zener
problem, with the Hamiltonian
\begin{equation}
  H_{\rm LZ}(t)=-\left[\frac{1}{2}-f\left(\frac{t}{\tau}\right)\right]
  h\sigma^z-\alpha \sigma^x.
   \label{eq:H_LZ}
\end{equation}
The energy gap of $H_{\rm LZ}(t)$ has the minimum $2\alpha$ at
$f(s)=1/2$. If the annealing time $\tau$ is not large enough to satisfy
(\ref{eq:adiabatic condition for tau}), non-adiabatic transitions occur.
The Landau-Zener theorem \cite{LandauL, Zener} provides the
excitation probability $P_{\rm ex}(\tau)=\bigl|\langle
1(1)|\tilde{\psi}(1)\rangle\bigr|^2$ as
\begin{equation}
  P_{\rm ex}(\tau) =\exp\left[-\frac{\pi \alpha^2 \tau}{f'(s^*) h}\right],
   \label{eq:p1_LZ}
\end{equation}
where $s^*$ denotes the solution of $f(s^*)=1/2$.  On the other hand, if
$\tau$ is sufficiently large, the system evolves adiabatically.
Then the excitation probability has the upper bound
(\ref{eq:UB}), which is estimated as
\begin{equation}
  P_{\rm ex}(\tau) \lesssim
   \frac{4h^2\alpha^2}{\tau^{2m}(h^2+4\alpha^2)^{m+2}}
   \left[\left|\frac{{\rm d}^m f}{{\rm d}s^m}(0)\right|
    +\left|\frac{{\rm d}^m f}{{\rm d}s^m}(1)\right|\right]^2.
   \label{eq:p2_LZ}
\end{equation}

\begin{figure}
  \begin{center}
   \includegraphics[scale=0.7]{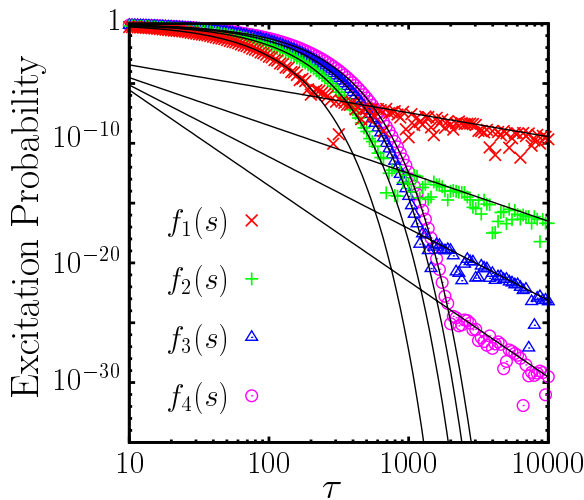}
  \end{center}
   \caption{The annealing-time dependence of the
   excitation probability for the two-level system (\ref{eq:H_LZ}) using
   schedules (\ref{eq:f1}) to (\ref{eq:f4}). The
   curved and straight lines show (\ref{eq:p1_LZ}) and
   (\ref{eq:p2_LZ}) for each annealing schedule, respectively. The
   parameters in (\ref{eq:H_LZ}) are chosen to be $h=2$ and
   $\alpha=0.2$.}
   \label{fig:LZ}
\end{figure}

We numerically solved the Schr\"{o}dinger equation (\ref{eq:Schrodinger
equation}) for this system (\ref{eq:H_LZ}) with the Runge-Kutta method
\cite{NRecipes}. Figure \ref{fig:LZ} shows the result for the excitation
probability with annealing schedules (\ref{eq:f1})-(\ref{eq:f4}). The
initial state is the ground state of $H_{\rm LZ}(0)$. The parameters are
chosen to be $h=2$ and $\alpha=0.2$. The curved and straight lines show
(\ref{eq:p1_LZ}) and (\ref{eq:p2_LZ}), respectively.  In the small
and large $\tau$ regions, the excitation probability perfectly fits to
those two expressions.

\subsubsection{Spin glass model}
\label{subseb:spinglass}
We next carried out simulations of a rather large system, the Ising spin
system with random interactions. The quantum fluctuations are introduced
by the uniform transverse field. Thus, the potential and kinetic energy
terms are defined by
\begin{gather}
  H_{\rm pot}=-\sum_{\langle ij \rangle}J_{ij}\sigma_i^z\sigma_j^z
   -h\sum_{i=1}^{N} \sigma_i^z,\\
  H_{\rm kin}=-\Gamma\sum_{i=1}^{N}\sigma_i^x.
\end{gather}
The initial state, namely the ground state of $H_{\rm kin}$, is the
all-up state along the $x$ axis.

The difference between the obtained approximate energy and the true
ground state energy (exact solution) is the residual energy $E_{\rm res}$.
It is a useful measure of the error rate
of QA. It has the same behavior as the excitation probability because it
is rewritten as
\begin{align}
  E_{\rm res} &\equiv\langle \tilde{\psi}(1)| H_{\rm pot}
   |\tilde{\psi}(1)\rangle-\varepsilon_0(1)\\
  &=\sum_{j> 0} \Delta_j(1)
  \left|\langle j(1) | \tilde{\psi}(1)\rangle \right|^2.
\end{align}
Therefore $E_{\rm res}$ is expected to be asymptotically proportional
to $\tau^{-2m}$ using the improved annealing schedules.

\begin{figure}
  \begin{center}
   \includegraphics[scale=0.7]{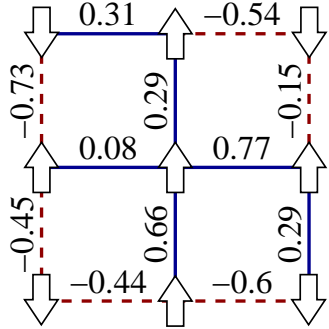}
   \caption{Configuration
   of random interactions $\{J_{ij}\}$ on the $3\times 3$ square lattice
   which we investigated, and spin configuration of the target state.
   The solid and dashed lines indicate
   ferromagnetic and antiferromagnetic interactions, respectively. }
   \label{fig:sg_int}
  \end{center}
\end{figure}

\begin{figure}
  \begin{center}
   \includegraphics[scale=0.7]{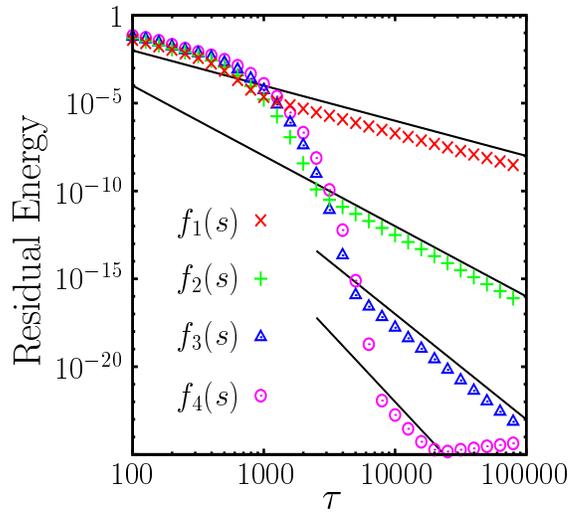}
   \caption{The annealing-time dependence of the residual energy for the
   two-dimensional spin glass model with improved annealing schedules. The
   solid lines denote functions proportional to $\tau^{-2m}$
   $(m=1,2,3,4)$. The parameter values are $h=0.1$ and $\Gamma=1$.}
   \label{fig:sg}
  \end{center}
\end{figure}

We investigated the two-dimensional square lattice of size $3\times
3$. The quenched random coupling constants $\{J_{ij}\}$ are chosen from
the uniform distribution between $-1$ and $+1$, as shown in
Fig. \ref{fig:sg_int}. The parameters are $h=0.1$ and $\Gamma=1$.
Figure \ref{fig:sg} shows the $\tau$ dependence of the residual energy
using the annealing schedules (\ref{eq:f1})-(\ref{eq:f4}). Straight lines
representing $\tau^{-2m}$ $(m=1,2,3,4)$ are also shown for comparison.
The data clearly indicates the $\tau^{-2m}$-law for large $\tau$.
The irregular behavior around $E_{\rm res}\approx 10^{-25}$ comes from
numerical rounding errors.

\subsubsection{Database search problem}
As another example, we apply the improved annealing schedule to the 
database search
problem of an item in an unsorted database. Consider $N$ items,
among which one is marked. The goal of this problem is to find the
marked item in a minimum time. The pioneering quantum algorithm proposed
by Grover \cite{Grover} solves this task in time of order $\sqrt{N}$,
whereas the classical algorithm tests $N/2$ items on average. Farhi {\it
et al.} \cite{FarhiGGS} proposed a QAE algorithm and
Roland and Cerf \cite{RolandC} found a QAE-based algorithm with the same
computational complexity as Grover's algorithm. Although their schedule is
optimal in the sense that the excitation probability by the adiabatic
transition is equal to a small constant at each time, it has the
$\tau^{-2}$ error rate. We show that annealing schedules with the
$\tau^{-2m}$ error rate can be constructed by a slight modification of
their optimal schedule.

Let us consider the Hilbert space which has the basis states
$|i\rangle$ $(i=1,2,\cdots, N)$, and the marked state is denoted by
$|m\rangle$. Suppose that we can construct the Hamiltonian
(\ref{eq:Hamiltonian}) with two terms,
\begin{equation}
  H_{\rm pot}=1-|m\rangle \langle m|,
\end{equation}
\begin{equation}
  H_{\rm kin}=1-\frac{1}{N}\sum_{i=1}^{N}\sum_{j=1}^{N}|i\rangle\langle j|.
\end{equation}
The Hamiltonian $H_{\rm pot}$ can be applied without the explicit knowledge
of $|m\rangle$, the same assumption as in Grover's algorithm.
The initial state is a superposition of all basis states,
\begin{equation}
  |\psi(0)\rangle = \frac{1}{\sqrt{N}}\sum_{i=1}^{N}|i\rangle,
\end{equation}
which does not depend on the marked state. The energy gap between the
ground state and the first excited state,
\begin{equation}
  \Delta_1(s)=\sqrt{1-4\frac{N-1}{N}f(s)[1-f(s)]},
\end{equation}
has a minimum at $f(s)=1/2$. The highest eigenvalue
$\varepsilon_2(s)=1$ is $(N-2)$-fold degenerate.

\begin{figure}
  \begin{center}
   \includegraphics[scale=0.8]{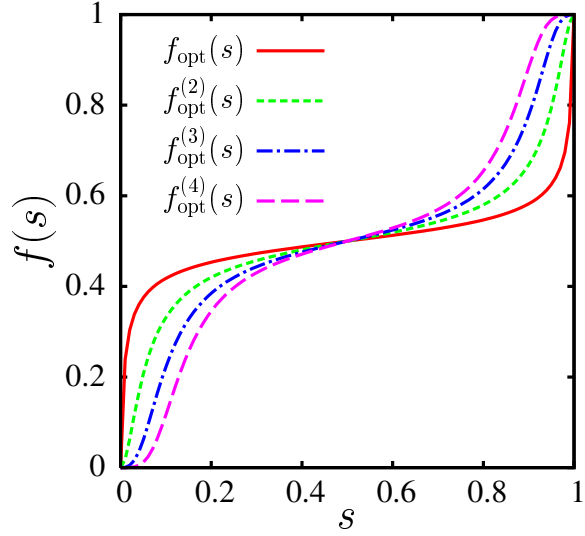}
   \caption{The optimal annealing schedules for
   the database search problem ($N=64$). The solid line denotes the
   original optimal schedule (\ref{eq:opt}) and the dashed lines are
   for the modified schedules.}  \label{fig:opt}
  \end{center}
\end{figure}

To derive the optimal annealing schedule, we briefly review the results
reported by Roland and Cerf \cite{RolandC}. When the energy gap is small
({\em i.e.} for $f(s)\approx 1/2$),
non-adiabatic transitions are likely to occur. Thus we need to change the
Hamiltonian carefully. On the other hand, when the energy gap is not very
small, too slow a change wastes time.
Thus the speed of parameter change should be adjusted adaptively to the
instantaneous energy gap. This is realized by tuning the annealing schedule
to satisfy the adiabaticity condition (\ref{eq:adiabatic condition for tau})
in each infinitesimal time interval, that is,
\begin{equation}
  \frac{\left|A_1(s)\right|}{\tau}=\delta,
\end{equation}
where $\delta$ is a small constant.  In the database search problem,
this condition is rewritten as
\begin{equation}
  \frac{\sqrt{N-1}}{\tau N \Delta_1(s)^3}
   \frac{\rmd f}{\rmd s}=\delta.
   \label{eq:db_adia}
\end{equation}
After integration under boundary conditions $f(0)=0$ and $f(1)=1$, we
obtain
\begin{equation}
  f_{\rm opt}(s)=\frac{1}{2}+\frac{2s-1}{2\sqrt{N-(N-1)(2s-1)^2}}.
   \label{eq:opt}
\end{equation}
As plotted by a solid line in Fig. \ref{fig:opt}, this function changes
most slowly when the energy gap takes the minimum value.  It is noted that the
annealing time is determined by the small constant $\delta$ as
\begin{equation}
  \tau=\frac{\sqrt{N-1}}{\delta},
\end{equation}
which means that the computation time is of order $\sqrt{N}$ similarly
to Grover's algorithm.

\begin{figure}
  \begin{center}
   \includegraphics[scale=0.8]{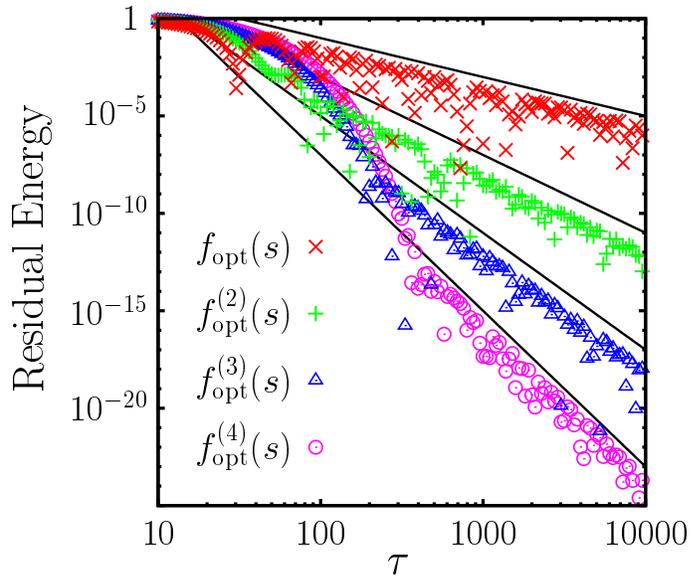}
   \caption{The annealing-time dependence of the residual
   energy for the database search problem ($N=64$) with the optimal
   annealing schedules described in Fig.\ \ref{fig:opt}. The solid lines
   represent functions proportional to $\tau^{-2m}$ $(m=1,2,3,4)$.}
   \label{fig:db}
  \end{center}
\end{figure}

The optimal annealing schedule (\ref{eq:opt}) shows the $\tau^{-2}$
error rate because its derivative is non-vanishing at $s=0$ and 1.  It is
easy to see from (\ref{eq:opt}) that the simple replacement of $s$
with $f_m(s)$ fulfils the condition for the $\tau^{-2m}$ error
rate. We carried out numerical simulations for $N=64$ with such
annealing schedules, $f_{\rm opt}^{(m)}(s)\equiv f_{\rm
opt}\left(f_m(s)\right)$, as plotted by dashed lines in
Fig.\ \ref{fig:opt}. As shown in Fig.\ \ref{fig:db}, the residual energy
with $f_{\rm opt}^{(m)}(s)$ is proportional to $\tau^{-2m}$.  The
characteristic time $\tau_c$ for the $\tau^{-2m}$ error rate to show up
increases with $m$: Since the modified optimal schedule $f_{\rm
opt}^{(m)}(s)$ has a steeper slope at $s=1/2$ than $f_{\rm opt}(s)$, a
longer annealing time is necessary to satisfy the adiabaticity condition
(\ref{eq:db_adia}). Nevertheless, the difference in slopes of $f_{\rm
opt}^{(m)}(s)$ is only a factor of $\mathcal{O}(1)$, and therefore $\tau_c$ is
still scaled as $\sqrt{N}$.
Significant qualitative reduction of errors has been achieved without
compromising computational complexity apart from a numerical factor.

\subsection{Imaginary-time Schr\"{o}dinger Dynamics}
\label{subsec:imaginary}

So far, we have concentrated on QA following the real-time (RT)
Schr\"{o}dinger dynamics. From the point of view of physics, it is
natural that the time evolution of a quantum system obeys the real-time
Schr\"{o}dinger equation. Since our goal is to find the solution of
optimization problems, however, we need not stick to physical
reality. We therefore investigate QA following the {\it imaginary-time} (IT)
Schr\"{o}dinger dynamics here to further reduce errors.

The IT evolution tends to filter out the excited states. Thus, it is
expected that QA with the IT dynamics can find the optimal solution more
efficiently than RT-QA. Stella {\it et al}. \cite{StellaST2005} have
investigated numerically the performance of IT-QA and conjectured that (i) the IT
error rate is not larger than in the RT, and that (ii) the asymptotic
behavior of the error rate for $\tau\rightarrow\infty$ is identical for
IT-QA and RT-QA. We prove their conjectures through the IT version of the
adiabatic theorem.

\subsubsection{Imaginary-time Schr\"{o}dinger equation}

The IT Schr\"{o}dinger equation is obtained by the
transformation $t \rightarrow -\rmi t$ in the time derivative of the
original RT Schr\"{o}dinger equation:
\begin{equation}
  -\frac{\rmd}{\rmd t}\ket{\Psi(t)}=H(t)\ket{\Psi(t)}.
\end{equation}
The time dependence of the Hamiltonian does not change. If the Hamiltonian
is time-independent, we easily see that the excitation amplitude
decreases exponentially relative to the ground state,
\begin{equation}
  \ket{\Psi(t)}=\sum_j c_j e^{-{\rm i}t\varepsilon_j} \ket{j}
   \longrightarrow
   \sum_j c_j e^{-t\varepsilon_j} \ket{j}
   =e^{-t\varepsilon_0}\sum_j c_j e^{-t(\varepsilon_j-\varepsilon_0)} \ket{j}.
\end{equation}
However, it
is not obvious that this feature survives in the time-dependent situation.

An important aspect of the IT Schr\"{o}dinger equation is
non-unitarity. The norm of the wave function is not conserved. Thus, we
consider the normalized state vector
\begin{equation}
  \ket{\psi(t)}\equiv\frac{1}{\sqrt{\langle{\Psi(t)} |\Psi(t)\rangle}}
   \ket{\Psi(t)}.
\end{equation}
The equation of motion for this normalized state vector is
\begin{equation}
  -\frac{\rmd}{\rmd t}\ket{\psi(t)}=
  \bigl[ H(t)-\left\langle H(t)\right\rangle \bigr] \ket{\psi(t)},
\end{equation}
where we defined the expectation value of the Hamiltonian
\begin{equation}
  \left\langle H(t)\right\rangle
   \equiv\bra{\psi(t)}H(t)\ket{\psi(t)}.
\end{equation}
The above equation is not linear but norm-conserving, which makes the
asymptotic expansion easy. In terms of the dimensionless time
$s=t/\tau$, the norm-conserving IT Schr\"{o}dinger equation is written
as
\begin{equation}
  -\frac{\rmd}{\rmd s}\ket{\tilde{\psi}(s)}=
  \tau \Bigl[ \tilde{H}(s)
  -\left\langle \tilde{H}(s)\right\rangle \Bigr]
  \ket{\tilde{\psi}(s)}.
  \label{eq:norm-conserving IT Eq.}
\end{equation}

\subsubsection{Asymptotic expansion of the excitation probability}
To prove the conjecture by Stella {\it et al.}, we derive the asymptotic
expansion of the excitation probability. The following Theorem provides
us with the imaginary-time version of the adiabatic theorem.
\begin{thm}
  Under the same hypothesis as in Theorem \ref{theorem:adiabatic}, the state
  vector following the norm-conserving IT Schr\"{o}dinger equation
  (\ref{eq:norm-conserving IT Eq.}) has the asymptotic form in the limit
  of large $\tau$ as
  \begin{equation}
   \ket{\tilde{\psi}(s)}=\sum_{j}c_j(s) \ket{j(s)},
  \end{equation}
  \begin{equation}
   c_0(s)\approx 1+O\left(\tau^{-2}\right),
    \label{eq:IT_gsAmplitude}
  \end{equation}
  \begin{equation}
   c_{j\neq 0}(s) \approx \frac{A_j(s)}{\tau}+O\left(\tau^{-2}\right).
    \label{eq:IT_exAmplitude}
  \end{equation}
\end{thm}

\begin{proof}
   The norm-conserving IT Schr\"{o}dinger equation (\ref{eq:norm-conserving
  IT Eq.}) is rewritten as the equation of motion for $c_j(s)$ as
  \begin{equation}
   \frac{\rmd c_j}{\rmd s}=\sum_{k\neq j}
    \frac{c_k(s)}{\varepsilon_j(s)-\varepsilon_k(s)}
    \bra{j(s)}\frac{\rmd \tilde{H}(s)}{\rmd s}\ket{k(s)}
   -\tau c_j(s)\left[\varepsilon_j(s)
   -\sum_{l} \varepsilon_l(s) |c_l(s)|^2\right].
  \end{equation}
  To remove the second term on the right-hand side, we define
  \begin{equation}
   \tilde{c}_j(s)\equiv\exp\left(\tau\int_{0}^{s}\rmd \tilde{s} \left[
   \varepsilon_j(\tilde{s})
   -\sum_{l}\varepsilon_l(\tilde{s})|c_l(\tilde{s})|^2\right]\right)
    c_j(s),
  \end{equation}
  and obtain the equation of motion for $\tilde{c}_j(s)$ as
  \begin{equation}
   \frac{\rmd \tilde{c}_j}{\rmd s}
   =\sum_{k\neq j}\tilde{c}_k(s)
    \frac{\rme^{\tau \left\{\phi_j(s)-\phi_k(s) \right\}}}
    {\varepsilon_j(s)-\varepsilon_k(s)}
   \bra{j(s)}\frac{\rmd \tilde{H}(s)}{\rmd s}\ket{k(s)},
  \end{equation}
  where we defined $\phi_j(s)\equiv \int_{0}^{s}\rmd s' \varepsilon_j(s')$
  for convenience.

  Integration of this equation yields the integral equation for
  $\tilde{c}_j(s)$. It is useful to introduce the following quantity,
  \begin{equation}
   \delta(s)\equiv \int_{0}^{s}\rmd \tilde{s} \sum_{l\neq 0}
   \left[\varepsilon_l(\tilde{s})-\varepsilon_0(\tilde{s})\right]
   |c_l(\tilde{s})|^2.
  \end{equation}
  Since the norm of the wave function is conserved, $\sum_{l}
  |c_l(s)|^2=1$ and therefore
  \begin{equation}
   \sum_{l}\varepsilon_l(s) |c_l(s)^2|=\varepsilon_0(s)
   +\sum_{l\neq 0}\left[\varepsilon_l(s)-\varepsilon_0(s)\right]|c_l(s)^2|.
  \end{equation}
  Thus, the definition of $\tilde{c}_j(s)$ is written as
  \begin{equation}
   \tilde{c}_j(s)=\rme^{-\tau\delta(s)}
   \rme^{\tau[\phi_j(s)-\phi_0(s)]}c_j(s).
  \end{equation}
  Finally we obtain the integral equation for $c_j(s)$:
  \begin{gather}
   c_0(s)= \rme^{\tau \delta(s)}
    +\rme^{\tau \delta(s)}\int_{0}^{s}\rmd \tilde{s}\,
    \rme^{-\tau\delta(\tilde{s})}
    \sum_{l\neq 0} c_l(\tilde{s})
    \frac{\bra{0(\tilde{s})}\frac{\rmd \tilde{H}}{\rmd \tilde{s}}\ket{l(\tilde{s})}}
    {\varepsilon_0(\tilde{s})-\varepsilon_l(\tilde{s})}, \\
   c_{j\neq 0}(s)= \rme^{\tau \delta(s)}\rme^{-\tau[\phi_j(s)-\phi_0(s)]}
    \int_{0}^{s}\rmd \tilde{s}\,\rme^{-\tau\delta(\tilde{s})}
    \rme^{\tau[\phi_j(\tilde{s})-\phi_0(\tilde{s})]} \sum_{k\neq j} c_k(\tilde{s})
    \frac{\bra{j(\tilde{s})}\frac{\rmd \tilde{H}}{\rmd \tilde{s}}\ket{k(\tilde{s})}}
    {\varepsilon_j(\tilde{s})-\varepsilon_k(\tilde{s})},\label{eq:IT integral eq2}
  \end{gather}
  where we used the initial condition $c_0(0)=1$ and $c_{j\neq 0}=0$.

  The next step is the asymptotic expansion of these integral equations for
  large $\tau$. It is expected that $c_0(s)=1$ and $c_{j\neq 0}(s)=0$ for
  $\tau\rightarrow\infty$ because of the following
  argument: Since the coefficient $c_0(s)$ is less than unity, $\delta(s)$
  should be $\mathcal{O}(\tau^{-1})$ at most and 
$\rme^{\tau\delta(s)}=\mathcal{O}(1)$. The
  second factor on the right-hand side of (\ref{eq:IT integral eq2}) is
  small exponentially with $\tau$ because $\phi_j(s)-\phi_0(s)$ is
  positive and an increasing function of $s$. Thus, $c_{j\neq0}(s)
  \rightarrow 0$ and then $c_0(s)\rightarrow 1$ owing to the norm
  conservation law.

  Therefore we estimate the next term of order $\tau^{-1}$ under the
  assumption that $c_0(s)\gg c_{j\neq 0}(s)$. Since $\delta(s)$ is
  proportional to the square of $c_{j\neq 0}(s)$, we have
  $\rme^{\tau\delta(s)}\approx 1$. Thus, the $\rme^{\pm\tau\delta(s)}$ factors
  can be ignored in the $\tau^{-1}$ term estimation of
  (\ref{eq:IT integral eq2}).  Consequently, evaluation of the integral equations
  yields
  \begin{equation}
   c_{j\neq 0}(s) \approx \rme^{-\tau [\phi_j(s)-\phi_0(s)]}
    \int_{0}^{s}\rmd \tilde{s}\,
    \rme^{\tau[\phi_j(\tilde{s})-\phi_0(\tilde{s})]}
    \frac{\bra{j(\tilde{s})}\frac{\rmd \tilde{H}}{\rmd \tilde{s}}\ket{0(\tilde{s})}}
    {\Delta_j(\tilde{s})} +\mathcal{O}\left(\tau^{-2}\right).
  \end{equation}
  The excitation amplitude is estimated by integration by parts as
  \begin{equation}
   c_{j\neq 0}(s)\approx\frac{1}{\tau}
   \left[ A_j(s)-\rme^{-\tau(\phi_j(s)-\phi_0(s))}A_j(0)\right]
   +\mathcal{O}\left(\tau^{-2}\right),
  \end{equation}
  where $A_j(s)$ is defined by (\ref{eq:definition of A_j(s)}). The
  second term in the square brackets is vanishingly small, which is a
  different point from the RT dynamics. From the above expression, we
  find $\delta(s)=\mathcal{O}(\tau^{-2})$, that is $\rme^{\tau\delta(s)}\approx
  1+\mathcal{O}(\tau^{-1})$. Therefore, we obtain (\ref{eq:IT_gsAmplitude})
  and (\ref{eq:IT_exAmplitude}), which is consistent with the assumption
  $c_0(s)\gg c_{j\neq 0}(s)$.
\end{proof}

\begin{rem}
  The excitation probability at the end of a QA process is
  proportional to $\tau^{-2}$ in the large $\tau$ limit:
  \begin{equation}
  \bigl|\langle j(1)|\tilde{\psi}(1)\rangle\bigr|^2 \approx \frac{1}{\tau^2}
  \left|A_j(1)\right|^2 +\mathcal{O}\left(\tau^{-3}\right).
  \end{equation}
  Its difference from the upper bound for the RT dynamics (\ref{eq:upper
  bound for excitation probability}) is only in the absence of $A_j(0)$. In
  the IT dynamics, this term decreases exponentially because of the factor
  $\rme^{-\tau(\phi_j(s)-\phi_0(s))}$. This result proves the conjecture
  proposed by Stella {\it et al.} \cite{StellaST2005}, that is,
  \begin{gather}
  \epsilon_{\rm IT}(\tau) \leq \epsilon_{\rm RT}(\tau),\\
  \epsilon_{\rm IT}(\tau) \approx \epsilon_{\rm RT}(\tau)
  \qquad (\tau\rightarrow\infty).
  \end{gather}
  Strictly speaking, the right-hand sides in the above equations denote
  the upper bound for the error rate for RT-QA, not the error rate itself.
  In some systems, for example, the two level system, the error rate
  oscillates because $A_j(0)$ and $A_j(1)$ may cancel in
  (\ref{eq:excitation amplitude}), and becomes smaller than that of IT-QA
  at some $\tau$. However, QA for ordinary optimization problems has
  different energy levels at initial and final times, and thus such a
  cancellation seldom occurs.
\end{rem}

\subsubsection{Numerical verification}
\begin{figure}
  \begin{center}
   \includegraphics[scale=0.8]{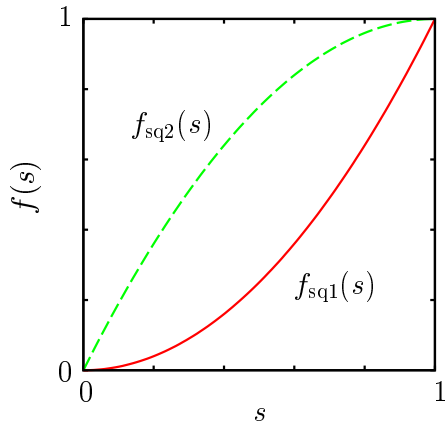}
   \caption{The annealing schedules defined in
   (\ref{eq:as_sq}). $f_{\rm sq1}(s)$ and $f_{\rm sq2}(s)$ have a
   vanishing slope
   at the initial time $s=0$ and the final time $s=1$, respectively.}
  \label{fig:as_sq}
  \end{center}
\end{figure}

\begin{figure}
  \begin{center}
   \includegraphics[scale=0.8]{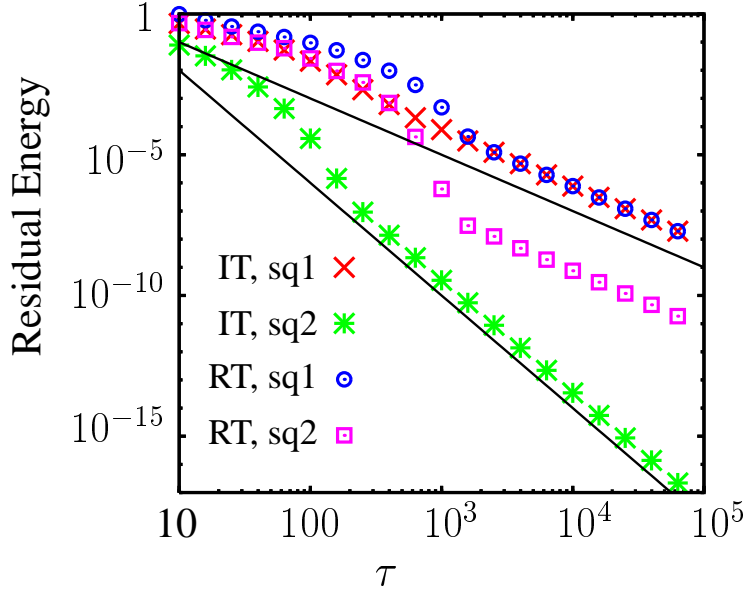}
   \caption{The annealing-time dependence of the residual energy for IT-
   and RT-QA with annealing schedules $f_\text{sq1}(s)$ and
   $f_\text{sq2}(s)$. The system is the spin-glass model presented in
   Sec.~\ref{subseb:spinglass}. The solid lines stand for functions proportional
   to $\tau^{-2}$ and $\tau^{-4}$. The parameters are $h=0.1$ and
   $\Gamma=1$.}
  \label{fig:sg_resE_im}
  \end{center}
\end{figure}

We demonstrate a numerical verification of the above results by
simulations of the IT- and RT-Schr\"{o}dinger equations. For this
purpose, we consider the following annealing schedules (Fig.\
\ref{fig:as_sq}):
\begin{equation}
  f_\text{sq1}(s)=s^2, \qquad f_\text{sq2}(s)=s(2-s).
   \label{eq:as_sq}
\end{equation}
The former has a zero slope at the initial time $s=0$ and the latter
at $s=1$. Thus, the $A_j(0)$ and $A_j(1)$ terms vanish with
$f_\text{sq1}(s)$ and $f_\text{sq2}(s)$, respectively. Since the error
rate for IT-QA depends only on $A_j(1)$, IT-QA with $f_\text{sq2}(s)$
should show the $\tau^{-4}$ error rate, while RT-QA with
$f_\text{sq2}(s)$ exhibits the $\tau^{-2}$-law. On the other hand, RT-QA
and IT-QA with $f_\text{sq1}(s)$ should have the same error rate for
large $\tau$. Figure \ref{fig:sg_resE_im} shows the residual energy with
two annealing schedules for the spin-glass model presented in
Sec.~\ref{subseb:spinglass}, which explicitly supports our results.

\section{Convergence condition of QA -- Quantum Monte Carlo evolution}
\label{sec:QMC}

So far, we have discussed QA with the Schr\"{o}dinger dynamics. When we
solve the Schr\"{o}dinger equation on the classical computer,
the computation time and memory increase exponentially with the system
size. Therefore, some approximations are necessary to simulate QA
processes for large-size problems. In most numerical studies, stochastic
methods are used. In this section, we investigate two types of
quantum Monte Carlo methods and prove their convergence theorems,
following \cite{MoritaN06}.

\subsection{Inhomogeneous Markov chain}
Since we prove the convergence of stochastic processes, it is
useful to recall various definitions and theorems for inhomogeneous
Markov processes \cite{AartsK}.  We denote the space of discrete
states by $\mathcal{S}$ and assume that the size of $\mathcal{S}$ is
finite.  A Monte Carlo step is characterized by the {\em transition
probability} from state $x (\in \mathcal{S})$ to state $y
(\in\mathcal{S})$ at time step $t$:
\begin{equation}
  \label{eq:G}
  G(y,x;t)= \begin{cases}
             P(y,x)A(y,x;t) & (x\neq y)\\
             1-\sum_{z\in\mathcal{S}}P(z,x)A(z,x;t) & (x=y),
             \end{cases}
\end{equation}
where $P(y,x)$ and $A(y,x;t)$ are called the {\em generation probability}
and the {\em acceptance probability}, respectively.
The former is the probability to generate the next candidate state $y$
from the present state $x$. We assume that this probability does not depend
on time and satisfies the following conditions:
\begin{gather}
  \forall x,y\in\mathcal{S} : P(y,x)=P(x,y)\geq 0 ,\\
  \forall x\in\mathcal{S} : P(x,x)=0 ,\\
  \forall x\in\mathcal{S} : \sum_{y\in\mathcal{S}}P(y,x)=1,
\end{gather}
\begin{gather}
  \forall x,y\in\mathcal{S}, \exists n>0, \exists
   z_1,\cdots,z_{n-1}\in{\mathcal S}:
  \prod_{k=0}^{n-1}P(z_{k+1},z_k)>0, z_0=x, z_n=y.
\end{gather}
The last condition represents irreducibility of $\mathcal{S}$, that is,
any state in $\mathcal{S}$ can be reached from any other state in
$\mathcal{S}$.

We define $\mathcal{S}_x$ as the neighborhood of $x$, {\em i.e.}, the set
of states that can be reached by a single step from $x$:
\begin{equation}
  \mathcal{S}_x = \{y\mid y\in\mathcal{S}, P(y,x)>0\}.
\end{equation}
The acceptance probability $A(y,x;t)$ is the probability to accept the
candidate $y$ generated from state $x$. The matrix $G(t)$, whose
$(y, x)$ component is given by (\ref{eq:G}), $[G(t)]_{y,x}=G(y,x;t)$,
is called the {\em transition matrix}.

Let $\mathcal{P}$ denote the set of probability distributions on
$\mathcal{S}$.  We regard a probability distribution $p\
(\in\mathcal{P})$ as the column vector with the component
$[p]_x=p(x)$. The probability distribution at time $t$, started from an
initial distribution $p_0\ (\in\mathcal{P})$ at time $t_0$, is written as
\begin{equation}
  p(t,t_0)= G^{t,t_0} p_0\equiv G(t-1)G(t-2)\cdots G(t_0)p_0.
\end{equation}

A Markov chain is called {\em inhomogeneous} when the transition
probability depends on time.  In the following sections, we will prove
that inhomogeneous Markov chains associated with QA are ergodic under
appropriate conditions.  There are two kinds of ergodicity, weak and
strong.  {\em Weak ergodicity} means that the probability distribution
becomes independent of the initial conditions after a sufficiently long
time:
\begin{equation}
  \forall t_0\geq 0: \lim_{t\rightarrow\infty} \sup \{
   \| p(t,t_0)-p'(t,t_0)\| \,|\, p_0, p'_0\in\mathcal{P}\}=0,
   \label{eq:def of weak ergodicity}
\end{equation}
where $p(t,t_0)$ and $p'(t,t_0)$ are the probability distributions with
different initial distributions $p_0$ and $p'_0$. The norm is defined by
\begin{equation}
  \| p \|=\sum_{x\in\mathcal{S}} |p(x)|.
\end{equation}
{\em Strong ergodicity} is the property that the probability
distribution converges to a unique distribution irrespective of the
initial state:
\begin{equation}
  \label{eq:SE}
  \exists r\in\mathcal{P}, \forall t_0\geq 0:
   \lim_{t\rightarrow\infty} \sup \{
   \|p(t,t_0)-r\| \,|\, p_0\in\mathcal{P} \}=0.
\end{equation}

The following two Theorems provide conditions for weak and strong
ergodicity of an inhomogeneous Markov chain \cite{AartsK}.
For proofs see Appendix B.

\begin{thm}[Condition for weak ergodicity]
An inhomogeneous Markov chain is weakly ergodic if and only if there
exists a strictly increasing sequence of positive numbers $\{t_i\},
(i=0,1,2,\dots)$, such that
\begin{equation}
  \sum_{i=0}^{\infty}\left[1-\alpha(G^{t_{i+1},t_i})
  \right]\longrightarrow\infty,\label{eq:theorem for weak-e}
\end{equation}
where $\alpha(G^{t_{i+1},t_i})$ is the {\it coefficient of ergodicity}
defined by
\begin{equation}
  \alpha(G^{t_{i+1},t_i})=1-\min\left\{ \sum_{z\in\mathcal{S}}
   \min\{G(z,x),G(z,y)\} \Big| x,y\in\mathcal{S} \right\}
  \label{eq:alpha}
\end{equation}
with the notation $G(z,x)=[G^{t_{i+1},t_i}]_{z,x}$.
\label{theorem:weak-e}
\end{thm}

The coefficient of ergodicity measures the variety of the transition
probability. If $G(z,x)$ is independent of a state $x$, $\alpha(G)$ is
equal to zero.

\begin{thm}[Condition for strong ergodicity]
An inhomogeneous Markov chain is strongly ergodic if the following
three conditions hold:
\begin{enumerate}
  \item the Markov chain is weakly ergodic,
  \item for all $t$ there exists a stationary state $p_t\in\mathcal{P}$
        such that $p_t=G(t)p_t$,
  \item $p_t$ satisfies
        \begin{equation}
         \label{eq:weak_e3}
         \sum_{t=0}^{\infty}\|p_t-p_{t+1}\|<\infty.
        \end{equation}
\end{enumerate}
Moreover, if $\displaystyle p=\lim_{t\rightarrow\infty}p_t$, then $p$ is
equal to the probability distribution $r$ in (\ref{eq:SE}).
\label{theorem:strong-e}
\end{thm}

We note that the existence of the limit is guaranteed by
(\ref{eq:weak_e3}). This inequality implies that the probability
distribution $p_t(x)$ is a Cauchy sequence:
\begin{equation}
  \forall \varepsilon>0, \exists t_0>0, \forall t, t' >t_0:
  |p_t(x)-p_{t'}(x)|< \varepsilon.
\end{equation}

\subsection{Path-integral Monte Carlo method}

Let us first discuss convergence conditions for the implementation of
quantum annealing by the path-integral Monte Carlo (PIMC) method
\cite{Suzuki,LandauB}.  The basic idea of PIMC is to apply the Monte Carlo
method to the classical system obtained from the original quantum system
by the path-integral formula.  We first consider the example of ground
state search of the Ising spin system whose quantum fluctuations are
introduced by adding a transverse field. The total Hamiltonian is
defined in  (\ref{eq:TFIM Hamiltonian}).  Although we only treat the
two-body interaction for simplicity in this section, the existence of
arbitrary many-body interactions between the $z$ components of Pauli matrix
and longitudinal random magnetic field $\sum h_i \sigma_i^z$, in
addition to the above Hamiltonian, would not change the following
argument.

In the path-integral method, the $d$-dimensional transverse-field Ising
model (TFIM) is mapped to a
$(d+1)$-dimensional classical Ising system so that the quantum system
can be simulated on the classical computer.  In numerical simulations,
the Suzuki-Trotter formula \cite{Trotter,Suzuki} is usually employed to
express the partition function of the resulting classical system,
\begin{equation}
  Z(t)\approx \sum_{\{S_i^{(k)}\}} \exp\left(\frac{\beta}{M}\sum_{k=1}^M
  \sum_{\langle ij\rangle}J_{ij}\sigma_i^{(k)}\sigma_j^{(k)}
  +\gamma(t)\sum_{k=1}^{M}\sum_{i=0}^{N}\sigma_i^{(k)}\sigma_i^{(k+1)}\right),
   \label{eq:ST}
\end{equation}
where $M$ is the length along the extra dimension (Trotter number) and
$\sigma_i^{(k)} (=\pm 1)$ denotes a classical Ising spin at site $i$ on
the $k$th Trotter slice.
The nearest-neighbour interaction between adjacent Trotter slices,
\begin{equation}
  \gamma(t)=\frac{1}{2}\log\left(\coth\frac{\beta\Gamma(t)}{M}\right),
  \label{eq:gamma}
\end{equation}
is ferromagnetic.  This approximation (\ref{eq:ST}) becomes exact in the
limit $M\to\infty$ for a fixed $\beta=1/k_BT$.  The magnitude of this
interaction (\ref{eq:gamma}) increases with time $t$ and tends to
infinity as $t\rightarrow\infty$, reflecting the decrease of $\Gamma
(t)$.  We fix $M$ and $\beta$ to arbitrary large values, which
corresponds to the actual situation in numerical simulations.  Therefore
the Theorem presented below does not directly guarantee the convergence
of the system to the true ground state, which is realized only after
taking the limits $M\to\infty$ and $\beta\to \infty$.  We will rather
show that the system converges to the thermal equilibrium represented by
the right-hand side of (\ref{eq:ST}), which can be chosen arbitrarily
close to the true ground state by taking $M$ and $\beta$ large enough.

With the above example of TFIM in mind, it will be convenient to treat a
more general expression than (\ref{eq:ST}),
\begin{equation}
  Z(t)=\sum_{x\in\mathcal{S}} \exp
  \left(-\frac{F_0(x)}{T_0}-\frac{F_1(x)}{T_1(t)}\right).
  \label{eq:Z_PIMC}
\end{equation}
Here $F_0(x)$ is the cost function whose global minimum is the desired
solution of the combinatorial optimization problem. The temperature
$T_0$ is chosen to be sufficiently small.  The term $F_1(x)$ derives
from the kinetic energy, which is the transverse field in the TFIM.
Quantum fluctuations are tuned by the extra temperature factor $T_1(t)$,
which decreases with time.  The first term $-F_0(x)/T_0$ corresponds to
the interaction term in the exponent of (\ref{eq:ST}), and the second
term $-F_1(x)/T_1(t)$ generalizes the transverse-field term in
(\ref{eq:ST}).

For the partition function (\ref{eq:Z_PIMC}), we define
the acceptance probability of PIMC as
\begin{gather}
  A(y,x;t)=g\left(\frac{q(y;t)}{q(x;t)}\right),
  \label{eq:A_PIMC} \\
  q(x;t)=\frac{1}{Z(t)}
  \exp\left(-\frac{F_0(x)}{T_0}-\frac{F_1(x)}{T_1(t)}\right).
  \label{eq:BD}
\end{gather}
This $q(x;t)$ is the equilibrium Boltzmann factor at a given fixed $T_1(t)$.
The function $g(u)$ is the {\em acceptance function}, a monotone increasing
function satisfying $0\leq g(u)\leq 1$ and $g(1/u)=g(u)/u$ for $u\geq 0$.
For instance, for the heat bath and the Metropolis methods, we have
\begin{gather}
  g(u)=\frac{u}{1+u}, \\
  g(u)=\min\{1,u\},
\end{gather}
respectively. The conditions mentioned above for $g(u)$ guarantee that
$q(x;t)$ satisfies the detailed balance condition,
$G(y,x;t)q(x;t)=G(x,y;t)q(y;t)$. Thus, $q(x;t)$ is the stationary
distribution of the homogeneous Markov chain defined by the transition
matrix $G(t)$ with a fixed $t$.  In other words, $q(x;t)$ is the right
eigenvector of $G(t)$ with eigenvalue 1.

\subsubsection{Convergence theorem for PIMC-QA}

We first define a few quantities.
The set of local maximum states of $F_1$ is written as $\mathcal{S}_m$,
\begin{equation}
  \mathcal{S}_m=\left\{x\,|\,x\in\mathcal{S},\ \forall y\in \mathcal{S}_x,\
   F_1(y)\leq F_1(x)\right\}.
\end{equation}
We denote by $d(y,x)$ the minimum number of steps
necessary to make a transition from $x$ to $y$.
Using this notation we define the minimum number of maximum steps
needed to reach any other state from an arbitrary state in the set
$\mathcal{S}\setminus\mathcal{S}_m$,
\begin{equation}
  R=\min\Bigl\{\max\left\{d(y,x)\,|\,y\in\mathcal{S}\right\}
  \bigm| x\in\mathcal{S}\setminus\mathcal{S}_m \Bigr\}.
  \label{def:R}
\end{equation}
Also, $L_0$ and $L_1$ stand for the maximum changes of $F_0(x)$ and $F_1(x)$,
respectively, in a single step,
\begin{gather}
  L_0=\max\Bigl\{\left|F_0(x)-F_0(y)\right| \bigm| P(y,x)>0,\
  x,y\in\mathcal{S}\Bigr\}, \\
  L_1=\max\Bigl\{\left|F_1(x)-F_1(y)\right| \bigm| P(y,x)>0,\
  x,y\in\mathcal{S}\Bigr\}.
\end{gather}
Our main results are summarized in the following Theorem and Corollary.

\begin{thm}[Strong ergodicity of the system (\ref{eq:Z_PIMC})]
The inhomogeneous Markov chain generated by (\ref{eq:A_PIMC}) and
(\ref{eq:BD}) is strongly ergodic and converges to the equilibrium state
corresponding to the first term of the right-hand side of (\ref{eq:BD}),
$\exp(-F_0(x)/T_0)$, if
\begin{equation}
  T_1(t)\geq \frac{RL_1}{\log(t+2)}.
  \label{eq:AS_PIMC}
\end{equation}
\label{theorem:PIMC}
\end{thm}

Application of this Theorem to the PIMC implementation of QA represented by
(\ref{eq:ST}) immediately yields the following Corollary.

\begin{cor}[Strong ergodicity of QA-PIMC for TFIM]
The inhomogeneous Markov chain generated by the Boltzmann factor on the
right-hand side of (\ref{eq:ST}) is strongly ergodic and converges to the
equilibrium state corresponding to the first term on the right-hand side
of (\ref{eq:ST}) if
\begin{equation}
  \Gamma(t)\geq\frac{M}{\beta}\tanh^{-1}\frac{1}{(t+2)^{2/RL_1}}.
\end{equation}
\label{corollary:PIMC}
\end{cor}

\begin{rem}
  For sufficiently large $t$, the above inequality reduces to
  \begin{equation}
   \Gamma(t)\geq \frac{M}{\beta}(t+2)^{-2/RL_1}.  \label{eq:power-decay}
  \end{equation}
  This result implies that a power decay of the transverse field is
  sufficient to guarantee the convergence of quantum annealing of TFIM by
  the PIMC.
  Notice that $R$ is of $\mathcal{O}(N^0)$ and $L_1$ is of $\mathcal{O}(N)$.
  Thus (\ref{eq:power-decay}) is qualitatively similar to (\ref{eq:adiabatic AS}).
\end{rem}

To prove strong ergodicity it is necessary to prove weak ergodicity
first.  The following Lemma is useful for this purpose.

\begin{lem}[Lower bound on the transition probability]
\label{lemma:1}
The elements of the transition matrix defined by (\ref{eq:G}),
(\ref{eq:A_PIMC}) and (\ref{eq:BD}) have the following lower bound:
\begin{equation}
  P(y,x)>0 \Rightarrow \forall t>0 :
  G(y,x;t)\geq w\, g(1) \exp\left(-\frac{L_0}{T_0}-\frac{L_1}{T_1(t)}\right),
  \label{eq:LB1}
\end{equation}
and
\begin{equation}
  \exists t_1>0, \forall x\in\mathcal{S}\setminus\mathcal{S}_m,
  \forall t\geq t_1
   : G(x,x;t)\geq w\, g(1)
  \exp\left(-\frac{L_0}{T_0}-\frac{L_1}{T_1(t)}\right).
  \label{eq:LB2}
\end{equation}
\end{lem}
Here, $w$ stands for the minimum non-vanishing value of $P(y,x)$,
\begin{equation}
  w = \min\left\{P(y,x)\,|\, P(y,x)>0,\ x,y\in\mathcal{S}\right\}.
\end{equation}

\begin{proof}
[{\bf Proof of Lemma \ref{lemma:1}}]
The first part of Lemma \ref{lemma:1} is proved straightforwardly.
Equation (\ref{eq:LB1}) follows directly from the definition of the
transition probability and the property of the acceptance function $g$.
When $q(y;t)/q(x;t)<1$, we have
\begin{equation}
   G(y,x;t)\geq w\, g\left(\frac{q(x;t)}{q(y;t)}\right)
  \frac{q(y;t)}{q(x;t)}\geq w\, g(1)
  \exp\left(-\frac{L_0}{T_0}-\frac{L_1}{T_1(t)}\right).
\end{equation}
On the other hand, if $q(y;t)/q(x;t)\geq1$,
\begin{equation}
  G(y,x;t)\geq w\, g(1)\geq w\, g(1)
  \exp\left(-\frac{L_0}{T_0}-\frac{L_1}{T_1(t)}\right),
\end{equation}
where we used the fact that both $L_0$ and $L_1$ are positive.

Next, we prove (\ref{eq:LB2}). Since $x$ is not a member of
$\mathcal{S}_m$, there exists a state $y\in\mathcal{S}_x$ such that
$F_1(y)-F_1(x)>0$. For such a state $y$,
\begin{equation}
  \lim_{t\rightarrow\infty} g\left(\exp\left(
  -\frac{F_0(y)-F_0(x)}{T_0}-\frac{F_1(y)-F_1(x)}{T_1(t)}\right)\right)
  =0,
\end{equation}
because $T_1(t)$ tends to zero as $t\rightarrow\infty$ and $0\leq g(u)\leq
u$. Thus, for all $\varepsilon>0$, there exists $t_1>0$  such that
\begin{equation}
  \forall t>t_1 : g\left(\exp\left(-\frac{F_0(y)-F_0(x)}{T_0}
  -\frac{F_1(y)-F_1(x)}{T_1(t)}\right)\right) < \varepsilon .
\end{equation}
We therefore have
\begin{align}
  \sum_{z\in\mathcal{S}}P(z,x)A(z,x;t)&=P(y,x)A(y,x;t)+
   \sum_{z\in\mathcal{S}\setminus\{y\}} P(z,x)A(z,x;t) \nonumber \\
  &<P(y,x)\varepsilon+\sum_{z\in\mathcal{S}\setminus\{y\}} P(z,x) \nonumber \\
  &=1-(1-\varepsilon)P(y,x),
\end{align}
and consequently,
\begin{equation}
  G(x,x;t)>(1-\varepsilon) P(y,x)>0.
\end{equation}
Since the right-hand side of (\ref{eq:LB2}) can be arbitrarily small for
sufficiently large $t$, we obtain the second part of Lemma
\ref{lemma:1}.
\end{proof}

\begin{proof}
[{\bf Proof of weak ergodicity implied in Theorem \ref{theorem:PIMC}}]
Let us introduce the following quantity
\begin{equation}
  x^*=\arg \min\Bigl\{\max\left\{d(y,x)\,|\,y\in\mathcal{S}\right\}
  \bigm| x\in\mathcal{S}\setminus\mathcal{S}_m \Bigr\}.
\end{equation}
Comparison with the definition of $R$ in (\ref{def:R}) shows that
the state $x^*$ is reachable by at most $R$ transitions from any states.

Now, consider the transition probability from an arbitrary state $x$ to
$x^*$. From the definitions of $R$ and $x^*$, there exists at least one
transition route within $R$ steps:
\begin{equation*}
  x\equiv x_0\neq x_1\neq x_2\neq \cdots \neq x_l
  =x_{l+1}=\cdots =x_R\equiv x^* .
\end{equation*}
Then Lemma \ref{lemma:1} yields that, for sufficiently large $t$, the
transition probability at each time step has the following lower bound:
\begin{equation}
  G(x_{i+1},x_i;t-R+i)\geq w g(1)
  \exp\left(-\frac{L_0}{T_0}-\frac{L_1}{T_1(t-R+i)}\right).
   \label{eq:G-bound1}
\end{equation}
Thus, by taking the product of (\ref{eq:G-bound1}) from $i=0$ to $i=R-1$,
we have
\begin{align}
  G^{t,t-R}(x^*,x)&\geq
  G(x^*,x_{R-1};t-1)G(x_{R-1},x_{R-2};t-2)\cdots G(x_1,x;t-R)
  \nonumber \\
  &\geq \prod_{i=0}^{R-1} w\, g(1)
  \exp\left(-\frac{L_0}{T_0}-\frac{L_1}{T_1(t-R+i)}\right)
  \nonumber \\
  &\geq w^R g(1)^R
  \exp\left(-\frac{RL_0}{T_0}-\frac{RL_1}{T_1(t-1)}\right),
   \label{eq:G-bound2}
\end{align}
where we have used monotonicity of $T_1(t)$.
Consequently, it is possible to find an integer $k_0\geq 0$ such that,
for all $k>k_0$, the coefficient of ergodicity satisfies
\begin{equation}
  1-\alpha(G^{kR,kR-R})\geq w^R g(1)^R
  \exp\left(-\frac{RL_0}{T_0}-\frac{RL_1}{T_1(kR-1)}\right),
  \label{eq:1-minus-alpha}
\end{equation}
where we eliminate the sum over $z$ in (\ref{eq:alpha}) by replacing it
with a single term for $z=x^*$.
We now substitute the annealing schedule (\ref{eq:AS_PIMC}).  Then weak
ergodicity is immediately proved from Theorem \ref{theorem:weak-e}
because we obtain
\begin{equation}
  \sum_{k=1}^{\infty} (1-\alpha(G^{kR,kR-R}))
  \geq w^R g(1)^R \exp\left(-\frac{RL_0}{T_0}\right)
  \sum_{k=k_0}^{\infty}\frac{1}{kR+1}\longrightarrow\infty.
  \qedhere
\end{equation}
\end{proof}

\begin{proof}
[{\bf Proof of Theorem \ref{theorem:PIMC}}]
To prove strong ergodicity, we refer to Theorem \ref{theorem:strong-e}.
The condition 1 has already been proved.  As has been mentioned, the
Boltzmann factor (\ref{eq:BD}) satisfies $q(t)=G(t)q(t)$, which is the
condition 2.  Thus the proof will be complete if we prove the
condition 3 by setting $p_t=q(t)$. For this purpose, we first prove
that $q(x;t)$ is monotonic for large $t$:
\begin{equation}
  \forall t\geq 0, \forall x\in\mathcal{S}_1^{\rm min} :
  q(x;t+1)\geq q(x;t),
  \label{eq:MBD1}
\end{equation}
\begin{equation}
  \exists t_1>0, \forall t\geq t_1, \forall
   x\in\mathcal{S}\setminus\mathcal{S}_1^{\rm min} :
  q(x;t+1)\leq q(x;t),
  \label{eq:MBD2}
\end{equation}
where $\mathcal{S}_1^{\rm min}$ denotes the set of global minimum states
of $F_1$.

To prove this monotonicity, we use the following notations for simplicity:
\begin{gather}
  A(x)=\exp\left(-\frac{F_0(x)}{T_0}\right), \quad
  B=\sum_{x\in{\mathcal S}_1^{\rm min}} A(x), \\
  \Delta(x)=F_1(x)-F_1^{\rm min}.
\end{gather}
If $x\in\mathcal{S}_1^{\rm min}$, the Boltzmann distribution can be
rewritten as
\begin{equation}
  q(x;t)=\frac{A(x)}
  {\displaystyle B +\sum_{y\in\mathcal{S}\setminus{\mathcal S}_1^{\rm min}}
  \exp\left(-\frac{\Delta(y)}{T_1(t)}\right)A(y)}.
\end{equation}
Since $\Delta(y)> 0$ by definition, the denominator decreases with
time. Thus, we obtain (\ref{eq:MBD1}).

To prove (\ref{eq:MBD2}), we consider the derivative of $q(x;t)$ with
respect to $T_1(t)$,
\begin{equation}
  \frac{\partial q(x;t)}{\partial T_1(t)}=
  \frac{A(x)\left[\displaystyle B \Delta(x)
   +\sum_{y\in{\mathcal S}\setminus \mathcal{S}_1^{\rm min}}(F_1(x)-F_1(y))
   \exp\left(-\frac{\Delta(y)}{T_1(t)}\right)A(y)\right]}
  {\displaystyle T(t)^2 
\exp\left(\frac{\Delta(x)}{T_1(t)}\right)\left[\displaystyle
  B+\sum_{y\in{\mathcal S}\setminus \mathcal{S}_1^{\rm min}}
  \exp\left(-\frac{\Delta(y)}{T_1(t)}\right)A(y)\right]^2}.
\end{equation}
Only $F_1(x)-F_1(y)$ in the numerator has the possibility of being
negative. However, the first term $B\Delta(x)$ is
larger than the second one for sufficient large $t$ because
$\exp\left(-\Delta(y)/T_1(t)\right)$ tends to zero as
$T_1(t)\rightarrow\infty$. Thus there exists $t_1>0$ such that $\partial
q(x;t)/\partial T(t)>0$ for all $t>t_1$. Since $T_1(t)$ is a decreasing
function of $t$, we have (\ref{eq:MBD2}).

Consequently, for all $t>t_1$, we have
\begin{align}
  \|q(t+1)-q(t)\|
  &=\sum_{x\in\mathcal{S}_1^{\rm min}}\left[
  q(x;t+1)-q(x;t)\right]
  -\sum_{x\not\in\mathcal{S}_1^{\rm min}}\left[
  q(x;t+1)-q(x;t)\right] \nonumber \\
  &=2\sum_{x\in\mathcal{S}_1^{\rm min}}\left[
  q(x;t+1)-q(x;t)\right],
  \label{eq:qq-difference}
\end{align}
where we used $\|q(t)\|=\sum_{x\in\mathcal{S}_1^{\rm min}}q(x;t)
+\sum_{x\not\in\mathcal{S}_1^{\rm min}}q(x;t)=1$. We then obtain
\begin{equation}
  \sum_{t=t_1}^{\infty}\|q(t+1)-q(t)\|
  =2\sum_{x\in\mathcal{S}_1^{\rm min}}\left[
  q(x;\infty)-q(x;t_1)\right]\leq 2\|q(x;\infty)\|=2.
\end{equation}
Therefore $q(t)$ satisfies the condition 3:
\begin{align}
  \sum_{t=0}^{\infty}\|q(t+1)-q(t)\| &=
  \sum_{t=0}^{t_1-1}\|q(t+1)-q(t)\|+\sum_{t=t_1}^{\infty}
  \|q(t+1)-q(t)\| \nonumber \\
  &\leq \sum_{t=0}^{t_1-1}\left[\|q(t+1)\|+\|q(t)\|\right]+2 \nonumber\\
  &=2t_1+2<\infty,
\end{align}
which completes the proof of strong ergodicity.
\end{proof}

\subsubsection{Generalized transition probability}
In Theorem \ref{theorem:PIMC}, the acceptance probability is defined by
the conventional Boltzmann form, (\ref{eq:A_PIMC}) and (\ref{eq:BD}).
However, we have the freedom to choose any transition (acceptance)
probability as long as it is useful to achieve our objective since our
goal is not to find finite-temperature equilibrium states but to
identify the optimal state.  There have been attempts to accelerate the
annealing schedule in SA by modifying the transition probability.  In
particular Nishimori and Inoue \cite{NishimoriI} have proved weak
ergodicity of the inhomogeneous Markov chain for classical simulated
annealing using the probability of Tsallis and Stariolo \cite{TsallisS}.
There the property of weak ergodicity was shown to hold under the
annealing schedule of temperature inversely proportional to a power of
time steps.  This annealing rate is much faster than the log-inverse law
for the conventional Boltzmann factor.

A similar generalization is possible for QA-PIMC by using
the following modified acceptance probability
\begin{equation}
  A(y,x;t)=g\left(u(y,x;t)\right),
\end{equation}
\begin{equation}
  u(y,x;t)=\rme^{-[F_0(y)-F_0(x)]/T_0}
  \left[1+(q-1)\frac{F_1(y)-F_1(x)}{T_1(t)}\right]^{1/(1-q)},
\end{equation}
where $q$ is a real number.  In the limit $q\rightarrow 1$, this
acceptance probability reduces to the Boltzmann form.  Similarly to the
discussions leading to Theorem \ref{theorem:PIMC}, we can prove that the
inhomogeneous Markov chain with this acceptance probability is weakly
ergodic if
\begin{equation}
  T_1(t)\geq \frac{b}{(t+2)^c}, \qquad 0< c\leq \frac{q-1}{R},
  \label{eq:T_G_PIMC}
\end{equation}
where $b$ is a positive constant.  We have to restrict ourselves to the
case $q>1$ for a technical reason as was the case previously
\cite{NishimoriI}.  We do not reproduce the proof here because it is
quite straightforward to generalize the discussions for Theorem
\ref{theorem:PIMC} in combination with the argument of
\cite{NishimoriI}.  The result (\ref{eq:T_G_PIMC}) applied to the TFIM
is that, if the annealing schedule asymptotically satisfies
\begin{equation}
  \Gamma(t)\geq \frac{M}{\beta}\exp\left(-\frac{2(t+2)^c}{b}\right),
\end{equation}
the inhomogeneous Markov chain is weakly ergodic. Notice that this
annealing schedule is faster than the power law of
(\ref{eq:power-decay}).  We have been unable to prove strong ergodicity
because we could not identify the stationary distribution for a fixed
$T_1(t)$ in the present case.

\subsubsection{Continuous systems}

In the above analyses we treated systems with discrete degrees of
freedom.  Theorem \ref{theorem:PIMC} does not apply directly to a
continuous system.  Nevertheless, by discretization of the continuous
space we obtain the following result.

Let us consider a system of $N$ distinguishable particles in a continuous
space of finite volume with the Hamiltonian
\begin{equation}
  H=\frac{1}{2m(t)}\sum_{i=1}^{N} \boldsymbol{p}_i^2+V(\{\boldsymbol{r}_i\}).
  \label{eq:H}
\end{equation}
The mass $m(t)$ controls the magnitude of quantum fluctuations.  The
goal is to find the minimum of the potential term, which is achieved by
a gradual increase of $m(t)$ to infinity according to the prescription
of QA.
After discretization of the continuous space (which is necessary anyway
in any computer simulations with finite precision) and an application of
the Suzuki-Trotter formula, the equilibrium partition function acquires
the following expression in the representation to diagonalize spatial
coordinates
\begin{equation}
   Z(t)\approx \Tr\exp\left(
  -\frac{\beta}{M}\sum_{k=1}^{M}V\left(\{\boldsymbol{r}_i^{(k)}\}\right)
  -\frac{Mm(t)}{2\beta}\sum_{i=1}^{N}\sum_{k=1}^{M}
   \left|\boldsymbol{r}_i^{(k+1)}-\boldsymbol{r}_i^{(k)}\right|^2 \right)
\end{equation}
with the unit $\hbar=1$.  Theorem \ref{theorem:PIMC} is
applicable to this system under the identification of $T_1(t)$ with
$m(t)^{-1}$.  We therefore conclude that a logarithmic increase of the
mass suffices to guarantee strong ergodicity of the
potential-minimization problem under spatial discretization.

The coefficient corresponding to the numerator of the right-hand side of
(\ref{eq:AS_PIMC}) is estimated as
\begin{equation}
  RL_1 \approx M^2NL^2/\beta,
  \label{eq:RL_PIMC}
\end{equation}
where $L$ denotes the maximum value of
$\left|\boldsymbol{r}_i^{(k+1)}-\boldsymbol{r}_i^{(k)}\right|$.  To
obtain this coefficient, let us consider two extremes. One is that any
states are reachable at one step. By definition, $R=1$ and $L_1\approx
M^2 N L^2/\beta$, which yields (\ref{eq:RL_PIMC}). The other case is
that only one particle can move to the nearest neighbor point at one
time step. With $a$ $(\ll L)$ denoting the lattice spacing, we have
\begin{equation}
  L_1 \approx \frac{M}{2\beta}\left[L^2-(L-a)^2\right]
  \approx\frac{ML a}{\beta}.
\end{equation}
Since the number of steps to reach any configurations is
estimated as $R\approx NML/a$, we again obtain (\ref{eq:RL_PIMC}).

\subsection{Green's function Monte Carlo method}

The path-integral Monte Carlo simulates only the equilibrium behavior
at finite temperature because its starting point is the equilibrium
partition function.  Moreover, it follows an artificial time evolution
of Monte Carlo dynamics, not the natural Schr\"odinger dynamics.  An
alternative approach to improve these points is the Green's function
Monte Carlo (GFMC) method \cite{LandauB,CeperleyA,TrivediC,StellaS2007}.
The basic idea is to solve the imaginary-time Schr\"{o}dinger equation
by stochastic processes. In the present section we derive sufficient
conditions for strong ergodicity in GFMC.

The evolution of states by the imaginary-time Schr\"{o}dinger equation
starting from an initial state $|\psi_0\rangle$ is expressed as
\begin{equation}
  |\psi(t)\rangle = {\rm T}\exp\left(-\int_{0}^{t}\rmd t' H(t')\right)
   |\psi_0\rangle,
\end{equation}
where T is the time-ordering operator.  The right-hand side can be
decomposed into a product of small-time evolutions,
\begin{equation}
  |\psi(t)\rangle =\lim_{n\rightarrow\infty}
   \hat{G}_0(t_{n-1})\hat{G}_0(t_{n-2})
   \cdots\hat{G}_0(t_1)\hat{G}_0(t_0)|\psi_0\rangle,
   \label{eq:GFMC1}
\end{equation}
where $t_k=k\Delta t$, $\Delta t=t/n$ and $\hat{G}_0(t)=1-\Delta t\cdot
H(t)$.  In the GFMC, one approximates the right-hand side of this
equation by a product with large but finite $n$ and replaces
$\hat{G}_0(t)$ with $\hat{G}_1(t)=1-\Delta t (H(t)-E_T)$, where $E_T$ is
called the reference energy to be taken approximately close to the final
ground-state energy.  This subtraction of the reference energy simply
adjusts the standard of energy and changes nothing physically.  However,
practically, this term is important to keep the matrix elements positive
and to accelerate convergence to the ground state as will be explained
shortly.

To realize the process of (\ref{eq:GFMC1}) by a stochastic method,
we rewrite this equation in a recursive form,
\begin{equation}
  \psi_{k+1}(y) = \sum_{x} \hat{G}_1(y,x;t_k)\psi_k(x),
  \label{eq:psi}
\end{equation}
where $\psi_k(x)=\langle x|\psi_k\rangle$ and $|x\rangle$ denotes a
basis state.  The matrix element of Green's function is given by
\begin{equation}
  \hat{G}_1(y,x;t)=\left\langle y\right|1-\Delta t\left[H(t)-E_T\right]|x\rangle.
  \label{eq:G1_GFMC}
\end{equation}
Equation (\ref{eq:psi}) looks similar to a Markov process
but is significantly different in several ways.
An important difference is that the Green's function is not normalized,
$\sum_{y}\hat{G}_1(y,x;t)\neq 1$.
In order to avoid this problem, one decomposes the Green's function into a
normalized probability $G_1$ and a weight $w$:
\begin{equation}
   \hat{G}_1(y,x;t)=G_1(y,x;t) w(x;t),
\end{equation}
where
\begin{equation}
   G_1(y,x;t)\equiv \frac{\hat{G}_1(y,x;t)}{\sum_y \hat{G}_1(y,x;t)}, \quad
   w(x;t)\equiv \frac{\hat{G}_1(y,x;t)}{G_1(y,x;t)}.
\end{equation}
Thus, using (\ref{eq:psi}), the wave function at time $t$ is written as
\begin{align}
  \psi_n(y) &=\sum_{\{x_k\}}\delta_{y,x_n}w(x_{n-1};t_{n-1})
  w(x_{n-2};t_{n-2})\cdots w(x_0;t_0) \nonumber \\
  &\quad\times G_1(x_n,x_{n-1};t_{n-1}) G_1(x_{n-1},x_{n-2};t_{n-2})
  \cdots G_1(x_1,x_0;t_0) \psi_0(x_0).
  \label{eq:GFMC}
\end{align}

The algorithm of GFMC is based on this formula and is defined by a
weighted random walk in the following sense.  One first prepares an
arbitrary initial wave function $\psi_0(x_0)$, all elements of which are
non-negative.  A random walker is generated, which sits initially
($t=t_0$) at the position $x_0$ with a probability proportional to
$\psi_0(x_0)$.  Then the walker moves to a new position $x_1$ following
the transition probability $G_1(x_1,x_0;t_0)$.  Thus this probability
should be chosen non-negative by choosing parameters appropriately as
described later.
Simultaneously, the weight of this walker is updated by the rule
$W_{1}=w(x_{0};t_{0})W_0$ with $W_0=1$.
This stochastic process is repeated to $t=t_{n-1}$.
One actually prepares $M$ independent walkers and let those walkers
follow the above process.
Then, according to (\ref{eq:GFMC}), the wave function $\psi_n(y)$
is approximated by the distribution of walkers at the final step
weighted by $W_n$,
\begin{equation}
  \psi_n(y)= \lim_{M\rightarrow\infty}\frac{1}{M}
   \sum_{i=1}^{M}W_n^{(i)}\delta_{y,x_n^{(i)}},
\end{equation}
where $i$ is the index of a walker.

As noted above, $G_1(y,x;t)$ should be non-negative, which is achieved by
choosing sufficiently small $\Delta t$ ({\em i.e.} sufficiently large $n$)
and selecting $E_T$ within the instantaneous spectrum of the
Hamiltonian $H(t)$.
In particular, when $E_T$ is close to the instantaneous ground-state energy
of $H(t)$ for large $t$ ({\em i.e.} the final target energy), $\hat{G}_1(x,x;t)$
is close to unity whereas other matrix components of $\hat{G}_1(t)$
are small.
Thus, by choosing $E_T$ this way, one can accelerate convergence of GFMC
to the optimal state in the last steps of the process.

If we apply this general framework to the TFIM with the
$\sigma^z$-diagonal basis, the matrix elements of Green's function are
immediately calculated as
\begin{equation}
   \hat{G}_1(y,x;t)=
    \begin{cases}
     1-\Delta t\left[E_0(x)-E_T\right] & (x=y) \\
     \Delta t\, \Gamma(t) & \text{($x$ and $y$ differ by a single-spin flip)} \\
     0 & \text{(otherwise)},
    \end{cases}
   \label{eq:g-hat}
\end{equation}
where $E_0(x)=\langle x| \left(-\sum_{ij} J_{ij}\sigma_i^z\sigma_j^z
\right) |x\rangle$.
One should choose $\Delta t$ and $E_T$ such that
$1-\Delta t(E_0(x)-E_T)\geq 0$ for all $x$.
Since $w(x,t)=\sum_{y}\hat{G}_1(y,x;t)$, the weight is given by
\begin{equation}
  w(x;t)=1-\Delta t\left[E_0(x)-E_T\right]+N\Delta t\, \Gamma(t).
  \label{eq:w-GFMC}
\end{equation}
One can decompose this transition probability into the generation probability
and the acceptance probability as in (\ref{eq:G}):
\begin{equation}
  P(y,x)=
  \begin{cases}
   \frac{1}{N} & \text{(single-spin flip)} \\
   0 & \text{(otherwise)}
  \end{cases}
  \label{eq:P_GFMC}
\end{equation}
\begin{equation}
  A(y,x;t)=\frac{N\Delta t\, \Gamma(t)}
   {1-\Delta t\left[E_0(x)-E_T\right]+N\Delta t\, \Gamma(t)}.
  \label{eq:A_GFMC}
\end{equation}
We shall analyze the convergence properties of stochastic processes
under these probabilities for TFIM.

\subsubsection{Convergence theorem for GFMC-QA}

Similarly to the QA by PIMC, it is necessary to reduce the strength of
quantum fluctuations slowly enough in order to find the ground state in the GFMC.
The following Theorem provides a sufficient condition in this regard.
\begin{thm}[Strong ergodicity of QA-GFMC]
\label{theorem:GFMC}
The inhomogeneous Markov process of the random walker for the QA-GFMC of TFIM,
(\ref{eq:G}), (\ref{eq:P_GFMC}) and (\ref{eq:A_GFMC}), is strongly ergodic if
\begin{equation}
  \Gamma(t)\geq \frac{b}{(t+1)^c}, \qquad 0< c\leq \frac{1}{N}.
  \label{eq:AS_GFMC}
\end{equation}
\end{thm}

The lower bound of the transition probability given in the following Lemma
will be used in the proof of Theorem \ref{theorem:GFMC}.

\begin{lem}
\label{lemma:GFMC}
The transition probability of random walk in the GFMC defined by (\ref{eq:G}),
(\ref{eq:P_GFMC}) and (\ref{eq:A_GFMC}) has the lower bound:
\begin{gather}
  P(y,x)>0 \Rightarrow \forall t>0: G_1(y,x;t)\geq
  \frac{\Delta t\,\Gamma(t)}{1-\Delta t\left(E_{\rm min}-E_T\right)
  +N\Delta t\,\Gamma(t)},\\
  \exists t_1>0, \forall t>t_1: G_1(x,x;t)\geq
  \frac{\Delta t\,\Gamma(t)}{1-\Delta t\left(E_{\rm min}-E_T\right)
  +N\Delta t\,\Gamma(t)},
  \label{eq:L2_2}
\end{gather}
where $E_{\rm min}$ is the minimum value of $E_0(x)$
\begin{equation}
  E_{\rm min}=\min\{ E_0(x)|x\in \mathcal{S}\}.
\end{equation}
\end{lem}

\begin{proof}
[{\bf Proof of Lemma \ref{lemma:GFMC}}]

The first part of Lemma \ref{lemma:GFMC} is trivial because the
transition probability is an increasing function with respect to
$E_0(x)$ when $P(y,x)>0$ as seen in (\ref{eq:A_GFMC}).  Next, we prove
the second part of Lemma \ref{lemma:GFMC}.  According to (\ref{eq:g-hat})
and (\ref{eq:w-GFMC}), $G_1(x,x;t)$ is written as
\begin{equation}
  G_1(x,x;t)=1-\frac{N\Delta t\, \Gamma(t)}{1-\Delta t\left[E_0(x)-E_T\right]
  +N\Delta t\,\Gamma(t)}.
\end{equation}
Since the transverse field $\Gamma(t)$ decreases to zero with time, the
second term on the right-hand side tends to zero as $t \rightarrow
\infty$. Thus, there exists $t_1>0$ such that $G_1(x,x;t)>1-\varepsilon$
for $\forall \varepsilon>0$ and $\forall t>t_1$. On the other hand, the
right-hand side of (\ref{eq:L2_2}) converges to zero as
$t\rightarrow\infty$. We therefore have (\ref{eq:L2_2}).
\end{proof}

\begin{proof}
[{\bf Proof of Theorem \ref{theorem:GFMC}}]

We show that the condition (\ref{eq:AS_GFMC}) is sufficient to
satisfy the three conditions of Theorem \ref{theorem:strong-e}.

1. From Lemma \ref{lemma:GFMC}, we obtain a bound on the coefficient of
ergodicity for sufficiently large $k$ as
\begin{equation}
   1-\alpha(G_1^{kN,kN-N})\geq \left[
   \frac{\Delta t\, \Gamma(kN-1)}
   {1-\Delta t\left(E_{\rm min}-E_T\right)+N\Delta t\,\Gamma(kN-1)}\right]^N,
\end{equation}
in the same manner as we derived (\ref{eq:1-minus-alpha}), where we used $R=N$.
Substituting the annealing schedule (\ref{eq:AS_GFMC}), we can prove weak
ergodicity from Theorem \ref{theorem:weak-e} because
\begin{equation}
  \sum_{k=1}^{\infty}\left[1-\alpha(G_1^{kN,kN-N})\right]\geq
  \sum_{k=k_0}^{\infty}\frac{{b'}^N}{(kN)^{cN}}
\end{equation}
which diverges when $0<c\leq 1/N$.

2. The stationary distribution of the instantaneous transition
probability $G_1(y,x;t)$ is
\begin{equation}
  q(x;t)\equiv \frac{w(x;t)}{\sum_{x\in{\mathcal S}}w(x;t)}
  =\frac{1}{2^N}-\frac{\Delta t\, E_0(x)}
  {2^N \left[1+\Delta t\, E_T+N\Delta t\, \Gamma(t)\right]},
  \label{eq:q_GFMC}
\end{equation}
which is derived as follows. The transition probability defined by
(\ref{eq:G}), (\ref{eq:P_GFMC}) and (\ref{eq:A_GFMC}) is rewritten in terms
of the weight (\ref{eq:w-GFMC}) as
\begin{equation}
  G_1(y,x;t)=\begin{cases}\displaystyle
   1-\frac{N\Delta t\, \Gamma(t)}{w(x;t)} & (x=y) \\
   \displaystyle \frac{\Delta t\, \Gamma(t)}{w(x;t)} &
  (x\in\mathcal{S}_y; \text{single-spin flip}) \\
   0 & (\text{otherwise}).
  \end{cases}
\end{equation}
Thus, we have
\begin{align}
  \sum_{x\in\mathcal{S}} G_1(y,x;t)q(x;t) &=
  \left[1-\frac{N\Delta t\, \Gamma(t)}{w(y;t)}\right]\frac{w(y;t)}{A}
  +\sum_{x\in\mathcal{S}_y} \frac{\Delta t\, \Gamma(t)}{w(x;t)}
  \frac{w(x;t)}{A} \nonumber \\
  &=q(y;t)-\frac{N\Delta t\, \Gamma(t)}{A}
  +\frac{\Delta t\, \Gamma(t)}{A}\sum_{x\in\mathcal{S}_y}1,\label{eq:APP_C}
\end{align}
where $A$ denotes the normalization factor,
\begin{align}
  \sum_{x\in\mathcal{S}} w(x;t)
  &= \Tr\left[ 1-\Delta t\left(-\sum_{\langle ij \rangle}
  J_{ij}\sigma_i^z\sigma_j^z-E_T\right)
  +N\Delta t\, \Gamma(t) \right] \nonumber \\
  &=2^N \left[1+\Delta t\, E_T+N\Delta t\, \Gamma(t)
  \right],
\end{align}
where we used $\Tr \sum J_{ij}\sigma_i^z\sigma_j^z=0$.
Since the volume of $\mathcal{S}_y$ is $N$, (\ref{eq:APP_C}) indicates
that $q(x;t)$ is the stationary distribution of $G_1(y,x;t)$. The
right-hand side of (\ref{eq:q_GFMC}) is easily derived from the above
equation.

3. Since the transverse field $\Gamma(t)$ decreases monotonically
with $t$, the above stationary distribution $q(x;t)$ is an increasing
function of $t$ if $E_0(x)< 0$ and is decreasing if $E_0\geq 0$.
Consequently, using the same procedure as in (\ref{eq:qq-difference}), we
have
\begin{equation}
  \|q(t+1)-q(t)\|=2\sum_{E_0(x)<0} [q(x;t+1)-q(x;t)],
\end{equation}
and thus
\begin{equation}
  \sum_{t=0}^{\infty}\|q(t+1)-q(t)\|=2\sum_{E_0(x)<0}[q(x;\infty)-q(x;0)]
  \leq 2.
\end{equation}
Therefore the sum $\sum_{t=0}^{\infty}\|q(t+1)-q(t)\|$ is finite, which
completes the proof of the condition 3.
\end{proof}

\begin{rem}
  Theorem \ref{theorem:GFMC} asserts convergence of the distribution of
  random walkers to the equilibrium distribution (\ref{eq:q_GFMC}) with
  $\Gamma(t)\to 0$.  This implies that the final distribution is not
  delta-peaked at the ground state with minimum $E_0(x)$ but is a
  relatively mild function of this energy.  The optimality of the solution
  is achieved after one takes the weight factor $w(x;t)$ into account: The
  repeated multiplication of weight factors as in (\ref{eq:GFMC}), in
  conjunction with the relatively mild distribution coming from the
  product of $G_1$ as mentioned above, leads to the asymptotically
  delta-peaked wave function $\psi_n(y)$ because $w(x;t)$ is larger for
  smaller $E_0(x)$ as seen in (\ref{eq:w-GFMC}).
\end{rem}

\subsubsection{Alternative choice of Green's function}
So far we have used the Green's function defined in (\ref{eq:G1_GFMC}),
which is linear in the transverse field, allowing single-spin flips
only.  It may be useful to consider another type of Green's function
which accommodates multi-spin flips.  Let us try the following form of
Green's function,
\begin{equation}
   \hat{G}_2(t)=\exp\left(\Delta t\,\Gamma(t)\sum_{i}\sigma_i^x\right)
   \exp\left(\Delta t\sum_{ij}J_{ij}\sigma_i^z\sigma_j^z\right),
   \label{eq:Green2}
\end{equation}
which is equal to $\hat{G}_0(t)$ to the order $\Delta t$.
The matrix element of $\hat{G}_2(t)$ in the $\sigma^z$-diagonal basis is
\begin{equation}
   \hat{G}_2(y,x;t)=\cosh^N \left(\Delta t\,\Gamma(t)\right)
   \tanh^\delta \left(\Delta t\,\Gamma(t)\right)\rme^{-\Delta t\,E_0(x)},
\end{equation}
where $\delta$ is the number of spins in different states in $x$ and $y$.
According to the scheme of GFMC, we decompose $\hat{G}_2(y,x;t)$
into the normalized transition probability and the weight:
\begin{equation}
  G_2(y,x;t)=\left[
  \frac{\cosh (\Delta t\,\Gamma(t))}{\rme^{\Delta t\,\Gamma(t)}}\right]^N
  \tanh^\delta (\Delta t\,\Gamma(t)),
  \label{eq:G2}
\end{equation}
\begin{equation}
  w_2(x;t)=\rme^{\Delta t\,N\Gamma(t)}\rme^{-\Delta t\,E_0(x)}.
\end{equation}
It is remarkable that the transition probability $G_2$ is independent of
$E_0(x)$, although it depends on $x$ through $\delta$. Thus, the
stationary distribution of random walk is uniform. This property is lost
if one interchanges the order of the two factors in (\ref{eq:Green2}).

The property of strong ergodicity can be shown to hold in this case as well:
\begin{thm}[Strong ergodicity of QA-GFMC 2]
The inhomogeneous Markov chain generated by (\ref{eq:G2}) is strongly
ergodic if
\begin{equation}
\Gamma(t)\geq-\frac{1}{2\Delta t}\log\left(
  1-2b(t+1)^{-1/N}\right).
  \label{eq:AS2_GFMC}
\end{equation}
\end{thm}

\begin{rem}
  For sufficiently large $t$, the above annealing schedule is reduced to
  \begin{equation}
  \Gamma(t)\geq \frac{b}{\Delta t\, (t+1)^{1/N}}.
  \end{equation}
\end{rem}

Since the proof is quite similar to the previous cases, we just outline the
idea of the proof.
The transition probability $G_2(y,x;t)$ becomes smallest when $\delta=N$.
Consequently, the coefficient of ergodicity is estimated as
\begin{equation*}
   1-\alpha(G_2^{t+1,t})\geq \left[
   \frac{1-\rme^{-2\Delta t\,\Gamma(t)}}{2}\right]^N.
\end{equation*}
We note that $R$ is equal to 1 in the present case because any states are
reachable from an arbitrary state in a single step.
 From Theorem \ref{theorem:weak-e}, the condition
\begin{equation}
  \left[\frac{1-\rme^{-2\Delta t\,\Gamma(t)}}{2}\right]^N
  \geq \frac{b'}{t+1}
\end{equation}
is sufficient for weak ergodicity. From this, one obtains
(\ref{eq:AS2_GFMC}).  Since the stationary distribution of $G_2(y,x;t)$
is uniform as mentioned above, strong ergodicity readily follows from
Theorem \ref{theorem:strong-e}.

Similarly to the case of PIMC, we can discuss the convergence condition of
QA-GFMC in systems with continuous degrees of freedom.
The resulting sufficient condition is a logarithmic
increase of the mass as will be shown now.
The operator $\hat{G}_2$ generated by the Hamiltonian (\ref{eq:H}) is
written as
\begin{equation}
   \hat{G}_2(t)=\exp\left(-\frac{\Delta t}{2m(t)}
   \sum_{i=1}^{N}\boldsymbol{p}_i^2\right)
   \rme^{-\Delta t V(\{\boldsymbol{r}_i\})}.
\end{equation}
Thus, the Green's function is calculated in a discretized space as
\begin{equation}
  \hat{G}_2(y,x;t)\propto \exp\left(
  -\frac{m(t)}{2\Delta 
t}\sum_{i=1}^{N}\left|\boldsymbol{r}'_i-\boldsymbol{r}_i\right|^2
  -\Delta t V(\{\boldsymbol{r}_i\})\right),
\end{equation}
where $x$ and $y$ represent $\{\boldsymbol{r}_i\}$ and
$\{\boldsymbol{r}'_i\}$, respectively.  Summation over $y$, {\em i.e.}
integration over $\{\boldsymbol{r}'_i\}$, yields the weight $w(x;t)$,
from which the transition probability is obtained:
\begin{equation}
  w(x;t)\propto \rme^{-\Delta t V(\{\boldsymbol{r}_i\})},
\end{equation}
\begin{equation}
  G_2(y,x;t)\propto \exp\left( -\frac{m(t)}{2\Delta t}
  \sum_{i=1}^{N}\left|\boldsymbol{r}'_i-\boldsymbol{r}_i\right|^2\right).
\end{equation}
The lower bound for the transition probability depends exponentially on
the mass: $G_2(y,x;t)\geq \rme^{-Cm(t)}$. Since $1-\alpha(G_2^{t+1,t})$
has the same lower bound, the sufficient condition for weak ergodicity
is $\rme^{-Cm(t)}\geq (t+1)^{-1}$, which is rewritten as
\begin{equation}
  m(t)\leq C^{-1}\log(t+1).
\end{equation}
The constant $C$ is proportional to $NL^2/\Delta t$, where $L$ denotes
the maximum value of $|\boldsymbol{r}'-\boldsymbol{r}|$. The derivation of $C$ is
similar to (\ref{eq:RL_PIMC}), because $G_2(t)$ allows any transition to
arbitrary states at one time step.

\section{Summary and perspective}
\label{sec:summary}

In this paper we have studied the mathematical foundation of quantum annealing,
in particular the convergence conditions and the reduction of residual
errors.
In Sec.~\ref{sec:QA}, we have seen that the adiabaticity condition of
the quantum system representing quantum annealing leads to the
convergence condition, {\em i.e.}  the condition for the system to reach
the solution of the classical optimization problem as $t\to\infty$
following the real-time Schr\"odinger equation.
The result shows the asymptotic power decrease of the transverse field as the
condition for convergence.
This rate of decrease of the control parameter is faster than the logarithmic
rate of temperature decrease for convergence of SA.
It nevertheless does not mean the qualitative reduction of
computational complexity from classical SA to QA.
Our method deals with a very generic system that represents most of the
interesting problems including worst instances of difficult problems,
for which drastic reduction of computational complexity is hard to expect.

Section \ref{sec:SA} reviews the quantum-mechanical derivation of the
convergence condition of SA using the classical-quantum mapping without
an extra dimension in the quantum system.
The adiabaticity condition for the quantum system has been shown to be equivalent
to the quasi-equilibrium condition for the classical system at finite 
temperature,
reproducing the well-known convergence condition of SA.
The adiabaticity condition thus leads to the convergence condition of 
both QA and SA.
Since the studies of QAE often exploits the adiabaticity condition to derive
the computational complexity of a given problem, the adiabaticity may be
seen as a versatile tool traversing  QA, SA and QAE.

Section \ref{sec:finiteQA} is for the reduction of residual errors after
finite-time quantum evolution of real- and imaginary-time 
Schr\"odinger equations.
This is a different point of view from the usual context of QAE, where the
issue is to reduce the evolution time (computational complexity) with the
residual error fixed to a given small value.
It has been shown that the residual error can becomes significantly 
smaller by the
ingenious choice of the time dependence of coefficients in the 
quantum Hamiltonian.
This idea allows us to reduce the residual error for any given 
QAE-based algorithm
without compromising the computational complexity apart from
a possibly moderate numerical factor.

In Sec.~\ref{sec:QMC} we have derived the convergence condition of QA implemented
by Quantum Monte Carlo simulations of path-integral and Green function methods.
These approaches bear important practical significance because only
stochastic methods allow us to treat practical large-size problems on 
the classical computer.
A highly non-trivial result in this section is that the convergence condition
for the stochastic methods is essentially the same power-law decrease
of the transverse-field term as in the Schr\"odinger dynamics of 
Sec.~\ref{sec:QA}.
This is surprising since the Monte Carlo (stochastic) dynamics is completely
different from the Schr\"odinger dynamics.
Something deep may lie behind this coincidence and it should be an interesting
target of future studies.

The results presented and/or reviewed in this paper serve as the mathematical
foundation of QA.  We have also stressed the similarity/equivalence of
QA and QAE.  Even the classical SA can be viewed from the same framework
of quantum adiabaticity as long as the convergence conditions are concerned.
Since the studies of very generic properties of QA seem to have been almost
completed, fruitful future developments would lie in the investigations
of problems specific to each case of optimization task by analytical 
and numerical methods.

\section*{Acknowledgement}
We thank G.~E.~Santoro, E.~Tosatti and S.~Suzuki for discussions and
G.~Ortiz for useful comments and correspondence on the details of the proofs of
Theorems in Sec. \ref{sec:SA}.
Financial supports by CREST(JST), DEX-SMI and JSPS are gratefully acknowledged.

\appendix
\section{Hopf's inequality}
\label{sec:Hopf}

In this Appendix, we prove the inequality (\ref{eq:Hopf inequality}).
Although Hopf \cite{Hopf} originally proved this inequality for positive
linear integral operators, we concentrate on a square matrix for
simplicity.

Let $M$ be a strictly positive $m\times m$ matrix. The strict positivity
means that all the elements of $M$ are positive, namely, $M_{ij}>0$ for
all $i$, $j$, which will be denoted by $M>0$. Similarly, $M\geq 0$ means
that $M_{ij}\geq 0$ for all $i$, $j$. We use the same notation for a
vector, that is, $\Vv > 0$ means that all the elements $v_i$ are
positive.

The product of the matrix $M$ and an $m$-element column vector $\Vv$
is denoted as usual by $M\Vv$ and its $i$th element is
\begin{equation}
  (M\Vv)_i = \sum_{j=1}^{m} M_{ij}v_j.
\end{equation}
The strict positivity for $M$ is equivalent to
\begin{equation}
  M\Vv > 0\quad \text{if}\quad \Vv\geq 0, \quad \Vv\neq \Vzero,
  \label{eq_Hopf:strictly positive}
\end{equation}
where $\Vzero$ denotes the zero-vector. Of course, if $\Vv = \Vzero$,
then $M\Vv = \Vzero$.

Any real-valued vector $\Vv$ and any strictly positive vector $\Vp >0$
satisfy
\begin{equation}
  \min_i \frac{v_i}{p_i} \leq \min_i \frac{(M\Vv)_i}{(M\Vp)_i}
  \leq\max_i \frac{(M\Vv)_i}{(M\Vp)_i} \leq \max_i\frac{v_i}{p_i},
  \label{eq_Hopf:trivial inequality}
\end{equation}
because
\begin{gather}
  (M\Vv)_i-\left(\min_i \frac{v_i}{p_i}\right)(M\Vp)_i
  =\sum_{j=1}^{m}M_{ij}\left[
  v_j-\left(\min_i \frac{v_i}{p_i}\right)p_j\right]\geq 0,\\
  \left(\max_i \frac{v_i}{p_i}\right)(M\Vp)_i-(M\Vv)_i
  =\sum_{j=1}^{m}M_{ij}\left[
  \left(\max_i \frac{v_i}{p_i}\right)p_j -v_j \right]\geq 0.
\end{gather}
The above inequality implies that the difference between maximum and
minimum of $(M\Vv)_i/(M\Vp)_i$ is smaller than that of
$v_i/p_i$. Following \cite{Hopf}, we use the notation,
\begin{equation}
  \osc_i \frac{v_i}{p_i}\equiv
   \max_i \frac{v_i}{p_i}-\min_i\frac{v_i}{p_i},
\end{equation}
which is called the oscillation. For a complex-valued vector, we define
\begin{equation}
  \osc_i v_i = \sup_{|\eta|=1} \osc_i \text{Re}(\eta v_i).
\end{equation}
It is easily to derive, for any complex $c$,
\begin{equation}
  \osc_i\, (c v_i)=|c|\osc_i v_i.\label{eq_Hopf:osc of complex constant}
\end{equation}
We can also easily prove that, if $\osc_i v_i=0$, $v_i$ does not depend
on $i$.

We suppose that the simple ratio of matrix elements is bounded,
\begin{equation}
   \frac{M_{ik}}{M_{jk}}\leq \kappa \quad \text{for all $i$, $j$, $k$}.
\end{equation}
This assumption is rewritten by the product form as
\begin{equation}
  \frac{(M\Vv)_i}{(M\Vv)_j}\leq \kappa, \quad \Vv\geq 0, \quad
   \Vv\neq\Vzero,
   \label{eq_Hopf:upper bound}
\end{equation}
for all $i$, $j$ and such $\Vv$. The following Theorem states that the
inequality (\ref{eq_Hopf:trivial inequality}) is sharpened under the
above additional assumption (\ref{eq_Hopf:upper bound}).

\begin{thm}
  \label{Hopf theorem 1}
  If $M$ satisfies the conditions (\ref{eq_Hopf:strictly positive}) and
  (\ref{eq_Hopf:upper bound}), for any $\Vp>0$ and any complex-valued
  $\Vv$,
  \begin{equation}
   \osc_i \frac{(M\Vv)_i}{(M\Vp)_i}\leq \frac{\kappa-1}{\kappa+1}
  \osc_i \frac{v_i}{p_i}.\label{eq_Hopf:theorem 1}
  \end{equation}
\end{thm}

\begin{proof}
  We consider a real-valued vector $\Vv$ at first. For fixed $i$, $j$ and
  fixed $\Vp>0$, we define $X_k$ by
  \begin{equation}
   \frac{(M\Vv)_i}{(M\Vp)_i}-\frac{(M\Vv)_j}{(M\Vp)_j}=
   \sum_{k=1}^{m}X_k v_k.
  \end{equation}
  We do not have to know the exact form of $X_k=X_k(i,j,\Vp)$. When
  $\Vv=a\Vp$, the left-hand side of the above equation vanishes, which implies
  $\sum_{k}X_k p_k=0$. Thus, we have
  \begin{equation}
   \frac{(M\Vv)_i}{(M\Vp)_i}-\frac{(M\Vv)_j}{(M\Vp)_j}=
   \sum_{k=1}^{m}X_k (v_k-a p_k).
   \label{eq_Hopf:sum of X(v-ap)}
  \end{equation}
  Now we choose
  \begin{equation}
   a=\min_i \frac{v_i}{p_i}, \qquad b=\max_i\frac{v_i}{p_i}.
  \end{equation}
  Since $v_k-ap_k=(b-a)p_k-(bp_k-v_k)$, $v_k-ap_k$ takes its minimum $0$
  at $v_k=ap_k$ and its maximum $(b-a)p_k$ at $v_k=bp_k$. Therefore, the
  right-hand side of (\ref{eq_Hopf:sum of X(v-ap)}) with $\Vp$ given
  attains its maximum for
  \begin{equation}
   \Vv = a\Vp^{-}-b\Vp^{+}=a\Vp+(b-a)\Vp^+,
  \end{equation}
  where we defined
  \begin{equation}
   p^{-}_i=
    \begin{cases}
     p_i & (X_i\leq 0)\\
     0 & (X_i>0)
    \end{cases}
   ,\qquad p^{+}_i=
   \begin{cases}
    0 & (X_i\leq 0)\\
    p_i & (X_i>0)
   \end{cases}.
  \end{equation}
  Consequently, we have
  \begin{equation}
   \frac{(M\Vv)_i}{(M\Vp)_i}-\frac{(M\Vv)_j}{(M\Vp)_j}\leq
    \left[\frac{(M\Vp^+)_i}{(M\Vp)_i}-\frac{(M\Vp^+)_j}{(M\Vp)_j}
    \right](b-a).
    \label{eq_Hopf:[](b-a)}
  \end{equation}

  Since, by assumptions, $M>0$ and $\Vp>0$, we have
  \begin{equation}
   M\Vp^{-}\geq 0, \quad M\Vp^{+}\geq 0,\quad
    M\Vp=M\Vp^{-}+M\Vp^{+}>0.
  \end{equation}
  Moreover, $M\Vp^{-}>0$ if $\Vp^{-}\neq\Vzero$ and $M\Vp^{+}>0$ if
  $\Vp^{+}\neq\Vzero$. In either case, namely, $\Vp^{-}=\Vzero$ or
  $\Vp^{+}=\Vzero$, the expression in the square brackets of
  (\ref{eq_Hopf:[](b-a)}) vanishes because $\Vp^{+}$ is equal to
  either $\Vp$ or $\Vzero$. Thus, we may assume that both $M\Vp^{-}>0$
  and $M\Vp^{+}>0$. Therefore the expression in inequality
  (\ref{eq_Hopf:[](b-a)}) is rewritten as
  \begin{equation}
   \frac{(M\Vp^+)_i}{(M\Vp)_i}-\frac{(M\Vp^+)_j}{(M\Vp)_j}\leq
    \frac{1}{1+t}-\frac{1}{1+t'}, \quad
    t\equiv\frac{(M\Vp^{-})_i}{(M\Vp^{+})_i}, \quad
    t'\equiv\frac{(M\Vp^{-})_j}{(M\Vp^{+})_j}.
  \end{equation}
  Since, from the assumption (\ref{eq_Hopf:upper bound}), $t$ and $t'$ are
  bounded from $\kappa^{-1}$ to $\kappa$, we find $t'\leq \kappa t^2$,
  which yields
  \begin{equation}
   \frac{(M\Vp^+)_i}{(M\Vp)_i}-\frac{(M\Vp^+)_j}{(M\Vp)_j}\leq
    \frac{1}{1+t}-\frac{1}{1+\kappa^2 t}.
  \end{equation}
  For $t>0$, the right-hand side of the above inequality takes its
  maximum value $(\kappa-1)/(\kappa+1)$ at $t=\kappa^{-1}$. Finally, we
  obtain
  \begin{equation}
   \frac{(M\Vv)_i}{(M\Vp)_i}-\frac{(M\Vv)_j}{(M\Vp)_j}\leq
   \frac{\kappa-1}{\kappa+1} \osc_i \frac{v_i}{p_i}
  \end{equation}
  for any $i$, $j$. Hence it holds for the sup of the left-hand side,
  which yields (\ref{eq_Hopf:theorem 1}).

  For a complex-valued vector $\Vv$, we replace $v_i$ by
  $\text{Re}\,(\eta v_i)$. Since $M \text{Re}\, (\eta \Vv)
  =\text{Re}\,(\eta M\Vv)$, the same argument for the real vector case yields
  \begin{equation}
   \osc_i \text{Re}\left(\eta \frac{(M\Vv)_i}{(M\Vp)_i}\right)
    \leq \frac{\kappa-1}{\kappa+1}\osc_i\text{Re}
    \left(\eta\frac{v_i}{p_i}\right).
  \end{equation}
  Taking the sup with respect to $\eta$, $|\eta|=1$, on both sides, we
  obtain (\ref{eq_Hopf:theorem 1}).
\end{proof}

We apply this Theorem to the eigenvalue problem,
\begin{equation}
  M\Vv =\lambda \Vv.
   \label{eq_Hopf:eigenvalue equation}
\end{equation}
The Perron-Frobenius theorem states that a non-negative
square matrix, $M\geq 0$, has a real eigenvalue $\lambda_0$ satisfying
$|\lambda|\leq \lambda_0$ for any other eigenvalue $\lambda$. This
result is sharpened for a strictly positive matrix, $M>0$, as the
following Theorems.

\begin{thm}
  \label{Hopf theorem 2}
  Under the hypotheses (\ref{eq_Hopf:strictly positive}) and
  (\ref{eq_Hopf:upper bound}), the eigenvalue equation
  (\ref{eq_Hopf:eigenvalue equation}) has a positive solution
  $\lambda=\lambda_0>0$, $\Vv=\Vq>0$. Moreover, for any vector $\Vp$
  $(\Vp\geq0,\ \Vp\neq\Vzero)$, the sequence
  \begin{equation}
   \Vq_n = \frac{M^n \Vp}{(M^n \Vp)_k}
  \end{equation}
  with $k$ fixed, converges toward such $\Vq$.
\end{thm}

\begin{thm}
  \label{Hopf theorem 3}
  Under the same hypotheses,  (\ref{eq_Hopf:eigenvalue equation}) has no
  other non-negative solutions than $\lambda=\lambda_0$, $\Vv=c\Vq$. For
  $\lambda=\lambda_0$, (\ref{eq_Hopf:eigenvalue equation}) has no other
  solutions than $\Vv=c\Vq$.
\end{thm}

\begin{thm}
  \label{Hopf theorem 4}
  Under the same hypotheses, any (complex) eigenvalue
  $\lambda\neq\lambda_0$ of (\ref{eq_Hopf:eigenvalue equation})
  satisfies
  \begin{equation}
   |\lambda|\leq \frac{\kappa-1}{\kappa+1}\lambda_0.\label{eq_Hopf:theorem 4}
  \end{equation}
\end{thm}

\begin{rem}
  We note that the factor $(\kappa-1)/(\kappa+1)$ is the best possible if
  there is no further condition. For example.
  \begin{equation}
  M=\begin{pmatrix}
     \kappa&1\\
     1&\kappa
    \end{pmatrix}, \quad \kappa >0,
  \end{equation}
has eigenvalues $\lambda_0=\kappa+1$ and $\lambda=\kappa-1$.
\end{rem}

\begin{proof}
  [\bf Proof of Theorem \ref{Hopf theorem 2}]
  Let us consider two vectors $\Vp$, $\bar{\Vp}$ which are non-negative
  and unequal to $\Vzero$, and define
  \begin{equation}
   \Vp_{n+1}=M\Vp_n,\quad \bar{\Vp}_{n+1}=M\bar{\Vp}_n,\quad
    \Vp_0=\Vp,\quad \bar{\Vp}_0=\bar{\Vp}.
  \end{equation}
  From the hypothesis (\ref{eq_Hopf:strictly positive}), both $\Vp_n$ and
  $\bar{\Vp}_n$ are strictly positive for $n>0$. We find by repeated
  applications of Theorem \ref{Hopf theorem 1} that, for $n>1$,
  \begin{equation}
   \osc_i \frac{\bar{p}_{n,i}}{p_{n,i}}\leq\left(
    \frac{\kappa-1}{\kappa+1}\right)^{n-1}
   \osc_i\frac{\bar{p}_{1,i}}{p_{1,i}},
    \label{eq_Hopf:inequality for osc}
  \end{equation}
  where we used the notation $p_{n,i}=(\Vp_n)_i$. Consequently, there
  exists a finite constant $\lambda>0$, such that
  \begin{equation}
   \frac{\bar{p}_{n,i}}{p_{n,i}}\longrightarrow \lambda \qquad
    (n\longrightarrow \infty)
    \label{eq_Hopf:limit of ratio}
  \end{equation}
  for every $i$. We normalize the vectors $\Vp_n$, $\bar{\Vp}_n$ as
  \begin{equation}
   \Vq_n=\frac{\Vp_n}{p_{n,k}}, \quad
   \bar{\Vq}_n=\frac{\bar{\Vp}_n}{\bar{p}_{n,k}},
  \end{equation}
  with $k$ fixed. The hypothesis (\ref{eq_Hopf:upper bound}) implies that
  \begin{equation}
   \kappa^{-1}\leq q_{n,i}\leq \kappa,\quad
    \kappa^{-1}\leq \bar{q}_{n,i}\leq \kappa.\label{eq_Hopf:bound for q}
  \end{equation}
  Thus, we find that
  \begin{equation}
   \left|\bar{q}_{n,i}-q_{n,i}\right|
    =q_{n,i} \frac{p_{n,k}}{\bar{p}_{n,k}}
    \left|\frac{\bar{p}_{n,i}}{p_{n,i}}-\frac{\bar{p}_{n,k}}{p_{n,k}}\right|
    \leq \kappa \frac{p_{n,k}}{\bar{p}_{n,k}}
    \osc_i \frac{\bar{p}_{n,i}}{p_{n,i}}.\label{eq_Hopf:difference of q}
  \end{equation}

  Now we specialize to the case that $\bar{\Vp}=M\Vp=\Vp_1$, namely,
  \begin{equation}
   \bar{\Vp}_n=M\Vp_n=\Vp_{n+1}, \quad \bar{\Vq}_n=\Vq_{n+1}.
  \end{equation}
  Using (\ref{eq_Hopf:inequality for osc}) and (\ref{eq_Hopf:limit
  of ratio}), we estimate (\ref{eq_Hopf:difference of q}) for
  $q_{n+1,i}-q_{n,i}$, which implies that the sequence $\Vq_n$ converges
  to a limit vector $\Vq$. Because of (\ref{eq_Hopf:bound for q}),
  we have $\Vq>0$. Now (\ref{eq_Hopf:limit of ratio}) reads
  \begin{equation}
   \frac{p_{n+1,i}}{p_{n,i}}
    =\frac{(M\Vp_n)_i}{(\Vp_n)_i}=\frac{(M\Vq_n)_i}{(\Vq_n)_i}
    \longrightarrow \lambda_0.
  \end{equation}
  Consequently, $M\Vq=\lambda_0 \Vq$. For any other initial vector
  $\bar{\Vp}$, the sequence $\bar{\Vq}_n$ converges to the same limit as
  $\Vq_n$ because of (\ref{eq_Hopf:inequality for osc}),
  (\ref{eq_Hopf:limit of ratio}) and (\ref{eq_Hopf:difference of
  q}). Theorem \ref{Hopf theorem 2} is thereby proved.
\end{proof}

\begin{proof}
  [\bf Proof of Theorem \ref{Hopf theorem 3}]
  We assume that $\Vv\geq 0$, $\Vv\neq \Vzero$ is a solution of the
  eigenvalue equation (\ref{eq_Hopf:eigenvalue equation}). Since the
  hypothesis (\ref{eq_Hopf:strictly positive}) implies $M\Vv >0$, we have
  $\lambda>0$ and $\Vv >0$. We use this $\Vv$ as an initial vector $\Vp$
  in Theorem \ref{Hopf theorem 2} and apply the last part of this Theorem
  to
  \begin{equation}
   \frac{M^n\Vv}{(M^n\Vv)_k}=\frac{\lambda^n \Vv}{\lambda_n v_k}
    =\frac{\Vv}{v_k}.
  \end{equation}
  Hence, the limit $\Vq$ is equal to $\Vv/v_k$, that is, $\Vv=c\Vq$, and
  $\lambda=\lambda_0$. Therefore the first part of Theorem \ref{Hopf theorem
  3} is proved.

  Next, we take $\lambda_0>0$ and $\Vq >0$ from Theorem \ref{Hopf theorem
  2} and consider a solution of $M\Vv=\lambda\Vv$. The application of
  Theorem \ref{Hopf theorem 1} to $\Vq$ and $\Vv$ yields
  \begin{equation}
   \frac{|\lambda|}{\lambda_0}\osc_i\frac{v_i}{q_i}
    =\osc_i\frac{\lambda v_i}{\lambda_0 q_i}
    =\osc_i\frac{(M\Vv)_i}{(M\Vq)_i}
    \leq\frac{\kappa-1}{\kappa+1}\osc_i\frac{v_i}{q_i},
    \label{eq_Hopf:inequality in proof of theorem3}
  \end{equation}
  where we used (\ref{eq_Hopf:osc of complex constant}). If
  $\lambda=\lambda_0$, the above inequality implies that $\osc_i v_i/q_i
  =0$ or $\Vv=c\Vq$, which provides the second part of Theorem \ref{Hopf
  theorem 3}.
\end{proof}

\begin{proof}
  [\bf Proof of Theorem \ref{Hopf theorem 4}] We consider
  (\ref{eq_Hopf:inequality in proof of theorem3}). If $\lambda\neq
  \lambda_0$ and $\Vv\neq\Vzero$, $\Vv$ can not be equal to
  $c\Vq$. Therefore $\osc_i v_i/q_i> 0$, and then
  (\ref{eq_Hopf:inequality in proof of theorem3}) yields
  (\ref{eq_Hopf:theorem 4}).
\end{proof}

\section{Conditions for ergodicity}

In this Appendix, we prove Theorems \ref{theorem:weak-e} and
\ref{theorem:strong-e} which provide conditions for weak and strong
ergodicity of an inhomogeneous Markov chain \cite{AartsK}.

\subsection{Coefficient of ergodicity}
\label{sec_Ergo:coefficient of ergodicity}
Let us recall the definition of the coefficient of ergodicity
\begin{equation}
  \alpha(G)=1-\min_{x,y\in \mathcal{S}}\left\{
  \sum_{z\in\mathcal{S}}\min\{G(z,x),G(z,y)\}\right\}.
  \label{eq_Ergo:def of coefficient}
\end{equation}
First, we prove that this coefficient is rewritten as
\begin{equation}
  \alpha(G)=\frac{1}{2}\max_{x,y\in\mathcal{S}}\left\{
  \sum_{z\in\mathcal{S}} \left|G(z,x)-G(z,y)\right|\right\}.
  \label{eq_Ergo:coefficient of ergodicity}
\end{equation}

\begin{proof}
  [\bf Proof of (\ref{eq_Ergo:coefficient of ergodicity})]
  For fixed $x,y\in\mathcal{S}$, we define two subsets of $\mathcal{S}$ by
  \begin{gather}
   \begin{split}
    \mathcal{S}_G^+ =\{z\in\mathcal{S} \mid G(z,x)-G(z,y)> 0\},\\
   \mathcal{S}_G^- =\{z\in\mathcal{S} \mid G(z,x)-G(z,y)\leq 0\}.
   \end{split}
  \end{gather}
  Since the transition matrix satisfies $\sum_{y\in\mathcal{S}}G(y,x)=1$,
  we have
  \begin{align}
   \sum_{z\in\mathcal{S}_G^+}[G(z,x)-G(z,y)]
   &=\Bigl[1-\sum_{z\in\mathcal{S}_G^-}G(z,x)\Bigr]
   -\Bigl[1-\sum_{z\in\mathcal{S}_G^-}G(z,y)\Bigr]\nonumber\\
   &=-\sum_{z\in\mathcal{S}_G^-}[G(z,x)-G(z,y)] .
  \end{align}
  Thus, we find
  \begin{align}
   \frac{1}{2}\sum_{z\in\mathcal{S}}\left|G(z,x)-G(z,y)\right|
   &=\sum_{z\in\mathcal{S}_G^+}[G(z,x)-G(z,y)] \nonumber\\
   &=\sum_{z\in\mathcal{S}}\max\left\{0,G(z,x)-G(z,y)\right\}\nonumber\\
   &=\sum_{z\in\mathcal{S}}\left[G(z,x)-\min\{G(z,x),G(z,y)\}\right]\nonumber\\
   &=1-\sum_{z\in\mathcal{S}}\min\{G(z,x),G(z,y)\},
   \label{eq_Ergo:sum of dif G}
  \end{align}
  for any $x,y$. Hence taking the max with respect to $x,y$ on both
  sides, we obtain (\ref{eq_Ergo:coefficient of ergodicity}).
\end{proof}

To derive the conditions for weak and strong ergodicity, the following
Lemmas are useful.
\begin{lem}
  \label{lem_Ergo:bound for coefficient}
  Let $G$ be a transition matrix. Then the coefficient of ergodicity
  satisfies
  \begin{equation}
   0\leq \alpha(G)\leq 1.
  \end{equation}
\end{lem}
\begin{lem}
  \label{lem_Ergo:coefficient1}
  Let $G$ and $H$ be transition matrices on $\mathcal{S}$. Then the
  coefficient of ergodicity satisfies
  \begin{equation}
   \alpha(GH)\leq \alpha(G)\alpha(H).\label{eq_Ergo:coefficient1}
  \end{equation}
\end{lem}
\begin{lem}
  \label{lem_Ergo:coefficient2}
  Let $G$ be a transition matrix and $H$ be a square matrix on $\mathcal{S}$ such
  that
  \begin{equation}
   \sum_{z\in\mathcal{S}}H(z,x)=0,
  \end{equation}
  for any $x\in\mathcal{S}$. Then we have
  \begin{equation}
   \|GH\|\leq \alpha(G)\|H\|,
  \end{equation}
  where the norm of a square matrix defined by
  \begin{equation}
   \|A\|\equiv \max_{x\in\mathcal{S}}\left\{
   \sum_{z\in\mathcal{S}}|A(z,x)|\right\}.
  \end{equation}
\end{lem}

\begin{proof}
  [\bf Proof of Lemma \ref{lem_Ergo:bound for coefficient}] The
  definition of $\alpha(G)$ implies $\alpha(G)\leq 1$ because $G(y,x)\geq
  0$. From (\ref{eq_Ergo:coefficient of ergodicity}), $\alpha(G)\geq
  0$ is straightforward.
\end{proof}

\begin{proof}
  [\bf Proof of Lemma \ref{lem_Ergo:coefficient1}]

  Let us consider a transition matrix $G$, a column vector $a$ such that
  $\sum_{z\in\mathcal{S}}a(z)=0$, and their product $b=Ga$. We note that
  the vector $b$ satisfies $\sum_{z\in\mathcal{S}}b(z)=0$ because
  \begin{equation}
   \sum_{z\in\mathcal{S}}b(z)=\sum_{z\in\mathcal{S}}\sum_{y\in\mathcal{S}}
    G(z,y)a(y)=\sum_{y\in\mathcal{S}}a(y)
    \left[\sum_{z\in\mathcal{S}}G(z,y)\right]
    =\sum_{y\in\mathcal{S}}a(y)=0.
  \end{equation}
  We define subsets of $\mathcal{S}$ by
  \begin{gather}
   \begin{split}
    \mathcal{S}_a^+ =\{z\in\mathcal{S} \mid a(z)> 0\},\quad
    \mathcal{S}_a^- =\{z\in\mathcal{S} \mid a(z)\leq 0\}, \\
    \mathcal{S}_b^+ =\{z\in\mathcal{S} \mid b(z)> 0\},\quad
    \mathcal{S}_b^- =\{z\in\mathcal{S} \mid b(z)\leq 0\}.
   \end{split}
  \end{gather}
  Since $\sum_{z\in\mathcal{S}}a(z)=\sum_{z\in\mathcal{S}}b(z)=0$, we find
  \begin{gather}
   \sum_{z\in\mathcal{S}}\left|a(z)\right|
   =\sum_{z\in\mathcal{S}_z^+} a(z)
   -\sum_{z\in\mathcal{S}_z^-} a(z)
   =2\sum_{z\in\mathcal{S}_z^+} a(z)
   =-2\sum_{z\in\mathcal{S}_z^-} a(z), \\
   \sum_{z\in\mathcal{S}}\left|b(z)\right|
   =2\sum_{z\in\mathcal{S}_z^+} b(z)
   =-2\sum_{z\in\mathcal{S}_z^-} b(z).
  \end{gather}
  Therefore, we obtain
  \begin{align}
   \sum_{z\in\mathcal{S}}\left|b(z)\right|
   &=2\sum_{z\in\mathcal{S}_b^+}
    \sum_{u\in\mathcal{S}}G(z,u)a(u)\nonumber \\
   &=2\sum_{u\in\mathcal{S}_a^+}
    \left[\sum_{z\in\mathcal{S}_b^+}G(z,u)\right]a(u)
   +2\sum_{u\in\mathcal{S}_a^-}
    \left[\sum_{z\in\mathcal{S}_b^+}G(z,u)\right]a(u)\nonumber \\
   &\leq 2\max_{v\in\mathcal{S}}
    \left\{\sum_{z\in\mathcal{S}_b^+}G(z,v)\right\}
    \sum_{u\in\mathcal{S}_a^+}a(u)
   + 2\min_{w\in\mathcal{S}}
    \left\{\sum_{z\in\mathcal{S}_b^+}G(z,w)\right\}
    \sum_{u\in\mathcal{S}_a^-}a(u)\nonumber \\
   &=\max_{v,w\in\mathcal{S}}
    \left\{\sum_{z\in\mathcal{S}_b^+}[G(z,v)-G(z,w)]\right\}
    \sum_{u\in\mathcal{S}}\left|a(u)\right|\nonumber \\
   &\leq \max_{v,w\in\mathcal{S}}
    \left\{\sum_{z\in\mathcal{S}}\max\left\{0,G(z,v)-G(z,w)\right\}\right\}
    \sum_{u\in\mathcal{S}}\left|a(u)\right|\nonumber \\
   &=\frac{1}{2}\max_{v,w\in\mathcal{S}}
    \left\{\sum_{z\in\mathcal{S}}\left|G(z,v)-G(z,w)\right|\right\}
    \sum_{u\in\mathcal{S}}\left|a(u)\right|\nonumber \\
   &=\alpha(G)\sum_{u\in\mathcal{S}}\left|a(u)\right|,
   \label{eq_Ergo:upper bound for a}
  \end{align}
  where we used  (\ref{eq_Ergo:sum of dif G}) and
  (\ref{eq_Ergo:coefficient of ergodicity}).

  Next, we consider transition matrices $G, H$ and $F=GH$.  We take
  $a(z)=H(z,x)-H(z,y)$, and then (\ref{eq_Ergo:upper bound for a})
  is rewritten as
  \begin{equation}
   \sum_{z\in\mathcal{S}}|F(z,x)-F(z,y)|\leq \alpha(G)
    \sum_{u\in\mathcal{S}}|H(u,x)-H(u,y)|,
  \end{equation}
  for any $x,y$. Hence this inequality holds for the max of both sides
  with respect to $x,y$, which yields Lemma \ref{lem_Ergo:coefficient1}.
\end{proof}

\begin{proof}
  [\bf Proof of Lemma \ref{lem_Ergo:coefficient2}] Let us consider
  $F=GH$. We can take $a(z)=H(z,x)$ in (\ref{eq_Ergo:upper bound
  for a}) because of the assumption
  $\sum_{y\in\mathcal{S}}H(y,x)=0$. Thus we have
  \begin{equation}
   \sum_{z\in\mathcal{S}}|F(z,x)|\leq \alpha(G)
    \sum_{u\in\mathcal{S}}|H(u,x)|,
  \end{equation}
  for any $x$. Hence this inequality holds for the max of both sides with
  respect to $x$, which provides Lemma \ref{lem_Ergo:coefficient2}.
\end{proof}

\subsection{Conditions for weak ergodicity}

The following Theorem provides the reason why $\alpha(G)$ is called the
coefficient of ergodicity.

\begin{thm}
  \label{thm_ergo:weak erogidicity}
  An inhomogeneous Markov chain is weakly ergodic if and only if the
  transition matrix satisfies
  \begin{equation}
   \lim_{t\rightarrow\infty} \alpha\left(G^{t,s}\right)=0
    \label{eq_Ergo: limit alpha}
  \end{equation}
  for any $s>0$.
\end{thm}

\begin{proof}
  We assume that the inhomogeneous Markov chain generated by $G(t)$ is
  weakly ergodic. For fixed $x,y\in\mathcal{S}$, we define probability
  distributions by
  \begin{equation}
   p_x(z)=\begin{cases}
           1 & (z=x)\\
           0 & (\text{otherwise})
          \end{cases},\qquad
   p_y(z)=\begin{cases}
           1 & (z=y)\\
           0 & (\text{otherwise})
          \end{cases}.
  \end{equation}
  Since
  $p_x(t,s;z)=\sum_{u\in\mathcal{S}}G^{t,s}(z,u)p_x(u)=G^{t,s}(z,x)$ and
  $p_y(t,s;z)=G^{t,s}(z,y)$, we have
  \begin{align}
   \sum_{z\in\mathcal{S}}\left|G^{t,s}(z,x)-G^{t,s}(z,y)\right|
   &=\sum_{z\in\mathcal{S}}\left|p_x(t,s;z)-p_y(t,s;z)\right|\nonumber \\
   &\leq \sup\{\|p(t,s)-p'(t,s)\|\mid p_0, p'_0 \in\mathcal{P}\}.
  \end{align}
  Taking the max with respect to $x,y$ on the left-hand side, we obtain
  \begin{equation}
   2\alpha(G^{t,s})
    \leq \sup\{\|p(t,s)-p'(t,s)\|\mid p_0, p'_0 \in\mathcal{P}\}.
  \end{equation}
  Therefore the definition of weak ergodicity (\ref{eq:def of weak
  ergodicity}) yields (\ref{eq_Ergo: limit alpha}).

  We assume (\ref{eq_Ergo: limit alpha}). For fixed
  $p_0, q_0\in\mathcal{P}$, we define the transition probabilities by
  \begin{gather}
   H=(p_0, q_0,\cdots,q_0),\\
   F=G^{t,s}H=(p(t,s),q(t,s),\cdots,q(t,s)),
  \end{gather}
  where $p(t,s)=G^{t,s}p_0$, $q(t,s)=G^{t,s}q_0$. From
  (\ref{eq_Ergo:coefficient of ergodicity}), the coefficient of
  ergodicity for $F$ is rewritten as
  \begin{equation}
   \alpha(F)=\frac{1}{2}\sum_{z\in\mathcal{S}}\left|p(t,s;z)-q(t,s;z)\right|
    =\frac{1}{2}\|p(t,s)-q(t,s)\|.
  \end{equation}
  Thus Lemmas \ref{lem_Ergo:bound for coefficient} and
  \ref{lem_Ergo:coefficient1} yield
  \begin{equation}
   \|p(t,s)-q(t,s)\|\leq 2\alpha(G^{t,s})\alpha(H)\leq 2\alpha(G^{t,s}).
  \end{equation}
  Taking the sup with respect to $p_0, q_0\in\mathcal{S}$ and the limit
  $t\rightarrow\infty$, we obtain
  \begin{equation}
   \lim_{t\rightarrow\infty}
    \sup\{\|p(t,s)-q(t,s)\|\mid p_0, q_0 \in\mathcal{P}\}
    \leq 2\lim_{t\rightarrow\infty} \alpha(G^{t,s})=0,
  \end{equation}
  for any $s>0$. Therefore the inhomogeneous Markov chain generated by
  $G(t)$ is weakly ergodic.
\end{proof}

Next, we prove Theorem \ref{theorem:weak-e}. For this purpose, the
following Lemma is useful.
\begin{lem}
  \label{lem_Ergo:analysis}
  Let $a_0, a_1,\cdots, a_n,\cdots$ be a sequence such that $0\leq a_i< 1$
  for any $i$.
  \begin{equation}
   \sum_{i=0}^{\infty}a_i=\infty\quad \Longrightarrow \quad
    \prod_{i=n}^\infty (1-a_i)=0.\label{eq_Ergo:analysis}
  \end{equation}
\end{lem}
\begin{proof}
  Since $0\leq 1-a_i\leq {\rm e}^{-a_i}$, we have
  \begin{equation}
   0\leq \prod_{i=n}^{m}(1-a_i)\leq \prod_{i=n}^{m}{\rm e}^{-a_i}
    \leq \exp\left(-\sum_{i=n}^{m}a_i\right).
  \end{equation}
  In the limit $m\rightarrow\infty$, the right-hand side converges to
  zero because of the assumption
  $\sum_{i=0}^{\infty}a_i=\infty$. Therefore we obtain
  (\ref{eq_Ergo:analysis}).
\end{proof}

\begin{proof}
  [\bf Proof of Theorem \ref{theorem:weak-e}] 
  We assume that the inhomogeneous Markov chain generated by
  $G(t)$ is weakly ergodic. Theorem \ref{thm_ergo:weak erogidicity}
  yields
  \begin{equation}
   \lim_{t\rightarrow\infty} \left[1-\alpha(G^{t,s})\right]=1
  \end{equation}
  for any $s>0$. Thus, there exists $t_1$ such that
  $1-\alpha(G^{t_1,t_0})>1/2$ with $t_0=s$. Similarly, there exists
  $t_{n+1}$ such that $1-\alpha(G^{t_{n+1},t_n})>1/2$ for any
  $t_n>0$. Therefore,
  \begin{equation}
   \sum_{i=0}^{n}\left[1-\alpha(G^{t_{i+1},t_i})\right]>\frac{1}{2}(n+1).
  \end{equation}
  Taking the limit $n\rightarrow\infty$, we obtain (\ref{eq:theorem
  for weak-e}).

We assume (\ref{eq:theorem for weak-e}). Lemma
  \ref{lem_Ergo:analysis} yields
  \begin{equation}
   \prod_{i=n}^\infty
    \left\{1-\left[1-\alpha(G^{t_{i+1},t_i})\right]\right\}
    = \prod_{i=n}^\infty \alpha(G^{t_{i+1},t_i})=0.
    \label{eq_Ergo:prod_alpha}
  \end{equation}
  For fixed $s$ and $t$ such that $t>s\geq 0$, we define $n$ and $m$ 
by $t_{n-1}\leq
  s<t_{n}$, $t_{m}<t\leq t_{m+1}$. Thus, from Lemma
  \ref{lem_Ergo:coefficient1}, we obtain
  \begin{align}
   \alpha(G^{t,s})
   &\leq \alpha(G^{t,t_m})\alpha(G^{t_m,t_{m-1}})\cdots
    \alpha(G^{t_{n+1},t_n})\alpha(G^{t_n,s}) \nonumber \\
   &=\alpha(G^{t,t_m})\left[\prod_{i=n}^m \alpha(G^{t_{i+1},t_i})\right]
   \alpha(G^{t_n,s}).
  \end{align}
  In the limit $t\rightarrow\infty$, $m$ goes to infinity and then the
  right-hand side converges to zero because of
  (\ref{eq_Ergo:prod_alpha}). Thus we have
  \begin{equation}
   \lim_{t\rightarrow\infty} \alpha(G^{t,s})=0,
  \end{equation}
  for any $s$. Therefore, from Theorem \ref{thm_ergo:weak erogidicity},
the inhomogeneous Markov chain generated by $G(t)$ is weakly ergodic.
\end{proof}

\subsection{Conditions for strong ergodicity}
\label{Conditions for strong ergodicity}

The goal of this section is to give the proof of Theorem
\ref{theorem:strong-e}. Before that, we prove the following Theorem,
which also provides the sufficient condition for strong ergodicity.

\begin{thm}
  \label{them_Ergo:strong ergodicity}
  An inhomogeneous Markov chain generated by $G(t)$ is strongly ergodic
  if there exists the transition matrix $H$ on $\mathcal{S}$ such that
  $H(z,x)=H(z,y)$ for any $x,y,z\in\mathcal{S}$ and
  \begin{equation}
   \lim_{t\rightarrow\infty}\left\|G^{t,s}-H\right\|=0
    \label{eq_Ergo:assumption for G-H}
  \end{equation}
  for any $s>0$.
\end{thm}

\begin{proof}
We consider
  $p_0\in\mathcal{P}$ and $p(t,s)=G^{t,s}p_0$.  For fixed
  $u\in\mathcal{S}$, we define a probability distribution $r$ by
  $r(z)=H(z,u)$. We find
  \begin{align}
   \left\|p(t,s)-r\right\|
   &=\sum_{z\in\mathcal{S}}\left|\sum_{x\in\mathcal{S}}G^{t,s}(z,x)p_0(x)
   -H(z,u)\right| \nonumber \\
   &=\sum_{z\in\mathcal{S}}\left|\sum_{x\in\mathcal{S}}\left[
   G^{t,s}(z,x)-H(z,u)\right]p_0(x)\right| \nonumber \\
   &\leq \sum_{z\in\mathcal{S}}\sum_{x\in\mathcal{S}}
   \left|G^{t,s}(z,x)-H(z,u)\right|
   =\sum_{z\in\mathcal{S}}\sum_{x\in\mathcal{S}}
   \left|G^{t,s}(z,x)-H(z,x)\right| \nonumber \\
   &\leq \sum_{x\in\mathcal{S}}\left\|G^{t,s}-H\right\|
   =\left|\mathcal{S}\right|\left\|G^{t,s}-H\right\|.
  \end{align}
  Taking the sup with respect to $p_0\in\mathcal{P}$ and using the
  assumption (\ref{eq_Ergo:assumption for G-H}), we obtain
  \begin{equation}
   \lim_{t\rightarrow\infty} \sup\left\{ \left\|p(t,s)-r\right\|\mid
    p_0\in\mathcal{P}\right\}=0.
  \end{equation}
  Therefore, the inhomogeneous Markov chain generated by $G(t)$ is
  strongly ergodic.
\end{proof}

\begin{proof}
  [\bf Proof of Theorem \ref{theorem:strong-e}] We assume that the three
  conditions in Theorem \ref{theorem:strong-e} hold.  Since the
  condition 3 is rewritten as
  \begin{equation}
   \sum_{x\in\mathcal{S}}\sum_{t=0}^{\infty}
    \left|p_t(x)-p_{t+1}(x)\right|
    =\sum_{t=0}^{\infty}\left\|p_t-p_{t+1}\right\|<\infty,
  \end{equation}
  we have
  \begin{equation}
   \sum_{t=0}^{\infty}\left|p_t(x)-p_{t+1}(x)\right|<\infty
  \end{equation}
  for any $x\in\mathcal{S}$. Thus, the stationary state $p_t$ converges
  to $p=\lim_{t\rightarrow\infty}p_t$. Now, let us define a transition
  matrices $H$ and $H(t)$ by $H(z,x)=p(z)$ and $H(z,x;t)=p_t(z)$,
  respectively.  For $t>u>s\geq 0$,
  \begin{equation}
   \begin{split}
    \left\|G^{t,s}-H\right\|
    &\leq \left\|G^{t,u}G^{u,s}-G^{t,u}H(u)\right\| \\
    &\qquad+\left\|G^{t,u}H(u)-H(t-1)\right\|
    +\left\|H(t-1)-H \right\|.
   \label{eq_Ergo:G-H}
   \end{split}
  \end{equation}
  Thus, we evaluate each term on the right-hand side and show that
  (\ref{eq_Ergo:assumption for G-H}) holds.

  [{\it 1st term}] Lemma \ref{lem_Ergo:coefficient2} yields that
  \begin{align}
   \left\|G^{t,u}G^{u,s}-G^{t,u}H(u)\right\|
    &\leq \alpha (G^{t,u}) \left\|G^{u,s}-H(u)\right\| \nonumber \\
    &\leq 2\alpha (G^{t,u}),
  \end{align}
  where we used $\left\|G^{u,s}-H(u)\right\| \leq 2$. Since the Markov
  chain is weakly ergodic (condition 1), Theorem \ref{thm_ergo:weak
  erogidicity} implies that
  \begin{equation}
   \forall \varepsilon>0, \exists t_1>0, \forall t>t_1:
    \left\|G^{t,u}G^{u,s}-G^{t,u}H(u)\right\|
    <\frac{\varepsilon}{3}.
    \label{eq_Ergo:1st term}
  \end{equation}

  [{\it 2nd term}] Since $p_t =G(t) p_t$ (condition 2), we find
  \begin{equation}
   H(u) = G(u)H(u)=G^{u+1,u}H(u)
  \end{equation}
  and then
  \begin{equation}
   G^{t,u}H(u)=G^{t,u+1} H(u)
    =G^{t,u+1}\left[H(u)-H(u+1) \right] +G^{t,u+1} H(u+1).
  \end{equation}
  The last term on the right-hand side of the above equation is similarly
  rewritten as
  \begin{equation}
   G^{t,u+1}H(u+1)
    =G^{t,u+2}\left[H(u+1)-H(u+2) \right] +G^{t,u+2} H(u+2).
  \end{equation}
  We recursively apply these relations and obtain
  \begin{align}
   G^{t,u}H(u)&=\sum_{v=u}^{t-2}G^{t,v+1}\left[H(v)-H(v+1)\right]
    +G^{t,t-1}H(t-1) \nonumber \\
   &=\sum_{v=u}^{t-2}G^{t,v+1}\left[H(v)-H(v+1)\right]
    +H(t-1).
  \end{align}
  Thus the second term in (\ref{eq_Ergo:G-H}) is rewritten as
  \begin{align}
   \left\|G^{t,u}H(u)-H(t-1)\right\|
   &=\left\|\sum_{v=u}^{t-2}G^{t,v+1}\left[H(v)-H(v+1)\right]\right\|
    \nonumber \\
   &\leq\sum_{v=u}^{t-2} \left\|G^{t,v+1}\left[H(v)-H(v+1)\right]\right\|.
  \end{align}
  Lemmas \ref{lem_Ergo:bound for coefficient} and
  \ref{lem_Ergo:coefficient2} yield that
  \begin{equation}
   \left\|G^{t,v+1}\left[H(v)-H(v+1)\right]\right\|
    \leq \left\|H(v)-H(v+1)\right\|
   =\left\| p_v-p_{v+1}\right\|,
  \end{equation}
  where we used the definition of $H(t)$. Thus we obtain
  \begin{equation}
   \left\|G^{t,u}H(u)-H(t-1)\right\|
    \leq\sum_{v=u}^{t-2}\left\| p_v-p_{v+1}\right\|.
  \end{equation}
  Since $\sum_{t=0}^{\infty} \left\|p_t-p_{t+1}\right\|<\infty$
  (condition 3), for all $\varepsilon>0$, there exists $t_2>0$ such that
  \begin{equation}
   \forall t>\forall u \geq t_2:
    \sum_{v=u}^{t-2}\left\| p_v-p_{v+1}\right\|
    <\frac{\varepsilon}{3}.
  \end{equation}
  Therefore
  \begin{equation}
   \forall\varepsilon>0, \exists t_2>0, \forall t>\forall u \geq t_2:
    \left\|G^{t,u}H(u)-H(t-1)\right\| < \frac{\varepsilon}{3}.
    \label{eq_Ergo:2nd term}
  \end{equation}

  [{\it 3rd term}] From the definitions of $H$ and $H(t)$, they clearly
  satisfy
  \begin{equation}
   \lim_{t\rightarrow \infty}\left\|H(t)-H\right\|=0,
  \end{equation}
  which implies that
  \begin{equation}
   \forall\varepsilon>0, \exists t_3>0, \forall t>t_3:
    \|H(t-1)-H\|<\frac{\varepsilon}{3}.
    \label{eq_Ergo:3rd term}
  \end{equation}\medskip

  Consequently, substitution of (\ref{eq_Ergo:1st term}),
  (\ref{eq_Ergo:2nd term}) and (\ref{eq_Ergo:3rd term}) into
  (\ref{eq_Ergo:G-H}) yields that
  \begin{equation}
   \left\|G^{t,s}-H\right\|<\frac{\varepsilon}{3}
    +\frac{\varepsilon}{3}+\frac{\varepsilon}{3}<\varepsilon,
  \end{equation}
  for all $t>\max\{t_1,t_2,t_3\}$.
  Since $\varepsilon$ is arbitrarily small, (\ref{eq_Ergo:assumption
  for G-H}) holds for any $s>0$ and then the given Markov chain is
  strongly ergodic from Theorem \ref{them_Ergo:strong ergodicity}, which
  completes the proof of the first part of Theorem
  \ref{theorem:strong-e}.\\

  Next, we assume $p=\lim_{t\rightarrow \infty}p_t$. For any distribution
  $q_0$, we have $H q_0=p$ because
  \begin{equation}
   \sum_{x\in\mathcal{S}}H(z,x)q_0(x)=p(z)\sum_{x\in\mathcal{S}}q_0(x)
    =p(z).
  \end{equation}
  Thus, we obtain
  \begin{align}
   \|q(t,t_0)-p\|&=\left\|\left(G^{t,t_0}-H\right) q_0\right\|
   \leq \left\|G^{t,t_0}-H\right\|.
  \end{align}
  Hence it holds for the sup with respect to $q_0 \in \mathcal{P}$, which
  yields (\ref{eq:SE}) in the limit of $t\rightarrow\infty$. Theorem
  \ref{theorem:strong-e} is thereby proved.
\end{proof}


\begin{thebibliography}{99}
  \bibitem{GareyJ} M.~R.~Garey and D.~S.~Johnson: {\it Computers and
          Intractability: A Guide to the Theory of NP-Completeness}
          (Freeman, San Francisco, 1979)
  \bibitem{HartmannW} A.~K.~Hartmann and M.~Weigt: {\it Phase Transitions
          in Combinatorial Optimization Problems: Basics, Algorithms and
          Statistical Mechanics} (Wiley-VCH, Weinheim, 2005)
  \bibitem{Helsgaun}K.~Helsgaun: Euro. J. Op. Res. {\bf 126} (2000) 106.
  \bibitem{KirkpatrickGV} S.~Kirkpatrick, S.~D.~Gelett and M.~P.~Vecchi:
          Science {\bf 220} (1983) 671
  \bibitem{AartsK} E.~Aarts and J.~Korst: {\it Simulated Annealing and
          Boltzmann Machines: A Stochastic Approach to Combinatorial
          Optimization and Neural Computing} (Wiley, New York, 1984)
  \bibitem{Finnila}A. B. Finnila, M. A. Gomez, C. Sebenik, S. Stenson,
  and J. D. Doll: Chem. Phys. Lett. {\bf 219} (1994) 343
  \bibitem{KadowakiN} T.~Kadowaki and H.~Nishimori: Phys. Rev. E {\bf 58}
          (1998) 5355
  \bibitem{Kadowaki} T.~Kadowaki: {\it Study of Optimization Problems by
          Quantum Annealing} (Thesis, Tokyo Institute of
          Technology, 1999); quant-ph/0205020
  \bibitem{DasC} A.~Das and B.~K.~Charkrabarti: {\it Quantum Annealing
          and Related Optimization Methods} (Springer, Berlin,
          Heidelberg, 2005) Lecture Notes in Physics, Vol. 679
  \bibitem{SantoroT} G.~E.~Santoro and E.~Tosatti: J. Phys. A {\bf 39}
          (2006) R393
  \bibitem{DasC08} A. Das and B. K. Chakrabarti: arXiv:0801.2193
          (to be published in Rev. Mod. Phys.).
  \bibitem{ApolloniCdF1989} B.~Apolloni, C.~Carvalho and D.~de~Falco:
          Stoch. Proc. Appl. {\bf 33} (1989) 233
  \bibitem{ApolloniCdF1990} B.~Apolloni, N.~Cesa-Bianchi and D.~de~Falco:
          in {\it Stochastic Processes, Physics and Geometry},
          eds. S.~Albeverio {et al}. (World Scientific, Singapore, 1990) 97
  \bibitem{SantoroMTC} G.~E.~Santoro, R.~Marto\v{n}\'{a}k, E.~Tosatti and
          R.~Car: Science {\bf 295} (2002) 2427
  \bibitem{MartonakST2002} R.~Marto\v{n}\'{a}k, G.~E.~Santoro and
          E.~Tosatti: Phys. Rev. B {\bf 66} (2002) 094203
  \bibitem{SuzukiO} S.~Suzuki and M.~Okada: J. Phys. Soc. Jpn. {\bf 74}
          (2005) 1649
  \bibitem{SarjalaPA} M.~Sarjala, V.~Pet\"{a}j\"{a} and M.~Alava:
          J. Stat. Mech. (2006) P01008
  \bibitem{SuzukiNS} S.~Suzuki, H.~Nishimori, and M.~Suzuki:
          Phys. Rev. E {\bf 75} (2007) 051112
  \bibitem{MartonakST2004} R.~Marto\v{n}\'{a}k, G.~E.~Santoro and
          E.~Tosatti: Phys. Rev. E {\bf 70} (2004) 057701
  \bibitem{StellaST2005} L.~Stella, G.~E.~Santoro and E.~Tosatti:
          Phys. Rev. B {\bf 72} (2005) 014303
  \bibitem{StellaST2006} L.~Stella, G.~E.~Santoro and E.~Tosatti:
          Phys. Rev. B {\bf 73} (2006) 144302
  \bibitem{DasCS} A.~Das, B.~K.~Chakrabarti and R.~B.~Stinchcombe:
          Phys. Rev. E {\bf 72} (2005) 026701
  \bibitem{Trotter} H.~F.~Trotter: Proc. Am. Math. Soc. {\bf 10} (1959) 545
  \bibitem{Suzuki} M.~Suzuki: Prog. Theor. Phys. {\bf 46} (1971) 1337
  \bibitem{LandauB} D.~P.~Landau and K.~Binder: {\it A Guide to Monte
          Carlo Simulations in Statistical Physics} (Cambridge, Cambridge
          University Press, 2000) Chap. 8
  \bibitem{FarhiGGS} E.~Farhi, J.~Goldstone, S.~Gutomann and M.~Sipser:
          quant-ph/0001106
  \bibitem{Mizel} A. Mizel, D. A. Lidar and M. Mitchel: Phys. Rev. Lett.
          {\bf 99} (2007) 070502.
  \bibitem{MoritaThesis} S. Morita: {\it Analytic Study of
          Quantum Annealing} (Thesis, Tokyo Institute of
          Technology, 2008).
  \bibitem{MoritaN07} S. Morita and H. Nishimori: J. Phys. Soc. Jpn. {\bf 76}
          (2007) 064002.
  \bibitem{Messiah} A.~Messiah: {\it Quantum Mechanics} (Wiley, New York,
          1976)
  \bibitem{Somma} R.~D.~Somma, C.~D.~Batista, and G.~Ortiz:
          Phys. Rev. Lett. {\bf 99} (2007) 030603
  \bibitem{Hopf} E.~Hopf: J. Math. Mech. {\bf 12} (1963) 683
  \bibitem{GemanG} S.~Geman and D.~Geman: IEEE Trans. Pattern
          Anal. Mach. Intell. {\bf PAMI-6} (1984) 721
  \bibitem{NishimoriI} H.~Nishimori and J.~Inoue: J. Phys. A:
          Math. Gen. {\bf 31} (1998) 5661
  \bibitem{NishimoriNono} H.~Nishimori and Y.~Nonomura: J. Phys. Soc. Jpn.
          {\bf 65} (1996) 3780
  \bibitem{Seneta} E. Seneta: {\it Non-negative Matrices and Markov
          Chains} (Springer, New York, 2006)
  \bibitem{Morita} S. Morita, J. Phys. Soc. Jpn. {\bf 76} (2007) 104001


  \bibitem{LandauL} L.~D.~Landau and E.~M.~Lifshitz: {\it Quantum
          Mechanics: Non-Relativistic Theory} (Pergamon Press, Oxford,
          1965)
  \bibitem{Zener} C.~Zener: Proc. R. Soc. London Ser. A {\bf 137} (1932)
          696
  \bibitem{MoritaN06} S. Morita and H. Nishimori, J. Phys. A: Math. and Gen.
         {\bf 39} (2006) 13903
  \bibitem{NRecipes} H.~W.~Press, A.~S.~Tuekolosky, T.~W.~Vettering and
          P.~B.~Flannery: {\it Numerical Recipes in C} (Cambridge
          University Press, Cambridge, 1992) 2nd ed.
  \bibitem{Grover} L.~K.~Grover: Phys. Rev. Lett. {\bf 79} (1997) 325
  \bibitem{RolandC} J.~Roland and N.~J.~Cerf: Phys. Rev. A {\bf 65}
          (2002) 042308
  \bibitem{TsallisS} C.~Tsallis and D.~A.~Stariolo: Physica A {\bf 233}
          (1996) 395
  \bibitem{CeperleyA} D.~M.~Ceperley and B.~J.~Alder:
          Phys. Rev. Lett. {\bf 45} (1980) 566
  \bibitem{TrivediC} N.~Trivedi and D.~M.~Ceperley: Phys. Rev. B {\bf 41}
          (1990) 4552
  \bibitem{StellaS2007} L.~Stella and G.~E.~Santoro: Phys. Rev. E {\bf
          75} (2007) 036703
\end{thebibliography}
\end{document}